\documentclass[superscriptaddress,amsmath,amssymb, aps, prx,longbibliography,twocolumn, footinbib]{revtex4-2}

\usepackage{upgreek}
\usepackage{color}
\usepackage{graphicx}
\usepackage{nccmath}
\usepackage{bm}
\usepackage{hyperref}
\hypersetup{colorlinks,breaklinks,
            urlcolor=[rgb]{0,0,0.36},
            linkcolor=[rgb]{0,0,0.36},
            citecolor=[rgb]{0,0,0.36}}
\usepackage[mathlines]{lineno}
\usepackage{dcolumn}
\usepackage{mathtools}  
\usepackage{siunitx}    
\usepackage{dsfont}
\usepackage{extarrows}
\usepackage{epsfig,tabularx,multirow,booktabs}

\usepackage{braket}
\usepackage{chemformula}

\usepackage[normalem]{ulem}

\definecolor{cadmiumgreen}{rgb}{0.0, 0.42, 0.24} 

\usepackage{pifont} 

\usepackage{adjustbox} 
\usepackage{tabularx} 
\newcolumntype{Y}{>{\centering\arraybackslash}X} 

\newcounter{SN}

\usepackage{array}
\newcolumntype{M}[1]{>{\centering\arraybackslash}m{#1}}
\makeatletter
\renewcommand\footnotesize{%
   \@setfontsize\footnotesize\@ixpt{8}%
   \abovedisplayskip 8\p@ \@plus2\p@ \@minus4\p@
   \abovedisplayshortskip \z@ \@plus\p@
   \belowdisplayshortskip 4\p@ \@plus2\p@ \@minus2\p@
   \def\@listi{\leftmargin\leftmargini
               \topsep 4\p@ \@plus2\p@ \@minus2\p@
               \parsep 2\p@ \@plus\p@ \@minus\p@
               \itemsep \parsep}%
   \belowdisplayskip \abovedisplayskip
}
\makeatother

\newcommand{\Harvard}{Department of Physics, Harvard University, Cambridge, MA 02138, USA.}

\newcommand{\MIT}{Department of Physics, Massachusetts Institute of Technology, Cambridge, MA 02139, USA.}

\newcommand{\MITEECS}{Department of Electrical Engineering, Massachusetts Institute of Technology, Cambridge, MA 02139, USA.}

\newcommand{\MaxPO}{Max Planck Institute of Quantum Optics, Hans-Kopfermann-Str. 1, 85748 Garching, Germany.}

\newcommand{\ETH}{Institute for Theoretical Physics, ETH Zurich, 8093 Zurich, Switzerland.}

\newcommand{\Princeton}{Department of Physics, Princeton University, Princeton, NJ 08544, USA.}

\begin{document}

\title{Accelerating analysis of Boltzmann equations using Gaussian mixture models:
Application to quantum Bose-Fermi mixtures}

\author{Pavel~E.~Dolgirev}\email[Correspondence to: ]{p\_dolgirev@g.harvard.edu}
\affiliation{\Harvard}
\author{Kushal~Seetharam}
\affiliation{\Harvard}
\affiliation{\MITEECS}
\author{M\'arton Kan\'asz-Nagy}
\affiliation{\MaxPO}
\author{Carsten~Robens}
\affiliation{\MIT}
\author{Zoe~Z.~Yan}
\affiliation{\Princeton}
\author{Martin~Zwierlein }
\affiliation{\MIT}
\author{Eugene~Demler}%
\affiliation{\ETH}

\begin{abstract}
    The Boltzmann equation is a powerful theoretical tool for modeling the collective dynamics of quantum many-body systems subject to external perturbations. Analysis of the equation gives access to linear response properties including collective modes and transport coefficients, but often proves intractable due to computational costs associated with multidimensional integrals describing collision processes. 
    Here, we present a method to resolve this bottleneck, enabling the study of a broad class of many-body systems that appear in fundamental science contexts and technological applications.
    Specifically, we demonstrate that a Gaussian mixture model can accurately represent equilibrium distribution functions, thereby allowing efficient evaluation of collision integrals.
    Inspired by cold atom experiments, we apply this method to investigate the collective behavior of a quantum Bose-Fermi mixture of cold atoms in a cigar-shaped trap, a system that is particularly challenging to analyze. We focus on monopole and quadrupole collective modes above the Bose-Einstein transition temperature, and find a rich phenomenology that spans interference effects between bosonic and fermionic collective modes, dampening of these modes, and the emergence of hydrodynamics in various parameter regimes. These effects are readily verifiable experimentally.
\end{abstract}

 \maketitle 

\section{Introduction}

The Boltzmann equation is ubiquitous in modern physics as it can model the non-equilibrium dynamics and near-equilibrium collective behavior of a wide range of quantum many-body systems. Phenomena such as hydrodynamics and turbulence in quantum liquids as well as transport properties in metals and semiconductors fall under its purview. 
Valid for systems where the length scales characterizing quasiparticle interactions are shorter than typical distances quasiparticles travel before colliding~\cite{Kardar2007particles}, the Boltzmann equation simplifies the typically challenging analysis of such interacting many-particle systems by accurately capturing the many-body dynamics with the evolution of single-particle distribution functions.
When studying the near-equilibrium dynamics of a system, this approach gives efficient access to its linear-response properties which characterize the possible quantum phases of matter. Solving the Boltzmann equation, even after it is linearized around equilibrium, however, can often be computationally challenging and precludes analysis of a system. Approaches to the problem must contend with non-Gaussian equilibrium distribution functions encoding quantum statistics as well as quantum correlations in the self-energies such as those arising from particle-exchange processes~\cite{Kardar2007particles,pitaevskii2012physical}. Even with modern computational resources, state-of-the-art techniques grant limited insight into scientifically interesting phenomenona such as spin transport in magnetic materials and electron hydrodynamics in 2D van der Waals (vdW) semiconductors. Existing solution methods also struggle to accurately compute transport properties in technologically relevant systems such as lithium-ion batteries~\cite{Morgan2022} and photovoltaic cells~\cite{Jermyn2019}.

\begin{figure*}[t!]
\centering
\includegraphics[width=0.7\linewidth]{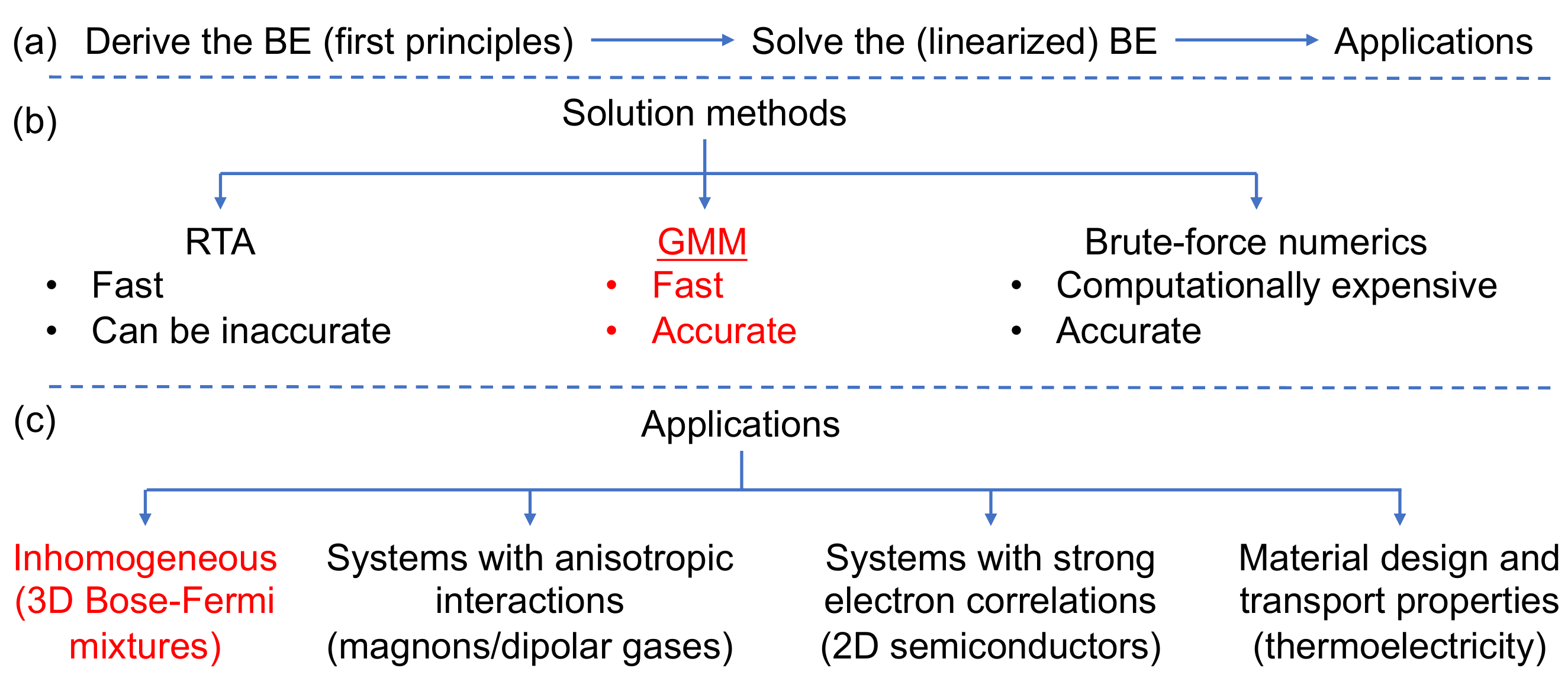} 
\caption{ Overview of the linearized Boltzmann equation and its applications. In this work, we develop a solution framework based on Gaussian mixture models (GMM) which enables efficient and accurate analysis of the Boltzmann equation. We demonstrate the method in the context of trapped Bose-Fermi mixtures of cold atoms -- see Fig.~\ref{fig::cartoon}.
}
\label{fig::BE_schematic}
\end{figure*}

State-of-the-art approaches make different levels of approximations to simplify the numerical task. The test particle method~\cite{PhysRevA.66.033606,griffin2009bose} assumes that the dynamics of distribution functions of a many-body system composed of around $10^{23}$ particles can be accurately estimated by sampling the trajectories of a much smaller number of $10^4\textrm{-}10^6$ particles. This is typically a safe approximation, but simulation of the system can remain challenging as collisions in the system correlate the trajectories and prevent their parallel computation. The method easily admits details of experimental protocols, but the computational cost often limits analysis of linear response properties as only a few parameter choices can be evaluated. Other methods also solve the Boltzmann equation numerically after simplifying the collision integrals via various assumptions~\cite{Vool2021, Varnavides2022, Wang2022}, but are limited in scope to where those assumptions are valid. The method of moments (MoM)~\cite{chapman1990mathematical,ziman2001electrons} takes an alternative approach to direct numerical simulation of the dynamics by variationally approximating the response of the system in the frequency domain. The MoM naturally captures collective behavior and, at the cost of some analytical work, reduces the problem to matrix inversion of a semi-analytical solution. It can, however, require additional work to incorporate certain experimental details such as complicated drive protocols. The computational bottleneck of the method comes from evaluating the matrix elements of the multidimensional collision integrals. Estimating these matrix elements via Monte Carlo sampling is often feasible for many 2D systems, albeit with difficulty, but fails for 3D systems.

In this work, we present two principal results. Our first result is the development of an efficient approach to analyze the linear response dynamics of many-body systems without resorting to commonly used relaxation time approximations. This method allows one to keep track of the exact collision integrals in Boltzmann equations and treat both 2D and 3D systems. Our second result leverages the above method to present an analysis of collective modes in quantum Bose-Fermi mixtures that are under active experimental investigation in ultracold atom systems~\cite{Yan2023}. We discover a surprising richness in dynamical regimes that this system manifests as the interactions between particles are tuned.

The method we develop allows for efficient computation of collision integrals, which we combine with the MoM framework, thus removing the bottleneck to solving Boltzmann equations. We term our approach GMM-MoM as it centers around using a Gaussian mixture model (GMM) representation of the equilibrium distribution functions around which the Boltzmann equation is linearized. We demonstrate that this representation provides an accurate approximation of the distribution functions and,  critical to efficient solution of the Boltzmann equation, enables semi-analytical computation of the multidimensional collision integrals.
This latter step is possible as the MoM basis functions are chosen to be polynomials which, when combined with a Gaussian integral kernel, are analytically computable via Wick's theorem~\cite{Kardar2007particles}. 
The efficiency of our approach results from the insight that both Bose-Einstein and Fermi-Dirac equilibrium distribution functions can be well approximated with a GMM consisting of only a few Gaussians (see Sec.~\ref{sec:GMM} for details). 
Our GMM-MoM approach enables efficient analysis of the linear response dynamics of a large range of 
systems that can be described using Boltzmann equations, thus allowing their characterization through quantities such as properties of collective modes and transport coefficients.
We release our fitting codebase online~\cite{GMM-MoM_codebase} and include tables of Gaussian fits to the equilibrium distribution functions in a broad range of parameter regimes; these fits are generic and can be directly applied to investigate a broad class of quantum many-body systems.

\subsection{Practical advantage and relevance of the GMM framework}

Typical approaches to solving the (linearized) Boltzmann equation either resort to approximations that make the analysis tractable at the cost of oversimplification, or employ brute-force numerical methods that are computationally costly and provide only limited information. By rendering multidimensional collision integrals efficiently computable, our GMM approach enables both fast and accurate analysis of the Boltzmann equation, thereby allowing investigation of broad classes of systems that have hereto remained intractable -- see Fig.~\ref{fig::BE_schematic}.

A ubiquitous approach to simplifying Boltzmann equations is the relaxation time approximation (RTA), often used to analyse transport properties of materials. This approximation assumes that all quantities of interest in the system relax to thermal equilibrium on the same time scale.
While the RTA can capture the behavior of relatively simple systems such as bulk crystalline silicon~\cite{Esfarjani2011}, it is often quantitatively inaccurate even in these contexts~\cite{Mingo2014}. The approximation also fails to  qualitatively describe the physics of 2D semiconductors~\cite{Cepellotti2016,Romano2020}, which foster both fundamental~\cite{Fritz2008,Muller2008,Kryhin2023a,Kryhin2023b} and technological interest~\cite{Novoselov2016}. A variety of other important systems -- including high-$T_c$ superconductors, resonantly interacting particles in neutron matter, and ultracold atoms at unitarity -- are also beyond the scope of RTA as these systems are expected to exhibit a clear separation of relaxation time scales for quantities such as heat, charge, spin, and momentum. The GMM framework introduced in this work obviates the need for the RTA while preserving computational efficiency. As a concrete demonstration of this capability, we benchmark the method on strongly-interacting systems in Appendix~\ref{appendix: benchmarking}. There, we compute the decay rate of a Bose polaron in the unitary limit and the viscosity of a two-component strongly-interacting Fermi fluid.

Two additional simplifications are commonly applied to the Boltzmann equation: the assumption of isotropic scattering and the assumption of position-independent collisions. The former precludes, for instance, analysis of magnetic systems, including simple ferromagnets, as magnon collision processes explicitly depend on the momenta of the scattering particles~\cite{Dyson1956}. The latter is strictly applicable only to spatially homogeneous systems, which precludes systems such as trapped gases, including dipolar atoms, that manifest unique transport regimes of fundamental interest~\cite{babadi2012collective}. The GMM framework is naturally suited to solve problems beyond these simplifications, as evidenced by our detailed analysis of trapped Bose-Fermi mixtures.

\begin{figure}[t!]
\centering
\includegraphics[width=1\linewidth]{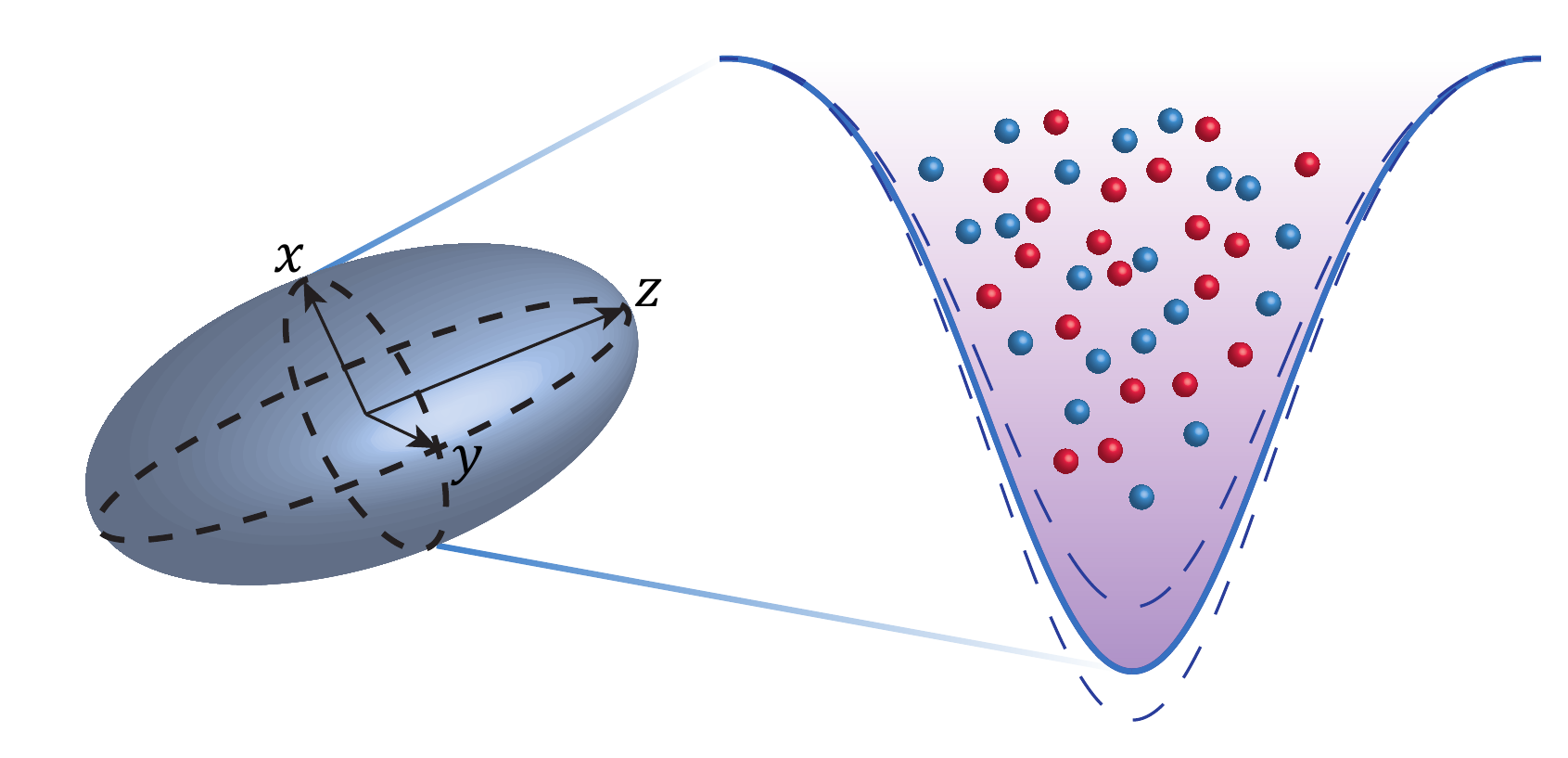} 
\caption{ Schematic of an atomic quantum Bose-Fermi mixture trapped into a cigar-shaped optical potential. A time-dependent (transverse) perturbation of the trap launches collective modes of the system, which characterize the equilibrium state of matter.
}
\label{fig::cartoon}
\end{figure}

In contexts where the typical approximations fail, one must resort to solving the full linearized Boltzmann equation. One such context is material design. Ab initio methods like density functional theory are combined with Boltzmann analysis to design transport properties relevant to a wide range of technologies including thermoelectric materials~\cite{Mingo2014}, lithium-ion batteries~\cite{Morgan2022}, and photovoltaic cells~\cite{Jermyn2019}. Accurate computation of transport coefficients often requires going beyond the RTA for most 2D materials~\cite{Cepellotti2016} as well as 3D materials like diamond where three-phonon or higher processes are relevant~\cite{Li2014}. Anisotropic and position-dependent scattering must also be accounted for in devices with nontrivial geometries~\cite{Li2014,Madsen2018,Varnavides2019}.

Remarkable progress has been made in brute force numerical methods in such systems; recent approaches have, for example, enabled solving the fully coupled electron-phonon transport kinetic equations relevant to thermoelectric and thermopower applications~\cite{Protik2020,Protik2022}. These state-of-the-art methods, however, are extremely time-consuming, as typical calculations take several months of CPU time~\cite{Protik2022}. We expect that the GMM framework introduced in this work can significantly ameliorate the cost of such computations.

\subsection{Analysis of Bose-Fermi mixtures}

Bose-Fermi mixtures are a salient class of systems, ubiquitous in both cold-atom and solid-state platforms, which exhibit rich collective dynamics and are challenging to analyze. Examples include electron-phonon and electron-magnon mixtures in both conventional and high-$T_c$ superconductors, \ch{He^3}-\ch{He^4} mixtures in dilution refrigerators, and quark-gluon plasmas in high-energy physics~\cite{ebner1971low,viverit2000zero,buchler2003supersolid,schaefer_phase_2005,lee2006doping,ludwig2011quantum,bertaina2013quantum,kinnunen2015induced,keimer2015quantum,zhang2017anomalous,gooth2018thermal,kagan2019fermi}. Bose-Fermi mixtures may also naturally appear in transition-metal-dichalcogenides (TMDs), two-dimensional semiconductors which exhibit novel physics promising for quantum optics applications; TMDs manifest optical excitons and charged fermions, and interactions between them can be tuned via physical processes similar to Feshbach resonances~\cite{sidler2017fermi,Wang2019,shimazaki2020strongly,schwartz2021electrically}.

As a concrete context, we focus on the case of Bose-Fermi mixtures of cold atomic gases trapped in a cigar-shaped optical potential~\cite{Inouye2004,Ferrier-Barbut2014,Delehaye2015,Yao2016,Roy2017,Wu2018,Huang2019}, illustrated in Fig.~\ref{fig::cartoon}. We choose this system to demonstrate our GMM-MoM method for three reasons. Firstly, quantum gases of bosonic and fermionic atoms have emerged as a leading quantum simulation platform to study several quantum many-body systems, as cold-atom setups provide a highly controllable setting to investigate different prototypical models~\cite{Bloch2008}. Until recently, they have primarily been used to study purely bosonic or fermionic systems due to the experimental difficulty of simultaneously trapping both types of atoms. 
The collective behavior of such bosonic and fermionic gases has been explored in several experiments, including precise measurements of collective modes~\cite{bartenstein2004collective,altmeyer2007precision,Tey2013}, observation of first and second sound modes~\cite{christodoulou2021observation,hilker2022first}, measurements of viscosities in the unitary regime~\cite{cao2011universal,joseph2015shear} and transport properties in strongly-interacting gases~\cite{strohmaier2007interaction,sommer2011universal,jepsen2020spin,jepsen2021transverse,Nichols2019,Patel2020,Yan2022}. Systems in the polaronic regime, where one gas is much more dilute than the other, have been predicted to exhibit novel collective behavior~\cite{Dolgriev2021} and non-equilibrium dynamics~\cite{Seetharam2021,Seetharam2021long}. Experiments have investigated the emergence of quasiparticles in such systems~\cite{Yan2019} and, more recently, have started to explore the collective behavior of mixtures of bosonic and fermionic cold atoms~\cite{Yan2023} beyond the polaronic regime, thus opening the door to studying Bose-Fermi mixtures in a controlled setting leveraging the entire cold-atom toolbox~\cite{Ferlaino2003,Zaccanti2006,Trautmann2018}. Secondly, on the theoretical side, Bose-Fermi mixtures beyond the zero-temperature impurity regime host potentially rich physics which has been unexplored in prior work due to the difficulty of numerical analysis and lack of experiment-guided motivation. Thirdly, the system is described by one of the most challenging types of Boltzmann equations to solve due to its spatial inhomogeneity, resulting from the asymmetric cigar-shaped optical trap, and the particle-conserving nature of collisions in the system. Making experimentally verifiable predictions in this context thus provides confidence in the GMM-MoM analysis method.

We analyze the collective behavior of the trapped Bose-Fermi cold-atom mixture across a variety of relative densities and interaction strengths. Motivated by modern cold-atom experiments, we study monopole and quadrupole collective modes, focusing on a quantum mixture above the Bose-Einstein transition temperature $T> T_c$; the gases are cold enough for particle statistics to matter and therefore form a quantum Bose-Fermi mixture. We find that this system exhibits a rich phenomenology which we summarize below for various regimes of relative densities of the atomic species.

(I) \textit{Single-component interacting Bose fluid (no fermions)}. The system exhibits a collisionless-to-hydrodynamic crossover as the boson-boson interaction strength is tuned. Both the monopole and quadrupole mode lifetimes manifest a non-monotonic behavior, being the shortest in the crossover region. We also find frequency mixing between transverse and longitudinal monopole modes for sufficiently strong interactions; these modes are distinct from each other due to the asymmetry of the cigar-shaped trap.

(II) \textit{Impurity regime (dilute fermions)}. 
When the fermions are dilute, they act as impurities; the bosons remain largely unaffected and behave as the single-component Bose fluid described above. The fermions and bosons have different monopole mode frequencies, and we can therefore understand the system in terms of coupled harmonic oscillators corresponding to the respective collective modes. When the boson-boson interaction is such that the boson fluid is not in its crossover regime and therefore manifests long-lived modes, we find typical Fano interference lineshapes encoding coherent mixing between the bosonic and fermionic oscillators (see Fig.~\ref{fig::figure-polaronic} and discussion in Sec.~\ref{subsec:Impurity regime}  for details). Upon increasing the boson-fermion coupling, initially sharp fermionic resonances become completely featureless in frequency, indicating that the fermions no longer have independent dynamics and instead follow the bosonic atoms. 
In addition to the above phenomenona, we find that the fermionic spectral functions exhibit Stokes and anti-Stokes sidebands due to the mixing between longitudinal and transverse monopole modes.

(III) \textit{Mixture regime (comparable densities of bosons and fermions)}. The Bose-Fermi mixture exhibits a collisionless-to-hydrodynamic crossover as the boson-fermion interaction strength is tuned. The emergent hydrodynamic behavior of the whole many-body system manifests as mode locking between bosons and fermions. Intriguingly, we find coherent mixing between bosonic and fermionic collective modes even when the bosons are hydrodynamic and fermions are collisionless.

Some of the phenomenology summarized above, such as the collisionless-to-hydrodynamic crossover of single-component Bose fluids, has been discussed in previous papers~\cite{PhysRevA.60.4851,griffin2009bose}. Other effects we uncover, however, such as the Fano lineshapes and the emergence of sidebands, are indicative of previously unexplored physics exhibited by Bose-Fermi mixtures. Additionally, 
many-body mode mixing and locking in the mixture regime has not been confirmed theoretically due to the collision integral computational bottleneck. The above results are therefore both a concrete window into the rich physics of Bose-Fermi cold-atom mixtures, as well as evidence of the power of the GMM-MoM method to analyze previously inaccessible many-body systems.

\section{Theoretical framework }
\label{sec:bose-fermi-framework}

This section is organized as follows: In Sec.~\ref{subsec:kin_eqs}, we first introduce the coupled Boltzmann kinetic equations to describe the dynamics of a trapped Bose-Fermi mixture for $T > T_c$. The equilibrium distribution functions are then discussed in Sec.~\ref{subsec:eqm}. Finally, Sec.~\ref{subsec:mom} is devoted to the method of moments for computing the linear response properties of the mixture, including the spectrum of collective modes. Our framework closely follows and substantially extends that in Ref.~\cite{babadi2012collective} on two-dimensional fermionic dipolar gases.
We end this section by writing down expressions~\eqref{eqn_coll_matr_in}-\eqref{eqn_coll_matr_fin} for the 15-dimensional collision integrals which represent the main challenge behind the method of moments in three spatial dimensions. These integrals, in general, are not feasible for the existing numerical tools such as direct Monte Carlo sampling that was proven efficient in two dimensions~\cite{babadi2012collective}. 
One of our main technical results is that we overcome this challenge and make the method of moments numerically tractable -- this will be the subject of the following sections. 

\subsection{Kinetic equations for $T> T_c$}
\label{subsec:kin_eqs}

We describe a trapped Bose-Fermi mixture using the following microscopic model (throughout the text, we set $\hbar = k_B = 1$):
\begin{align}
    \hat{H}  = \hat{H}_{\rm B} + \hat{H}_{\rm F} + \hat{H}_{\rm int}.
\end{align}
Here, the first term represents the bosonic Hamiltonian:
\begin{align}
    \hat{H}_{\rm B}
    & = \int d^3 \bm r \, \psi_{\rm B}^\dagger(\bm r)\Big[
    -\frac{\partial_{\bm r}^2}{2m_{\rm B} } + U_{\rm trap,B}(\bm r)
    \Big] \psi_{\rm B}(\bm r) \notag\\
    &\qquad\qquad
    + \frac{g_{\rm B}}{2}  \int d^3 \bm r \, \psi_{\rm B}^\dagger(\bm r)\psi_{\rm B}^\dagger(\bm r)\psi_{\rm B}(\bm r)\psi_{\rm B}(\bm r).
\end{align}
The second term is the fermionic one:
\begin{align}
    \hat{H}_{\rm F}
    & = \int d^3 \bm r \, \psi_{\rm F}^\dagger(\bm r)\Big[
    -\frac{\partial_{\bm r}^2}{2m_{\rm F}} + U_{\rm trap,F}(\bm r)
    \Big] \psi_{\rm F}(\bm r).
\end{align}
The two systems interact with each other through the third term, ${\hat H}_{\rm int}$, given by:
\begin{align}
   {\hat H}_{\rm int} = g_{\rm BF}  \int d^3 \bm r \, \psi_{\rm B}^\dagger(\bm r)\psi_{\rm B}(\bm r) \psi_{\rm F}^\dagger(\bm r)\psi_{\rm F}(\bm r).
\end{align}
Here $\psi_{\rm B/F}(\bm r)$ and $\psi_{\rm B/F}^\dagger(\bm r)$ encode the bosonic/fermionic annihilation and creation operators, respectively; they satisfy the canonical commutation (anti-commutation) relations. We assume that, as in cold-atom experiments, the contact coupling strengths  $g_{\rm B}$ and $g_{\rm BF}$ can be tuned over a broad range  using the physics of magnetic Feshbach resonances. For future reference,  $g_{\rm B} = 4\pi a_{\rm B}/m_{\rm B}$ and $g_{\rm BF} = 2\pi a_{\rm BF} (m_{\rm B} + m_{\rm F})/(m_{\rm B} m_{\rm F})$, where $a_{\rm B}$ and $a_{\rm BF}$ are the Bose-Bose and Bose-Fermi scattering lengths, respectively. In what follows, we consider the trapping potentials to have a cigar-like shape:
\begin{align}
    U_{\rm trap,B}(\bm r) = \frac{1}{2}m_{\rm B} \omega_\rho^2 (x^2 + y^2) + \frac{1}{2}m_{\rm B} \omega_z^2 z^2.\label{eqn:U_trap_B}
\end{align}
To relate to the MIT experiment~\cite{Yan2023}, we will consider the fermionic trapping potential to be $U_{\rm trap, F} = \lambda U_{\rm trap,B}(\bm r)$, where $\lambda\simeq 1$ is an experimental parameter that encodes the fact that the laser coupling to fermions is a bit different from that to bosons. At this stage, we do not need to specify concrete forms of trapping potentials and can let the formalism be generic.

We briefly remark that a possible microscopic derivation of Boltzmann equations uses the Keldysh technique of non-equilibrium Green's functions~\cite{kamenev2023field}. This description is closely related to an appropriate conserving Kadanaff-Baym framework~\cite{kadanoff2018quantum}. Both these approaches are usually too complicated to work with, either analytically or numerically, and require approximations. In systems such as quantum Bose-Fermi mixtures considered here, a natural simplification occurs because there is a separation of time and length
scales between fast microscopic degrees of freedom and macroscopic ones set by the trapping potential. This separation enables  one to make a systematic gradient expansion, which dramatically simplifies the kinetic equations. Further progress is possible if one is allowed to employ the quasiparticle approximation, where the complex dynamics of two-time Green's functions is reduced to that of single-particle distribution functions, thus, obtaining the Boltzmann equation~\cite{danielewicz1984quantum,babadi2013non}. The quantum statistics of the constituent particles manifests in both incoherent scatterings captured via collision integrals and in self-energies that might encode inherently quantum processes such as particle exchange. For more details on the derivation of Boltzmann equations, we refer the reader to Refs.~\cite{danielewicz1984quantum,kadanoff2018quantum,babadi2013non,kamenev2023field}. The primary goal of this paper is to develop a tool to solve Boltzmann equations rather than to provide a microscopic derivation of the equations of motion. To this end, below we consider a Bose-Fermi mixture above the BEC transition temperature $T> T_c$, where the kinetic equations are well-known.

Specifically, they read:
\begin{gather}
    \Big( \partial_t  + \frac{\bm p}{m_{\rm B}} \cdot \partial_{\bm r} - \partial_{\bm r} U_{\rm eff, B} \cdot \partial_{\bm p} \Big)n_{\rm B} = I_{\rm BB} + I_{\rm BF},\\
    \Big( \partial_t  + \frac{\bm p}{m_{\rm F}} \cdot \partial_{\bm r} - \partial_{\bm r} U_{\rm eff, F} \cdot \partial_{\bm p} \Big)n_{\rm F} = I_{\rm FB},
\end{gather}
where $n_{\rm B}(\bm p,\bm r, t)$ and $n_{\rm F}(\bm p,\bm r, t)$ are the bosonic and fermionic distribution functions, respectively. For the left-hand sides, we employ the Hartree-Fock approximation~\cite{hohenberg1965microscopic}:
\begin{gather}
    U_{\rm eff, B} = U_{\rm trap, B} + 2 g_{\rm B} n_{\rm B}(\bm r,t) + g_{\rm BF} n_{\rm F}(\bm r,t),\\
    U_{\rm eff, F} = U_{\rm trap, F} + g_{\rm BF} n_{\rm B}(\bm r,t),
\end{gather}
where $n_{\rm B/F} (\bm r,t) = \displaystyle\int \frac{d^3\bm p}{(2\pi)^3} n_{\rm B/F}(\bm p,\bm r,t)$ represents the bosonic/fermionic real-space density. We note that the Hartree-Fock self-energies do not depend on $\bm p$, which in turn implies that the effective masses are not affected by the contact interactions. For the right-hand sides, we use the Born-Markov approximation and write the Bose-Bose, Bose-Fermi, and Fermi-Bose collision integrals as:
\begin{widetext}
\begin{align}
    I_{\rm BB}(\bm p,\bm r,t) & = 2 g_{\rm B}^2 \int \frac{d^3 \bm p'}{(2\pi)^3}\frac{d^3 \bm p_1}{(2\pi)^3}\frac{d^3 \bm p'_1}{(2\pi)^3} \times (2\pi)^3\delta(\bm p + 
    \bm p' - \bm p_1 - \bm p_1') \times (2\pi)\delta(\varepsilon_{\rm B}(\bm p) + \varepsilon_{\rm B}(\bm p') - \varepsilon_{\rm B}(\bm p_1) -\varepsilon_{\rm B}(\bm p_1'))\notag\\
    &\qquad\qquad\qquad
    \times [ (1 + n_{\rm B}(\bm p))(1 + n_{\rm B}(\bm p'))n_{\rm B}(\bm p_1) n_{\rm B}(\bm p_1') - n_{\rm B}(\bm p) n_{\rm B}(\bm p')(1 + n_{\rm B}(\bm p_1))(1 + n_{\rm B}(\bm p_1')) ],\\
    I_{\rm BF}(\bm p,\bm r,t) & = g_{\rm BF}^2 \int \frac{d^3 \bm p'}{(2\pi)^3}\frac{d^3 \bm p_1}{(2\pi)^3}\frac{d^3 \bm p'_1}{(2\pi)^3} \times (2\pi)^3\delta(\bm p + 
    \bm p' - \bm p_1 - \bm p_1') \times (2\pi)\delta(\varepsilon_{\rm B}(\bm p) + \varepsilon_{\rm F}(\bm p') - \varepsilon_{\rm B}(\bm p_1) -\varepsilon_{\rm F}(\bm p_1'))\notag\\
    &\qquad\qquad\qquad
    \times [ (1 + n_{\rm B}(\bm p))(1 - n_{\rm F}(\bm p'))n_{\rm B}(\bm p_1) n_{\rm F}(\bm p_1') - n_{\rm B}(\bm p) n_{\rm F}(\bm p')(1 + n_{\rm B}(\bm p_1))(1 - n_{\rm F}(\bm p_1')) ],\\
    I_{\rm FB}(\bm p,\bm r,t) & = g_{\rm BF}^2 \int \frac{d^3 \bm p'}{(2\pi)^3}\frac{d^3 \bm p_1}{(2\pi)^3}\frac{d^3 \bm p'_1}{(2\pi)^3} \times (2\pi)^3\delta(\bm p + 
    \bm p' - \bm p_1 - \bm p_1') \times (2\pi)\delta(\varepsilon_{\rm F}(\bm p) + \varepsilon_{\rm B}(\bm p') - \varepsilon_{\rm F}(\bm p_1) -\varepsilon_{\rm B}(\bm p_1'))\notag\\
    &\qquad\qquad\qquad
    \times [ (1 - n_{\rm F}(\bm p))(1 + n_{\rm B}(\bm p'))n_{\rm F}(\bm p_1) n_{\rm B}(\bm p_1') - n_{\rm F}(\bm p) n_{\rm B}(\bm p')(1 - n_{\rm F}(\bm p_1))(1 + n_{\rm B}(\bm p_1')) ],
\end{align} 
\end{widetext}
where $\varepsilon_{\rm B/F}(\bm p) = p^2/(2m_{\rm B/F})$. Below we turn to analyse the linear-response properties of the Bose-Fermi mixture within this Boltzmann kinetic theory. Prior to that, we first compute the self-consistent equilibrium distribution functions needed for our subsequent analysis.
 
\subsection{The equilibrium state}
\label{subsec:eqm}

The equilibrium distribution functions are related to the effective potentials $U_{\rm eff,B/F}(\bm r)$ through ($\beta = 1/T$):
\begin{gather}
    n_{\rm B/F,{\rm eq}}(\bm p,\bm r) = \left[ \frac{1}{z_{\rm B/F}(\bm r)} \exp\Big(\frac{\beta p^2}{2m_{\rm B/F}}  \Big) \mp 1\right]^{-1},\label{eqn:eqm}
\end{gather}
where $z_{\rm B/F}(\bm r) \equiv \exp(\beta(\mu_{\rm B/F} - U_{\rm eff,B/F}(\bm r)))$ is the bosonic (fermionic) local fugacity and $\mu_{\rm B/F}$ is the corresponding chemical potential.
Since the effective potentials $U_{\rm eff,B/F}(\bm r)$ depend on $n_{\rm B/F,{\rm eq}}$, Eq.~\eqref{eqn:eqm} should be solved self-consistently. This latter task is relatively simple because the momentum integrals are evaluated analytically:
\begin{gather}
    \int \frac{d^3 \bm p}{(2\pi)^3} n_{\rm B,\rm eq} = \left(\frac{m_{\rm B}}{2\pi \beta}\right)^{\frac{3}{2}} \xi_{\frac{3}{2}}(z_{\rm B}(\bm r)),\\
    \int \frac{d^3 \bm p}{(2\pi)^3} n_{\rm F,\rm eq} = - \left(\frac{m_{\rm F}}{2\pi \beta}\right)^{\frac{3}{2}}\xi_{\frac{3}{2}}(-z_{\rm F}(\bm r)),
\end{gather}
where $\xi_{\frac{3}{2}}(x)$ is the polylogarithm function. The chemical potentials can be fixed via:
\begin{align}
    \int d \Gamma \, n_{\rm B/F,\rm eq} = \int d^3\bm r \int \frac{d^3 \bm p}{(2\pi)^3} n_{\rm B/F,\rm eq}  = N_{\rm B/F},
\end{align}
with $N_{\rm B/F}$ being the total number of Bose/Fermi particles. 

For convenience, throughout the rest of the paper we use the following dimensionless variables. As for the units of momentum and length, we choose $p_\rho = \sqrt{m_{\rm B} \omega_\rho}$ and $a_\rho = 1/p_\rho$, respectively; we fix the unit of energy to be $\omega_\rho$; the unit of mass is then naturally set by $m_{\rm B}$. In the dimensionless units, the trapping potentials read:
\begin{gather}
    \bar{U}_{\rm trap, B}(\bar{\rho},\bar{z}) =\frac{1}{2}\bar{\rho}^2 + \frac{\kappa}{2}\bar{z}^2, \\
    \bar{U}_{\rm trap, F}(\bar{\rho},\bar{z}) =\frac{\lambda}{2}\bar{\rho}^2 + \frac{\lambda \kappa}{2}\bar{z}^2, 
\end{gather}
where $\kappa = \omega_z^2/\omega_\rho^2$ is the anisotropy parameter, and bars on top of the symbols indicate that the corresponding variable has been appropriately rescaled. For instance, the dimensionless interactions strengths read: $\bar{g}_{\rm B} = g_{\rm B}/(a_\rho^3 \omega_\rho)$ and $\bar{g}_{\rm BF} = g_{\rm BF}/(a_\rho^3 \omega_\rho)$. We finally note that because the two trapping potentials depend only on the combination $\bar{\rho}^2 + \kappa \bar{z}^2$, the equilibrium self-consistency equations can be written as
\begin{align}
    z_{\rm B}(\tilde{r}) & = \exp\Big(\bar{\beta}\bar{\mu}_{\rm B} -\frac{\tilde{r}^2}{2} -\frac{{\bar \beta}}{(2\pi{\bar \beta})^{3/2}} \Big[ 2\bar{g}_{\rm B} \xi_{\frac{3}{2}}(z_{\rm B}) \notag\\
    &\qquad\qquad\qquad\quad
    - \bar{g}_{\rm BF} \Big(\frac{m_{\rm F}}{m_{\rm B}}\Big)^{\frac{3}{2}} \xi_{\frac{3}{2}}(-z_{\rm F}) \Big] \Big),\\
    z_{\rm F}(\tilde{r}) & = \exp\Big(\bar{\beta}\bar{\mu}_{\rm F} -\frac{\lambda\tilde{r}^2}{2} -  \frac{\bar{g}_{\rm BF} {\bar \beta}}{(2\pi{\bar \beta})^{3/2}}  \xi_{\frac{3}{2}}(z_{\rm B}) \Big),
\end{align}
where $\tilde{r}^2 = \bar{\beta} (\bar{\rho}^2 +  \kappa \bar{z}^2)$. We remark that these equations, which are solved numerically in practice, imply that the two fugacities depend only on the one-dimensional variable $\tilde{r}$ rather than on two independent variables $\bar{\rho}$ and $\bar{z}$. This observation is not crucial (and in principle, the trapping potential for fermions could be very different from the bosonic one) but facilitates our numerical calculations below.

\subsection{Linear response within the method of moments}
\label{subsec:mom}

Our primary goal now is to investigate linear response functions, related to collective modes, which represent small amplitude fluctuations on top of the equilibrium state. To this end, we write:
\begin{align}
    n_{\rm B}(\bm r, \bm p,t) = n_{\rm B,\rm eq}(\bm r, \bm p) + \Delta_{\rm B} (\bm r, \bm p) \Phi_{\rm B}(\bm r, \bm p,t),\label{eqn: n_B_expand}\\
    n_{\rm F}(\bm r, \bm p,t) = n_{\rm F,\rm eq}(\bm r, \bm p) + \Delta_{\rm F} (\bm r, \bm p) \Phi_{\rm F}(\bm r, \bm p,t),
\end{align}
where $ \Delta_{\rm B/F} (\bm r, \bm p) = n_{\rm B/F,\rm eq}(\bm r, \bm p)[1\pm n_{\rm B/F,\rm eq}(\bm r, \bm p)]$. In what follows, we assume that $\Phi_{\rm B}(\bm r, \bm p,t)$ and $\Phi_{\rm F}(\bm r, \bm p,t)$ are small and, as such, linearize the equations of motion. In the remainder of the text, we consider only dimensionless variables that have been properly rescaled and 
omit writing  bars on top of the corresponding symbols.

The method of moments is based on the idea of expanding small fluctuations $\Phi_{\rm B/F}(\bm r, \bm p,t)$ using a suitable basis set of functions $\{\phi_\alpha(\bm r, \bm p)\}_{\alpha = 1}^N$:
\begin{align}
    \Phi_{\rm B/F}(\bm r, \bm p,t) \approx \sum_{\alpha = 1}^N \Phi_{\rm B/F,\alpha}(t) \phi_\alpha(\bm r, \bm p).
\end{align}
Here $N$ is the total number of basis functions that we take into account, and, as such, this parameter controls the accuracy of our approximations. Motivated by the experiments, below we investigate linear response to perturbations of the bosonic and/or fermionic trapping potentials, which we also write as:
\begin{align}
    \delta U_{\rm B/F}(\bm r,\bm p, \omega) = \sum_{\alpha = 1}^N \delta U_{\rm B/F,\alpha}(\omega) \phi_\alpha(\bm r,\bm p),
\end{align}
where the frequency amplitudes $\delta U_{\rm B/F,\alpha}(\omega)$ are set by the actual experimental driving protocol. Plugging these expansions into the kinetic equations and subsequently projecting the linearized equations onto the basis functions $\{\phi_\alpha(\bm r, \bm p)\}$, we arrive at:
\begin{align}
    [-i{\omega} \hat{\rm M} + \hat{\rm H} - \hat{\rm S} - \hat{\rm I}] \Phi  = -{\beta} \hat{\rm H} \delta U ,
    \label{eqn:MoM}
\end{align}
where we defined $\Phi = (\Phi_{\rm B,1},\dots,\Phi_{\rm B,N}, \Phi_{\rm F,1},\dots,\Phi_{\rm F,N})^T$ and $\delta U = (\delta U_{\rm B,1},\dots,\delta U_{\rm B,N},\delta U_{\rm F,1}, \dots,\delta U_{\rm F,N})^T$ to be $2N$-dimensional vectors. The matrices ${\hat{\rm M}}$ and ${\hat{\rm H}}$ are diagonal in the Bose-Fermi space:
\begin{gather}
    {\hat{\rm M}} = \begin{bmatrix}  {\rm M}_{\rm B} & 0\\
    0 & {\rm M}_{\rm F}
    \end{bmatrix},\quad {\hat{\rm H}} = \begin{bmatrix}  {\rm H}_{\rm B} & 0\\
    0 & {\rm H}_{\rm F}
    \end{bmatrix},
\end{gather}
with the matrix elements given by: 
\begin{align}
    ({\rm M}_{\rm B/F})_{\alpha\beta} & = \int d{\Gamma}\,\Delta_{\rm B/F}\phi_\alpha(\bm r, \bm p) \phi_\beta(\bm r, \bm p) \notag\\
    & \equiv \langle\!\langle \phi_\alpha\phi_\beta \rangle\!\rangle_{\rm B/F},\label{eqn:M_def}\\
    ({\rm H}_{\rm B/F})_{\alpha\beta} &= \langle\!\langle \phi_\alpha \{\phi_\beta, {\cal H}_{\rm B/F}\} \rangle\!\rangle_{\rm B/F}, \label{eqn:def_H_matr}
\end{align}
where ${\cal H}_{\rm B} = p^2/2 + U_{\rm eff,B}$, ${\cal H}_{\rm F} = p^2 m_{\rm B}/2m_{\rm F} + U_{\rm eff,F},$ and  $\{ f,g \} = \partial_{\bm r} f\cdot \partial_{\bm p} g - \partial_{\bm p} f\cdot \partial_{\bm r} g$ is the  Poisson bracket. Throughout the derivations, we used the identity $\{n_{\rm B/F,\rm eq}, f \} = -{ \beta} \Delta_{\rm B/F}  \{{\cal H}_{\rm B/F}, f \}$. The matrix $\hat{\rm S}$ encodes the changes in self-energies arising from the changes in the bosonic and fermionic distribution functions:
\begin{align}
    \hat{\rm S} = \begin{bmatrix}  {\rm S}_{\rm B} & {\rm S}_{\rm BF}\\
    {\rm S}_{\rm FB} & 0
    \end{bmatrix},
    \label{eqn:Sigma_gen}
\end{align}
where
\begin{align}
    ({\rm S}_{\rm B})_{\alpha\beta} & = \langle\!\langle { \beta}    \Sigma_{\rm B}[\Delta_{\rm B} \phi_\beta] \{ \phi_\alpha , {\cal H}_{\rm B}\} \rangle\!\rangle_{\rm B}, \label{eqn:SB_gen}\\
    ({\rm S}_{\rm BF})_{\alpha\beta} & = \langle\!\langle { \beta}    \Sigma_{\rm BF}[\Delta_{\rm F} \phi_\beta] \{ \phi_\alpha , {\cal H}_{\rm B}\} \rangle\!\rangle_{\rm B},\\
    ({\rm S}_{\rm FB})_{\alpha\beta} & = \langle\!\langle { \beta}    \Sigma_{\rm BF}[\Delta_{\rm B} \phi_\beta] \{ \phi_\alpha , {\cal H}_{\rm F}\} \rangle\!\rangle_{\rm F}.
\end{align}
Here we defined $\Sigma_{\rm B}[f](\bm r,t) = 2 g_{\rm B} \displaystyle \int\frac{d^3 \bm p}{(2\pi)^3} f(\bm r,\bm p, t)$ and $\Sigma_{\rm BF}[f](\bm r,t) = g_{\rm BF} \displaystyle \int\frac{d^3 \bm p}{(2\pi)^3} f(\bm r,\bm p, t)$. Finally, the collision integral matrix $\hat{\rm I}$ is given by:
\begin{align}
    \hat{\rm I} = \begin{bmatrix}  {\rm I}_{\rm B} + {\rm J}_{\rm BB} & {\rm J}_{\rm BF}\\
    {\rm J}_{\rm FB} & {\rm J}_{\rm FF}
    \end{bmatrix},
\end{align}
where the corresponding matrix elements read:
\begin{widetext}
\begin{align}
    ({\rm I}_{\rm B})_{\alpha\beta} = & - \frac{{g}^2_{\rm B}}{2}  \int d^3\bm r\frac{d^3 \bm p}{(2\pi)^3} \frac{d^3 \bm p'}{(2\pi)^3}\frac{d^3 \bm p_1}{(2\pi)^3}\frac{d^3 \bm p'_1}{(2\pi)^3}  (2\pi)^3\delta(\bm p + 
    \bm p' - \bm p_1 - \bm p_1') \times (2\pi)\delta(\varepsilon_{\rm B}(\bm p) +  \varepsilon_{\rm B}(\bm p')  - \varepsilon_{\rm B}(\bm p_1) -\varepsilon_{\rm B}(\bm p_1'))\notag\\
    &
    \times (1 + n_{\rm B, eq}(\bm p))(1 + n_{\rm B, eq}(\bm p'))n_{\rm B, eq}(\bm p_1) n_{\rm B, eq}(\bm p_1') \times S[\phi_\alpha]S[\phi_\beta],\label{eqn_coll_matr_in}\\
    ({\rm J}_{\rm BB})_{\alpha\beta} = & - \frac{{g}^2_{\rm BF}}{2}  \int d^3\bm r\frac{d^3 \bm p}{(2\pi)^3} \frac{d^3 \bm p'}{(2\pi)^3}\frac{d^3 \bm p_1}{(2\pi)^3}\frac{d^3 \bm p'_1}{(2\pi)^3}  (2\pi)^3\delta(\bm p + 
    \bm p' - \bm p_1 - \bm p_1') \times (2\pi)\delta(\varepsilon_{\rm B}(\bm p) +  \varepsilon_{\rm F}(\bm p')  - \varepsilon_{\rm B}(\bm p_1) -\varepsilon_{\rm F}(\bm p_1'))\notag\\
    &
    \times (1 + n_{\rm B, eq}(\bm p))(1 - n_{\rm F, eq}(\bm p'))n_{\rm B, eq}(\bm p_1) n_{\rm F, eq}(\bm p_1') \times (\phi_\alpha(\bm p) - \phi_\alpha(\bm p_1))(\phi_\beta(\bm p) - \phi_\beta(\bm p_1)),\label{eqn_coll_matr_in_v2}\\
    ({\rm J}_{\rm BF})_{\alpha\beta} = & - \frac{{g}^2_{\rm BF}}{2}  \int d^3\bm r\frac{d^3 \bm p}{(2\pi)^3} \frac{d^3 \bm p'}{(2\pi)^3}\frac{d^3 \bm p_1}{(2\pi)^3}\frac{d^3 \bm p'_1}{(2\pi)^3}  (2\pi)^3\delta(\bm p + 
    \bm p' - \bm p_1 - \bm p_1') \times (2\pi)\delta(\varepsilon_{\rm B}(\bm p) +  \varepsilon_{\rm F}(\bm p')  - \varepsilon_{\rm B}(\bm p_1) -\varepsilon_{\rm F}(\bm p_1'))\notag\\
    &
    \times (1 + n_{\rm B, eq}(\bm p))(1 - n_{\rm F, eq}(\bm p'))n_{\rm B, eq}(\bm p_1) n_{\rm F, eq}(\bm p_1') \times (\phi_\alpha(\bm p) - \phi_\alpha(\bm p_1))(\phi_\beta(\bm p') - \phi_\beta(\bm p_1')),\\
    ({\rm J}_{\rm FB})_{\alpha\beta} = & ({\rm J}_{\rm BF})_{\beta\alpha},\\
    ({\rm J}_{\rm FF})_{\alpha\beta} = & - \frac{{g}^2_{\rm BF}}{2}  \int d^3\bm r\frac{d^3 \bm p}{(2\pi)^3} \frac{d^3 \bm p'}{(2\pi)^3}\frac{d^3 \bm p_1}{(2\pi)^3}\frac{d^3 \bm p'_1}{(2\pi)^3}  (2\pi)^3\delta(\bm p + 
    \bm p' - \bm p_1 - \bm p_1') \times (2\pi)\delta(\varepsilon_{\rm B}(\bm p) +  \varepsilon_{\rm F}(\bm p')  - \varepsilon_{\rm B}(\bm p_1) -\varepsilon_{\rm F}(\bm p_1'))\notag\\
    &
    \times (1 + n_{\rm B, eq}(\bm p))(1 - n_{\rm F, eq}(\bm p'))n_{\rm B, eq}(\bm p_1) n_{\rm F, eq}(\bm p_1') \times (\phi_\alpha(\bm p') - \phi_\alpha(\bm p_1'))(\phi_\beta(\bm p') - \phi_\beta(\bm p_1')), \label{eqn_coll_matr_fin}
\end{align}
\end{widetext}
where $S[f] = f + f' - f_1 - f_1'$, $\varepsilon_{\rm B} (\bm p) = p^2/2$, and $\varepsilon_{\rm F} (\bm p) = p^2 m_{\rm B}/(2m_{\rm F})$. 
The matrix ${\rm I}_{\rm B}$ describes changes in the bosonic distribution function due to boson-boson scatterings, ${\rm J}_{\rm BF}$ describes changes in the distribution function of bosons due boson-fermion scatterings, etc.

What we have achieved so far is that the initial infinite-dimensional (linear) problem is now reduced to a finite-dimensional one. To complete the discussion of dynamics, we also need to review an algorithm for evaluating time-dependent expectation values of operators, which can be measured in experiments. As a concrete example, we consider an arbitrary bosonic observable:
\begin{align}
    \langle{\cal O}_{\rm B}\rangle (t) \equiv \int d\Gamma\, n_{\rm B}(\bm r,\bm p,t) {\cal O}_{\rm B}(\bm r,\bm p)  = {\cal O}_{\rm B}^T {\rm M}_{\rm B} \Phi_{\rm B}(t),\notag 
\end{align}
where we assumed that $\langle {\cal O}_{\rm B}\rangle_{\rm eq} = 0$, used Eq.~\eqref{eqn: n_B_expand}, and expanded ${\cal O}_{\rm B}(\bm r,\bm p) = \sum_\alpha {\cal O}_{\rm B,\alpha}\phi_\alpha(\bm r,\bm p)$. More generally, we write:
\begin{align}
    \langle {\cal O}\rangle(\omega) = {\cal O}^T \hat{\rm M} \Phi(\omega + i0^+).
\end{align}
To evaluate $\Phi(\omega + i0^+)$, we diagonalize $\hat{\rm M}^{-1}(\hat{\rm H} - \hat{\rm S} - \hat{\rm I}) = i \hat{\rm V} \hat{\rm\Omega} \hat{\rm V}^{-1}$, with $\hat{\rm\Omega}$ being diagonal$\Rightarrow$
\begin{align}
    \Phi(\omega) = -i\beta \hat{\rm V} (\omega - \hat{\rm \Omega})^{-1}\hat{\rm V}^{-1} \hat{\rm M}^{-1} \hat{\rm H}\delta U(\omega).
\end{align}
From this, we finally get:
\begin{align}
    {\cal O}(\omega) = \sum_a \frac{r_a(\omega)}{\omega - {\rm \Omega}_a + i0^+}, \label{eqn:linear_response}
\end{align}
where 
\begin{align}
    r_a(\omega) = -i \beta (\hat{\rm V}^T\hat{\rm M} {\cal O})_a (\hat{\rm V}^{-1} \hat{\rm M}^{-1} \hat{\rm H}\delta U(\omega))_a.
\end{align}
Equation~\eqref{eqn:linear_response} is among the most useful results, as it defines the linear response within the method-of-moments. Physically, the ``residue'' $r_a(\omega)$ encodes the relative importance of the corresponding pole $\Omega_a$. We remark that the eigenvalues of $
\hat{\rm\Omega}$ are not necessarily real (but for stability of the equilibrium state, their imaginary parts must be negative), which implies that the matrix $\hat{\rm V}$ is generally not unitary.

The main challenge that we encounter in this section is the evaluation of the 15-dimensional collision integral matrix elements in Eqs.~\eqref{eqn_coll_matr_in}-\eqref{eqn_coll_matr_fin}. We remark that the energy and momentum conservation laws render these integrals to be 11-dimensional. Further simplifications can occur if one uses symmetries of the optical trap. Nevertheless, the resulting expressions are still too complicated for accurate numerical evaluations.
So far, this issue has been a major computational challenge, which is the primary reason why the method of moments has not been applied in its full capacity to three-dimensional problems. We note that in two spatial dimensions, such matrix elements encode only 5-dimensional integrals, which could be accurately evaluated using Monte Carlo sampling~\cite{babadi2012collective}.

In this work, we provide an efficient method for computing these collision integrals. Our approach is based on the observation that the integrals are hard to evaluate because of the complexity of the equilibrium Bose-Einstein and Fermi-Dirac distribution functions. To further appreciate this point, we note that in a classical system, where the distribution functions have Gaussian (Boltzmann-Maxwell) forms, one could compute the collision integrals analytically, by means of Wick's theorem~\cite{PhysRevA.60.4851}. With this in mind, in the following section, we offer an approximation to the equilibrium distribution functions that, on the one hand, can be made arbitrarily accurate and, on the other hand, also acquire a Gaussian-like structure, in turn enabling us to semi-analytically evaluate the collision matrix with essentially arbitrary precision.

\section{Gaussian mixture model}
\label{sec:GMM}

\begin{figure*}[ht!]
\centering
\includegraphics[width=1\linewidth]{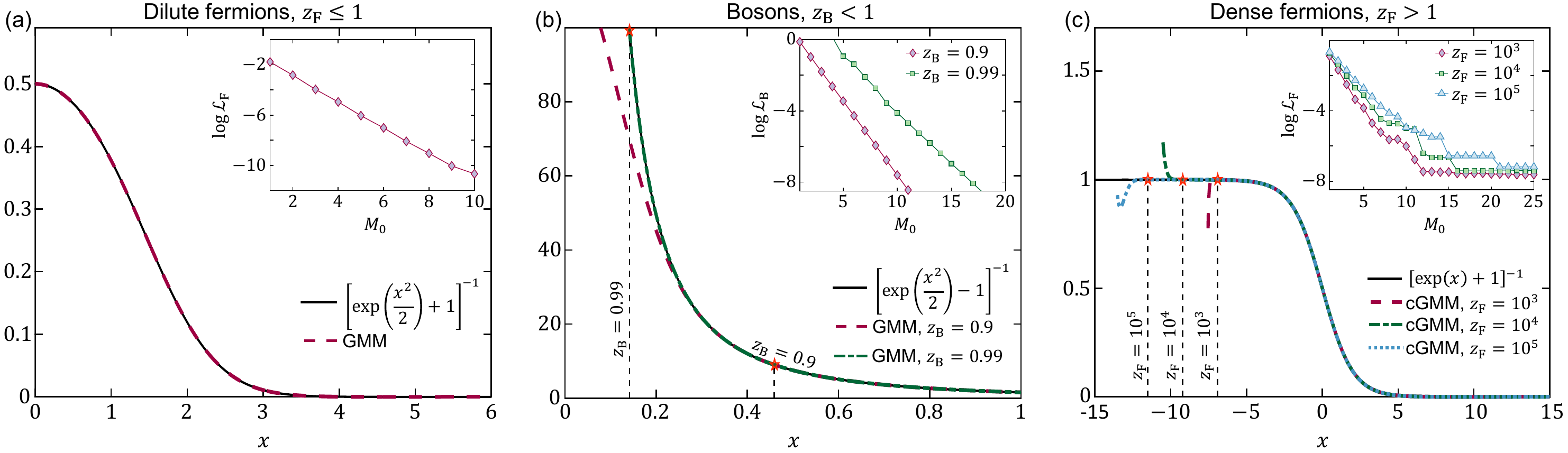} 
\caption{ Performance of the Gaussian mixture model (GMM). (a) The fermionic function $f_1(x)$ (solid line), Eq.~\eqref{eqn:f_1(x)}, that encodes the Fermi-Dirac distribution function for any $z_{\rm F} \leq 1$, and its GMM fit (dashed line) are not distinguishable by eye from each other. The inset shows the dependence of the absolute error of the fit ${\cal L}_{\rm F}$, Eq.~\eqref{eqn:L_F_1}, on the total number of Gaussians $M_0$ that are involved in the fitting: for $M_0 = 10$, we obtain ${\cal L}_{\rm F}\lesssim 10^{-10}$.
(b) Similar analysis as in (a) but now we consider the bosonic function $f_2(x)$ (solid line), Eq.~\eqref{eqn:f_2(x)}, in a range $x\in [x_{\rm min},\infty)$, where $x_{\rm min}$ (vertical dashed lines) is set by the maximal fugacity we consider: $x_{\rm min} \approx 0.46 $ for $z_{\rm B} = 0.9$ and  $x_{\rm min} \approx 0.14 $ for $z_{\rm B} = 0.99$. The GMM fits provide excellent approximations to $f_2(x)$ in their respective domains of applicability (dashed line for $z_{\rm B} = 0.9$ and solid dashed line for $z_{\rm B} = 0.99$): The inset shows that the bosonic error ${\cal L}_{\rm B}$, Eq.~\eqref{eqn:L_B}, can be as small as $10^{-8}$ with only about 10 (17) Gaussians for $z_{\rm B} = 0.9$ ($z_{\rm B} = 0.99$).
(c) The fermionic function $f_3(x)$ (solid line), Eq.~\eqref{eqn:f_3(x)}, that captures the Fermi-Dirac distribution function for any $1\leq z_{\rm F}\leq z_{\rm max}$, can be well fitted with the complex version of the GMM, Eq.~\eqref{eqn:cGMM}. Here, the domain $x\in[x_{\rm min},\infty)$ is determined by the maximal fermionic fugacity we consider: $x_{\rm min} = -\ln (z_{\rm max})$.
As shown in the inset, the absolute error of the fit ${\cal L}_{\rm F}$ decreases with increasing the total number $M_0$ of complex conjugate Gaussian pairs; for $M_0 \lesssim 25$, we obtain ${\cal L}_{\rm F} \simeq 10^{-8}$ at $z_{\rm max} = 10^5$.
}
\label{fig::GMM fitting}
\end{figure*}

In this section, we argue that locally, i.e., for a given real-space point $\bm r$, the equilibrium distribution functions (both bosonic and fermionic) can be accurately approximated using the following ansatz:
\begin{equation}
    n_{\rm GMM}(\bm r, \bm p) = \sum_{n = 1}^{M_0} a_n(\bm r) \exp\Big(-\frac{p^2}{2\gamma_n^2(\bm r)} \Big) \label{eqn:GGM_gen}.
\end{equation}
Here $M_0$ is the total number of Gaussians that we take into consideration, implying that we have $2\times M_0$ fitting parameters $a_n(\bm r)$ and $\gamma_n(\bm r)$ that, in general, are allowed to depend on the real-space coordinate $\bm r$. In the machine learning literature~\cite{bishop2006pattern}, the ansatz in Eq.~\eqref{eqn:GGM_gen} is referred to as Gaussian mixture model (GMM), though in contrast to the most generic GMM, here we set all the means to zero. In the following subsections, we first consider a few situations of increasing complexity, where we explicitly construct such GMM fittings and discuss their limitations. Then, in subsection~\ref{subsec:summary_fit}, we briefly summarize and further discuss the presented approach.

\subsection{Dilute fermions, $z_{\rm F} \leq 1$}

We begin by considering the Fermi-Dirac distribution function, which in the appropriately chosen units, can be written as:
\begin{align}
    n_{\rm F} (z, p) = \Big[ \frac{1}{z} \exp{\Big(\frac{p^2}{2}\Big)}  + 1\Big]^{-1}.
    \label{eqn:n_F_eq_v0}
\end{align}
The local fugacity $z = z(\bm r)$ is a monotonic function of the density, and, for reasons that will become apparent shortly, we first focus on the case $z \leq 1$, corresponding to dilute fermions. It is then suggestive to change variables as $z = \exp(-q^2/2)$ and consider the following one-variable function:
\begin{align}
    f_1(x) = \Big[ \exp{\Big(\frac{ x^2}{2}\Big)}  + 1\Big]^{-1}.
    \label{eqn:f_1(x)}
\end{align}
In other words, $f_1(x)$ is nothing but the Fermi-Dirac distribution for $z = 1$. If this function $f_1(x)$ can be accurately approximated with a GMM
\begin{align}
    f_1(x) \approx \sum_n \alpha_n \exp\Big(-\frac{x^2}{2\gamma_n^2} \Big),
    \label{eqn:f1_GMM}
\end{align}
we immediately get a GMM approximation for $n_{\rm F}(z,p)$ for any $z \leq 1$:
\begin{align}
    n_{\rm F} (z, p) &  = f_1(p^2 + q^2)  \notag\\
    & \approx \sum_n \alpha_n \exp\Big(-\frac{q^2}{2\gamma_n^2} \Big) \exp\Big(-\frac{p^2}{2\gamma_n^2} \Big)\notag\\
    & = \sum_n \alpha_n z^{1/\gamma_n^2} \exp\Big(-\frac{p^2}{2\gamma_n^2} \Big).
    \label{eqn:n_z<1_GMM}
\end{align}
Importantly, the quality of this fit for $n_{\rm F}(z,p)$ is entirely determined by the quality of the approximate form~\eqref{eqn:f1_GMM} for $f_1(x)$. Moreover, by comparing Eqs.~\eqref{eqn:GGM_gen} and~\eqref{eqn:n_z<1_GMM}, we observe that the amplitudes $a_n(\bm r)$ in Eq.~\eqref{eqn:GGM_gen} do explicitly depend on the local fugacity $z$ (and, as such, on the real-space coordinate $\bm r$), whereas the variances $\gamma_n$, which are the same for both Eq.~\eqref{eqn:GGM_gen} and Eq.~\eqref{eqn:n_z<1_GMM}, are universal, i.e., independent of $z$.

In our numerics, we optimize the coefficients $\alpha_n$ and $\gamma_n$ using standard fitting routines that eventually minimize the absolute error
\begin{align}
{\cal L}_{\rm F} = \max_x |f_1(x)- f_{1,\rm GMM}[\alpha_n,\gamma_n](x)|.
\label{eqn:L_F_1}
\end{align}
Specifically, we use the Levenberg-Marquardt optimization algorithm, which we supply with gradients, that is implemented in Matlab via the {\it lsqcurvefit} function. The result of such fitting is shown in Fig.~\ref{fig::GMM fitting}(a): Remarkably, we find that ${\cal L}_{\rm F}$ can be made as small as $\sim 10^{-10}$ with only $\sim 10$ Gaussians and can be further reduced by increasing the total number of Gaussians $M_0$ [see the inset of Fig.~\ref{fig::GMM fitting}(a)]. While $f_1(x)$ is expected to be well approximated with a GMM~\cite{kammler1976chebyshev}, what is a bit surprising is how few Gaussians turn out to be needed for such a high quality of fitting.

To further appreciate this point, we note that for $z < 1$, one can formally Taylor expand the fermionic distribution function:
\begin{align}
    n_{\rm F}(z,p) = \sum_{n = 1}^\infty (-1)^{n + 1} z^n \exp\Big( -\frac{1}{2} n p^2 \Big).
    \label{eqn:TS_fermions}
\end{align}
The fact that this expansion exists in the first place confirms the expectation that the GMM in Eq.~\eqref{eqn:GGM_gen} can, in principle, be made arbitrarily accurate. We note, however, that the results in Fig.~\ref{fig::GMM fitting}(a) clearly demonstrate that the GMM significantly outperforms the truncated Taylor series, especially for $z$ approaching unity, where more and more Gaussians in the Taylor series are needed to keep the same quality of the approximation. For instance, one needs to consider about $150$ Taylor series coefficients for $z = 0.9$ to get the same precision as one has from only $10$ GMM Gaussians. Importantly, the cost of computing the collision matrices scales as $\sim M_0^4$, which implies that the GMM dramatically facilitates otherwise too demanding calculations. We also remark that both the GMM in Eq.~\eqref{eqn:n_z<1_GMM} and the Taylor series in Eq.~\eqref{eqn:TS_fermions} share the property that their variances $\gamma_n$ do not depend on $z$.

We briefly comment that in our numerics, perhaps the most non-trivial step turns out to be the  initial choice of the fitting parameters for the optimization algorithm. Whenever available, one can start from the coefficients of the truncated Taylor series, which we find then lead to excellent GMM fits. In more complicated situations, where Taylor series is no longer an option, one can try to sequentially sample over random Gaussians with subsequent optimization. This approach gives fits of high precision, but it can be further improved by simple post-processing that then gives a more efficient fit with fewer Gaussians, as we discuss below.

\subsection{Bosons}

We switch to the analysis of the Bose-Einstein distribution function:
\begin{align}
    n_{\rm B} (z, p) = \Big[ \frac{1}{z} \exp{\Big(\frac{p^2}{2}\Big)}  - 1\Big]^{-1}. \label{eqn:n_B_eq_v0}
\end{align}
For bosons, the local fugacity has to be smaller than unity $z \leq 1$ so that essentially all the discussion from the preceding subsection applies here as well. One notable difference, however, is that the bosonic one-variable function $f_2(x)$, now defined as
\begin{align}
    f_2(x) \equiv  n_{\rm B} (x)\Big|_{z = 1} = \Big[ \exp{\Big(\frac{ x^2}{2}\Big)}  - 1\Big]^{-1}
    \label{eqn:f_2(x)},
\end{align}
diverges for $x\to 0$. In particular, this implies that the function $f_2(x)$ cannot be fitted with a GMM~\eqref{eqn:GGM_gen} in the entire range $x \in [0,\infty)$. In practice, this is not an issue because one can always choose a maximal bosonic fugacity $z_{\rm max}$ that cuts off the range to $x \in [x_{\rm min},\infty)$, with $x_{\rm min} = \sqrt{-2\ln{(z_{\rm max})}}$. Physically, by choosing such a cutoff, one limits the analysis to temperatures $T$ outside a small window near $T_c$ since the bosonic fugacity can approach unity only in the immediate vicinity of the BEC transition. For instance, based on mean-field theory, the choice of $z_{\rm max} = 0.9$ would correspond to temperatures $|T-T_c|/T_c \gtrsim 5\%$; and $z_{\rm max} = 0.99$ implies that $|T-T_c|/T_c \gtrsim 1\%$. One can also increase the cutoff $z_{\rm max}$ closer to unity if one wants to investigate the detailed behavior near $T_c$. This will come at the cost that the fitting near $T_c$ becomes more challenging and requires more Gaussians [see Fig.~\ref{fig::GMM fitting}(b)].

Figure~\ref{fig::GMM fitting}(b) shows the GMM fits to the bosonic distribution $n_{\rm B}(z,p)$: for $z=0.9$ and $z=0.99$, we can reduce the bosonic error 
\begin{align}
{\cal L}_{\rm B} 
& = \max_{x \geq x_{\rm min}} |f_2(x) - f_{\rm GMM}(x)|
\label{eqn:L_B}
\end{align}
to $10^{-8}$ with as few as $10$ and $17$ Gaussians, respectively. If needed, this precision can be further improved by increasing $M_0$ -- see the inset of Fig.~\ref{fig::GMM fitting}(b). Once the GMM fit at $z_{\rm max}$ is established
\begin{align}
    n_{\rm B}(z_{\rm max},p) \approx \sum_n a_{n,z_{\rm max}} \exp\Big(-\frac{p^2}{2\gamma_n^2} \Big),
    \label{eqn:n_F_z0_p}
\end{align}
we immediately get the same quality GMM approximation for any $z \leq z_{\rm max}$:
\begin{align}
    n_{\rm B}(z,p) \approx \sum_n a_{n,z_{\rm max}} (z/z_{\rm max})^{1/\gamma_n^2} \exp\Big(-\frac{p^2}{2\gamma_n^2} \Big).
    \label{eqn:n_F_z1_p}
\end{align}
Notably, as Fig.~\ref{fig::GMM fitting}(b) illustrates, the fitting at a given $z_{\rm max}$ can be extrapolated (using Eqs.~\eqref{eqn:n_F_z0_p} and~\eqref{eqn:n_F_z1_p}) to even get a reasonably good approximation for $z > z_{\rm max}$.

\subsection{Dense fermions, $z_{\rm F} > 1$}

We turn to discuss the fermionic distribution function $n_{\rm F}(z,p)$ in the regime $z >1$, corresponding to dense fermions. Following the preceding analyses, now it is suggestive to change variables as $z = \exp(q^2/2)$ and consider the function
\begin{align}
    f_3(x) = [\exp(x) + 1]^{-1}\label{eqn:f_3(x)}
\end{align}
so that $n_{\rm F}(z,p) = f_3((p^2 - q^2)/2)$. Since, in principle, one can have arbitrary large fermionic fugacities, it implies that generically $x\in (-\infty,\infty)$. In practice, however, the range of the variable $x$ is set by the maximal fermionic fugacity $z_{\rm max}$ that we consider, i.e., $x\in [x_{\rm min},\infty)$ with $x_{\rm min} = - \ln(z_{\rm max})$. Physically, $z_{\rm max}$ is essentially set by the fermionic density and local temperature.

To be consistent with the preceding subsections and to allow for negative values of $x$, a GMM fit to the function $f_3(x)$ translates as a sum of exponentials (rather then a sum of Gaussians), which we write as:
\begin{align}
    f_3(x) \approx \sum_n \alpha_n \exp\Big( - \frac{x}{\gamma_n^2} \Big).
    \label{eqn:cGMM}
\end{align}
In contrast to what we have developed so far, here we now allow for the coefficients $\alpha_n$ and $\gamma_n$ to be complex (and we only require $\text{Re}\,\gamma_n^2 > 0$), as we find that to fit the function $f_3(x)$ with purely real parameters, assumed in Eq.~\eqref{eqn:GGM_gen}, turns out to be numerically demanding, especially for very large fugacities $z \gg 1$. This extended class of functions incorporates exponentially decaying trigonometric functions, which can fit the initial plateau of $f_3(x)$ substantially more accurately and with fewer resources than real exponentials. Importantly, the collision integrals in Eqs.~\eqref{eqn_coll_matr_in}-\eqref{eqn_coll_matr_fin} can still be expressed analytically within this new representation. In what follows, we will refer to the model in Eq.~\eqref{eqn:cGMM} as complex GMM (cGMM). We finally remark that to get an approximation for $n_{\rm F}(z,p)$ for any $1\leq z \leq z_{\rm max}$, one can use Eq.~\eqref{eqn:n_z<1_GMM} and Eqs.~\eqref{eqn:n_F_z0_p}-\eqref{eqn:n_F_z1_p}, with the only difference that the corresponding coefficients are now complex.

Since the function $f_3(x)$ is real, we consider $M_0$ complex conjugate pairs so that we now have $4\times M_0$ fitting parameters. Figure~\ref{fig::GMM fitting}(c) summarizes the results of cGMM fits: we find that the absolute fermionic error ${\cal L}_{\rm F}$, defined similarly as above, of the fit at $z_{\rm max} = 10^5$ can be reduced to $\sim 10^{-8}$ with only about $M_0 = 21$ Gaussian pairs. As we show in the inset of Fig.~\ref{fig::GMM fitting}(c), this error ${\cal L}_{\rm F}$ decreases with increasing $M_0$. However, there are regions where ${\cal L}_{\rm F}$ plateaus at some values, meaning that the addition of new Gaussian pairs does not improve the quality of the fit, which could possibly be improved by using better initialization of the fitting parameters -- we leave this issue to future work.
Furthermore, this expectation is supported by the fact that a simple post-processing, where we remove as many redundant Gaussians as possible with subsequent re-optimization, gives a notably better result. specifically, we obtain the target precision ${\cal L}_{\rm F}\leq 10^{-7}$ at $z_{\rm max} = 10^5$ with only $M_0 = 17$ complex conjugate Gaussian pairs.

\subsection{Summary of the section}
\label{subsec:summary_fit}

We have numerically shown that one can achieve a highly accurate numerical representation of the equilibrium distribution functions using a mixture of Gaussians. Crucially, this representation makes the evaluation of collision integrals in Eqs.~\eqref{eqn_coll_matr_in}-\eqref{eqn_coll_matr_fin} numerically tractable, as we further discuss below. The quality of such fits is controlled solely by the total number of Gaussians (complex Gaussian pairs) $M_0$ used in the fitting model. In our numerics, we first get self-consistent equilibrium distribution functions  and then, for each real-space point $\bm r$, we obtain a GMM approximation following the procedures outlined in the preceding three subsections; the resulting absolute errors are ensured to be smaller than $10^{-7}$ in all regimes that we investigate in the remainder of the paper. 

We have also shown that not only the GMM dramatically outperforms the truncated Taylor series approximation, its complex analogue, cGMM, gives excellent fits in the regimes where the Taylor series expansion is not even available. Since only a few Gaussians are needed to get outstanding-quality fits, it renders calculations of the collision integrals numerically efficient and accurate.

The developed here approach is the central technical result of this paper, as it allows one to study a broad class of quantum many-body systems describable within some Boltzmann equation. In the following section, motivated by the most recent cold-atom experiments, we apply this framework to investigate monopole and quadrupole collective modes of a trapped Bose-Fermi mixture.

We note in passing that a cautious reader might argue that our framework applies specifically to systems with contact $s$-wave interactions.
Indeed, when interactions are explicitly non-local, the resulting Hartree-Fock self-energies, which enter into the equilibrium distribution functions through the self-consistency Eq.~\eqref{eqn:eqm}, can become momentum-dependent. In general, this makes the equilibrium forms in Eqs.~\eqref{eqn:n_F_eq_v0} and~\eqref{eqn:n_B_eq_v0} no longer applicable. While the full equilibrium distribution functions are generically more complicated, we expect that they still can be accurately approximated by 
the GMM or cGMM, upon a straightforward extension of our optimization routines.

\section{Collective modes of a trapped Bose-Fermi mixture}
\label{sec:bose-fermi}

The formalism developed in the preceding two sections enables us to efficiently evaluate all of the method-of-moment matrices in Eq.~\eqref{eqn:MoM} for three-dimensional systems. With the GMM and cGMM representations of the equilibrium distribution functions, the multidimensional collision integrals  no longer constitute computational obstacles. In practice, however, actual calculations turn out to be quite cumbersome, and, for this reason, detailed derivations, as well as our hybrid numerical-analytical scheme for the collision integrals,  are relegated to Appendices~\ref{Appendix_monopole} and~\ref{Appendix_quad}. In Appendix~\ref{appendix: benchmarking}, we present a series of benchmarks and applications of the collision integral calculations, for various physical systems of interest. 

Equipped with the method of moments, we turn to investigate the collective modes of a quantum Bose-Fermi mixture trapped in a cigar-shaped optical potential. In Sec.~\ref{subsec:details_mom}, we first briefly discuss the choice of basis functions and model parameters. Our main findings are then split into three parts: In Sec.~\ref{subsec:bosonic_fluid}, we study a single-component interacting bosonic fluid, with the primary result being the crossover from the collisionless regime to collision-dominated hydrodynamics. The physics of the quantum mixture, when both fluids are present and interact with each other, is rich so that we consider two limiting situations: the case of dilute fermions $N_{\rm F} \ll N_{\rm B}$, further refereed to as the impurity regime, is studied in Sec.~\ref{subsec:Impurity regime}, and the case with $N_{\rm F} \simeq N_{\rm B}$, the mixture regime, is discussed in Sec.~\ref{subsec:Mixture regime}. There we explore coherent mixing between the bosonic and fermionic monopole modes, their dampening due to incoherent scatterings, and the emergent hydrodynamics characterized by synchronization of the bosonic and fermionic responses.

\subsection{Basis functions and model parameters}
\label{subsec:details_mom}

So far we have not made any assumptions about the trapping potentials or the basis functions $\phi_{\alpha}(\bm r,\bm p)$. Below we consider cigar-shaped traps with cylindrical symmetry. In this case, the external trap modulations and the generated by this perturbation responses will have the same $z$-axis symmetry. This allows us to classify the collective modes and efficiently choose the corresponding basis functions $\phi_{\alpha}(\bm r,\bm p)$ -- see Ref.~\cite{babadi2012collective} for a related discussion in two-dimensional pancake-like geometry. In what follows, we investigate two types of collective excitations corresponding to the radial breathing monopole mode [Fig.~\ref{fig::bosons_monopole}a] and to the quadrupole mode [Fig.~\ref{fig::bosons_quad}a].

The monopole mode represents an excitation with zero $z$-axis angular momentum, allowing one to restrict the analysis to the following basis functions:
\begin{align}
    \phi_\alpha(\bm r,\bm p) = \rho^{2m_\alpha} p_\rho^{2n_\alpha} (\bm \rho \cdot\bm p_\rho )^{k_
    \alpha} \times z^{2p_\alpha} p_z^{2q_\alpha} (z p_z)^{r_\alpha}. \label{eqn: basis-mon}
\end{align}
Here $m_\alpha + n_\alpha + k_\alpha + p_\alpha + q_\alpha + r_\alpha \leq M$, $r_\alpha = \{0,1\}$, and $M$ sets the truncation order. At first order, we have 7 functions $\{1, \rho^2,p_\rho^2,\bm \rho \cdot \bm p_\rho,z^2,p_z^2,zp_z \}$, and this simplified choice of basis functions was shown to perfectly capture the monopole mode of a classical gas~\cite{PhysRevA.60.4851}. Specifically, since this gas 
follows the Boltzmann-Maxwell distribution, one can evaluate all of the method-of-moments matrices analytically, and the result of such an analysis agrees well with more precise molecular-dynamics simulations~\cite{PhysRevA.60.4851}. Besides, the results of Ref.~\cite{PhysRevA.60.4851} helped us test our collision integral computations -- see Appendix~\ref{appendix: benchmarking}. The situation we consider here, of a quantum system with more complicated distribution functions, is much more challenging and likely not feasible analytically because of the complexity of the collision integral matrix elements.
We provide expressions for all of the monopole matrices, for arbitrary $M \geq 1$,  in Appendix~\ref{Appendix_monopole}. Below we investigate the monopole spectral function, which for bosons is defined as:
\begin{align}
    {\cal A}^{\rm B}_{\rho^2}(\omega) = - \frac{1}{N_{\rm B}}\text{Im}\{\chi^{\rm B}_{\rho^2}(\omega)\},
\end{align}
where $\chi^{\rm B}_{\rho^2}(t) \equiv (\langle \rho^2 \rangle_t - \langle \rho^2 \rangle_{t=0})$ is the response of the transverse bosonic cloud size to a trap modulation of the form: $\delta U_{\rm B}(t) \propto \rho^2 \delta(t)$. Analogous spectral function can be introduced for fermions as well, but we return to this below.

The quadrupole mode is generated by a perturbation of the type $\delta U_{\rm B/F}\sim (x^2 - y^2)$, and the corresponding basis functions are given by $\xi_{ i = 1,2,3}(\bm r,\bm p)\times \phi_{\alpha}(\bm r,\bm p)$, where
\begin{align}
    \xi_1 = x^2 - y^2,\quad \xi_2 = x p_x - yp_y,\quad \xi_3 = p_x^2 - p_y^2.
\end{align}
We note that $2(\bm \rho \cdot \bm p_\rho)\times \xi_2 = p_\rho^2 \xi_1 + \rho^2\xi_3$, implying that a function proportional to $(\bm \rho \cdot \bm p_\rho)\times \xi_2$ should be removed from this basis set since it can be represented as a linear superposition of the rest of the basis functions~\cite{babadi2012collective}. For instance, this means that for $M = 1$ we have only 20 functions. In Appendix~\ref{Appendix_quad}, we analytically evaluate all of the corresponding matrices entering the linearized kinetic equations, and the resulting expressions are then used in our numerical simulations.
The quadrupole spectral function for bosons is defined as:
\begin{align}
    {\cal A}^{\rm B}_{x^2-y^2}(\omega) = - \frac{1}{N_{\rm B}} \text{Im}\{\chi^{\rm B}_{x^2-y^2}(\omega)\},
\end{align}
where $\chi^{\rm B}_{x^2-y^2}(t) \equiv( \langle x^2 \rangle_t - \langle y^2 \rangle_t )$ encodes the quadrupole response of the bosonic cloud to a trap modulation of the form: $\delta U_{\rm B}(t) \propto (x^2 - y^2) \delta(t)$.

Let us now briefly comment on the validity of the Boltzmann equation to describe the Bose-Fermi mixture. We need to check two conditions: 1) characteristic range of the inter-particle interactions is smaller than the typical travel distance between collisions; 2) it is sufficient to use scattering length to describe the low-energy part of the scattering amplitude, $f(k)=-a$, and neglect the full momentum dependence of $f(k)$. The lower bound on the travel distance is given by the distance between the atoms. Hence the former condition is guaranteed provided that the distance between bosons (which we always take to be denser than fermions) is larger than both scattering lengths. To discuss the second condition we recall that in three spatial dimensions the $s$-wave scattering amplitude can be written as~\cite{levinsen2011atom,landau2013quantum}:
\begin{align}
    f(k) = - \frac{1}{a^{-1} + ik - \frac{1}{2}r_0 k^2},
\end{align}
where $k$ is the relative wave vector of the scattering particles, $a$ is the scattering length, and $r_0$ is the effective range of interactions. The use of the Born-Markov approximation for the collision integrals is fully justified if one can neglect the momentum dependence of $f(k)$, i.e., one requires $k a \lesssim 1$. If one uses $k = n^{1/3}$ as the typical momentum set by the interparticle separation at low temperatures, then we find that both conditions for the validity of the Boltzmann approach are satisfied provided that $a_{\rm B} \lesssim 2\times 10^4 a_0$, with $a_0$ being the Bohr radius. 
For this estimate, we used the bosonic density at the center of the trap, $N_{\rm B} = 10^6$, and parameters of the MIT experiment~\cite{Yan2023} on a $^{40}\text{K}-^{23}$Na mixture: $\omega_z = 2\pi\times 12.2\,$Hz, $\omega_\rho = 2\pi\times 97\,$Hz, and $\lambda \approx 2.4$. If instead one considers the thermal momentum $k \simeq \sqrt{m T}$, then $a_{\rm B} \lesssim 4 \times 10^5 a_0$ for $T\simeq 500\,$nK (for comparison, the mean-field transition temperature is $T_c \approx 220\,$nK). The thermal momentum constraint becomes more essential at higher temperatures, away from the quantum regime.
In our numerical analyses below, we, therefore, restrict ourselves to $|a_{\rm B}|, |a_{\rm BF}| \leq 10^4 a_0$. In Appendix~\ref{appendix: benchmarking}, we discuss that our approach can be directly applied to systems with ultra-strong interactions, where one is required to keep the full $k$-dependence of the scattering amplitude.

\subsection{One-component interacting fluid}
\label{subsec:bosonic_fluid}

\begin{figure}[t!]
\centering
\includegraphics[width=1\linewidth]{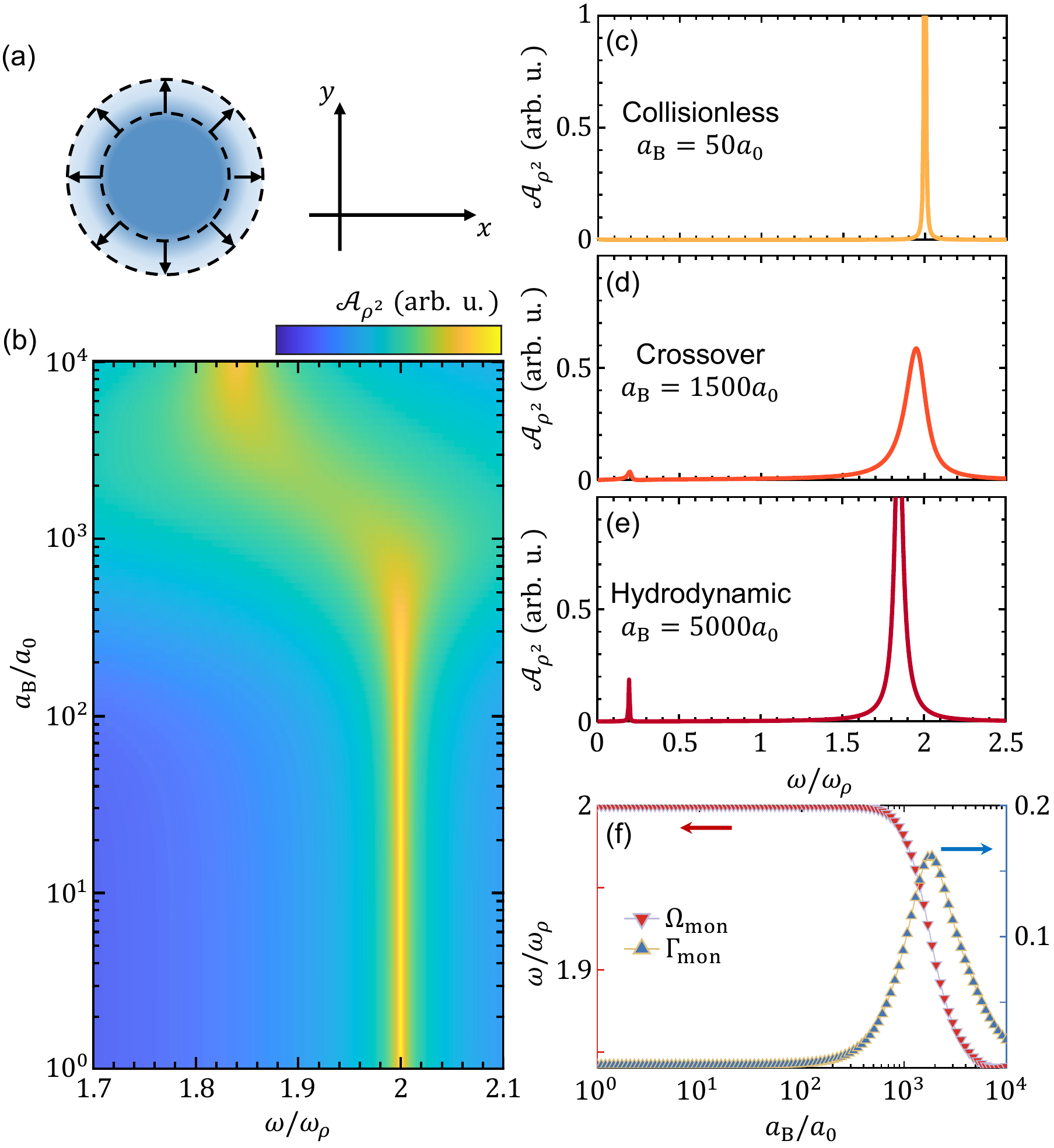} 
\caption{Collisionless-to-hydrodynamics crossover in a single-component bosonic fluid, the case of monopole mode schematically shown in (a). (b) The bosonic spectral response ${\cal A}_{\rho^2}$ as a function of frequency $\omega$ and for different interaction strengths, as encoded in $a_{\rm B}$. When the Bose-Bose interactions are weak, one gets a sharp response at $\omega \simeq 2\omega_\rho$; as $a_{\rm B}$ is increased, the system starts to exhibit a many-body hydrodynamic response, as evidenced by the fact that first, the spectral function completely broadens but then reforms again into a sharp peak at a frequency notably different from $2\omega_\rho$ [see also a few cuts shown in (c,d,e)]. This is further illustrated in (f) as the behavior of the position $\Omega_{\rm mon}$ and width $\Gamma_{\rm mon}$ of the peak in (b), obtained from a single Lorentzian fit. In the hydrodynamic regime, transverse and longitudinal monopole modes mix, resulting in the emergence of an additional weak peak at  the longitudinal frequency $\omega\simeq2\omega_z$ (e).
Parameters used: $\omega_z = 2\pi\times 12.2\,$Hz, $\omega_\rho = 2\pi\times 97\,$Hz, $T = 500\,$nK, $N_{\rm B} = 10^6$, and $M = 2$.  
}
\label{fig::bosons_monopole}
\end{figure}

We begin by investigating a single-component bosonic fluid, i.e., for now, we assume that there are no fermions. Figure~\ref{fig::bosons_monopole} summarizes our results for the monopole mode in this case. We find that when the coupling $g_{\rm B}$ (or equivalently $a_{\rm B}$) is weak, the fluid behaves as a non-interacting Bose gas, and, as such, the monopole spectral function ${\cal A}_{\rho^2}(\omega)$
exhibits a sharp peak in frequency at $\omega = 2\omega_{\rho}$. This regime is further referred to as collisionless [Fig.~\ref{fig::bosons_monopole}c]. 
Upon increasing $a_{\rm B}$ (using, for instance, a magnetic Feshbach resonance), we first observe that  the resonance in ${\cal A}_{\rho^2}(\omega)$ becomes dramatically broader [Fig.~\ref{fig::bosons_monopole}d] -- the intuition behind this behavior is that the fluid can no longer be considered as non-interacting, since various sound-like excitations of the system mix and, as such, can broaden the signal. Remarkably, however, upon further increase of $a_{\rm B} \gtrsim 5\times 10^3 a_0$, this peak reforms at a notably different frequency, and it becomes sharper with $a_{\rm B}$ [Fig.~\ref{fig::bosons_monopole}e].
We interpret this result as the system entering the hydrodynamic regime.
Interactions are so strong in this regime that the collective response of the fluid to a slow macroscopic perturbation is described via dynamics where the system is locally always in equilibrium. 
Let us also remark that the width of the monopole mode in the hydrodynamic regime [Fig.~\ref{fig::bosons_monopole}f] is related to the viscosity of the fluid. In other words, by measuring collective modes in optical traps, one can extract information about transport coefficients~\cite{jensen1989transport,massignan2005viscous,bruun2007shear,rupak2007shear,schafer2007ratio,ketterle2008making,sommer2011universal,baur2013collective,Schafer2009,cao2011universal,elliott2014anomalous,choudhury2020collective,yan2020bose}, and the method we have developed here enables one to accurately evaluate them and compare to such experiments. As a concrete demonstration of this capability of our approach, we analyse in Appendix~\ref{appendix: benchmarking} the shear viscosity of a two-component strongly-interacting Fermi fluid.

\begin{figure}[t!]
\centering
\includegraphics[width=1\linewidth]{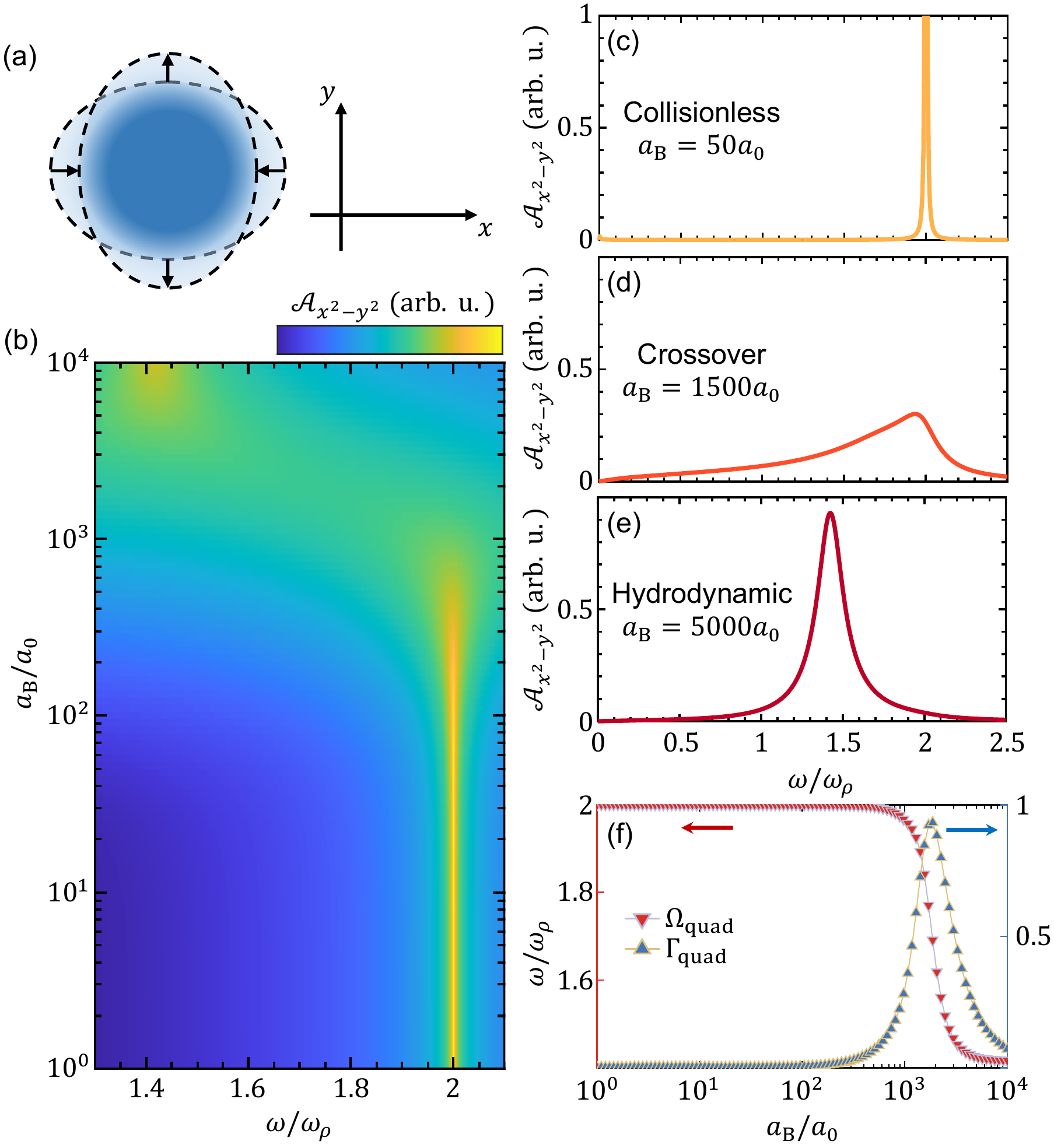} 
\caption{ The same analysis as in Fig.~\ref{fig::bosons_monopole} but for the quadrupole mode. (b) The quadrupole spectral function ${\cal A}_{x^2 - y^2}(\omega)$, similarly to ${\cal A}_{\rho^2}(\omega)$, exhibits a crossover from the collisionless regime, $a_{\rm B}\lesssim 10^3 a_0$, to a hydrodynamic one, $a_{\rm B}\gtrsim 5 \times 10^3 a_0$.
Just as it was for the monopole case, for weak interactions, the function ${\cal A}_{x^2 - y^2}(\omega)$ peaks in frequency at $\omega \simeq 2\omega_\rho$. As the interactions are increased, this peak 
first significantly broadens [this broadening is much stronger compared to the one seen in the crossover regime for the monopole mode in Fig.~\ref{fig::bosons_monopole}] and then reforms at around $\omega \simeq \sqrt{2}\omega_\rho$ -- this frequency is notably different from the corresponding monopole resonance in the hydrodynamic regime. 
}
\label{fig::bosons_quad}
\end{figure}

Noteworthy, we find that when the interactions are strong [Fig.~\ref{fig::bosons_monopole}d,e], the spectral function ${\cal A}_{\rho^2}(\omega)$ exhibits an additional weak resonance at $\omega\simeq 2\omega_z$. This resonance comes from the longitudinal monopole mode, which is far detuned from the transverse one for strongly anisotropic cigar-shaped traps with $\kappa \ll 1$.
The two modes mix with each other for $a_{\rm B} \neq 0$, explaining the appearance of the longitudinal resonance even when the trap perturbation is transverse.

Figure~\ref{fig::bosons_quad} summarizes our results for the quadrupole mode in an interacting bosonic fluid. Here we also find a similar collisionless-to-hydrodynamics crossover in the parameter range consistent with the above discussion. In the collisionless regime, the quadrupole spectral function ${\cal A}_{x^2 - y^2}(\omega)$, similarly to ${\cal A}_{\rho^2}(\omega)$, also exhibits a resonance at $\omega \simeq 2\omega_\rho$. Interestingly, as the hydrodynamic regime is approached, the quadrupole resonance red-shifts to a frequency notably different from the corresponding monopole peak. Typically, in cold-atom experiments, trapping potentials may be anisotropic,  which gives rise to a weak mixing between the monopole and quadrupole modes -- this feature might facilitate experiments, as these modes can be seen in a single measurement. Indeed, upon tuning the temperature or interaction strength, one could observe how the two modes, that are almost degenerate in the collisionless regime, split from each other in the hydrodynamic one. Another interesting aspect of the quadrupole mode is that it is much broader than the monopole one [see Fig~\ref{fig::bosons_quad}f]. This has to do with the fact that the phase space, associated with the scattering processes that contribute to the quadrupole linewidth, is significantly larger than the corresponding phase space associated with the monopole mode.

To sum up, we conclude that the overall behavior of the monopole and quadrupole modes is qualitatively similar to each other, except the lifetime of the quadrupole mode can be significantly shorter. For this reason, when considering the Bose-Fermi mixture below, we focus on the monopole mode.

\begin{figure*}[t!]
\centering
    \includegraphics[width=0.75\linewidth]{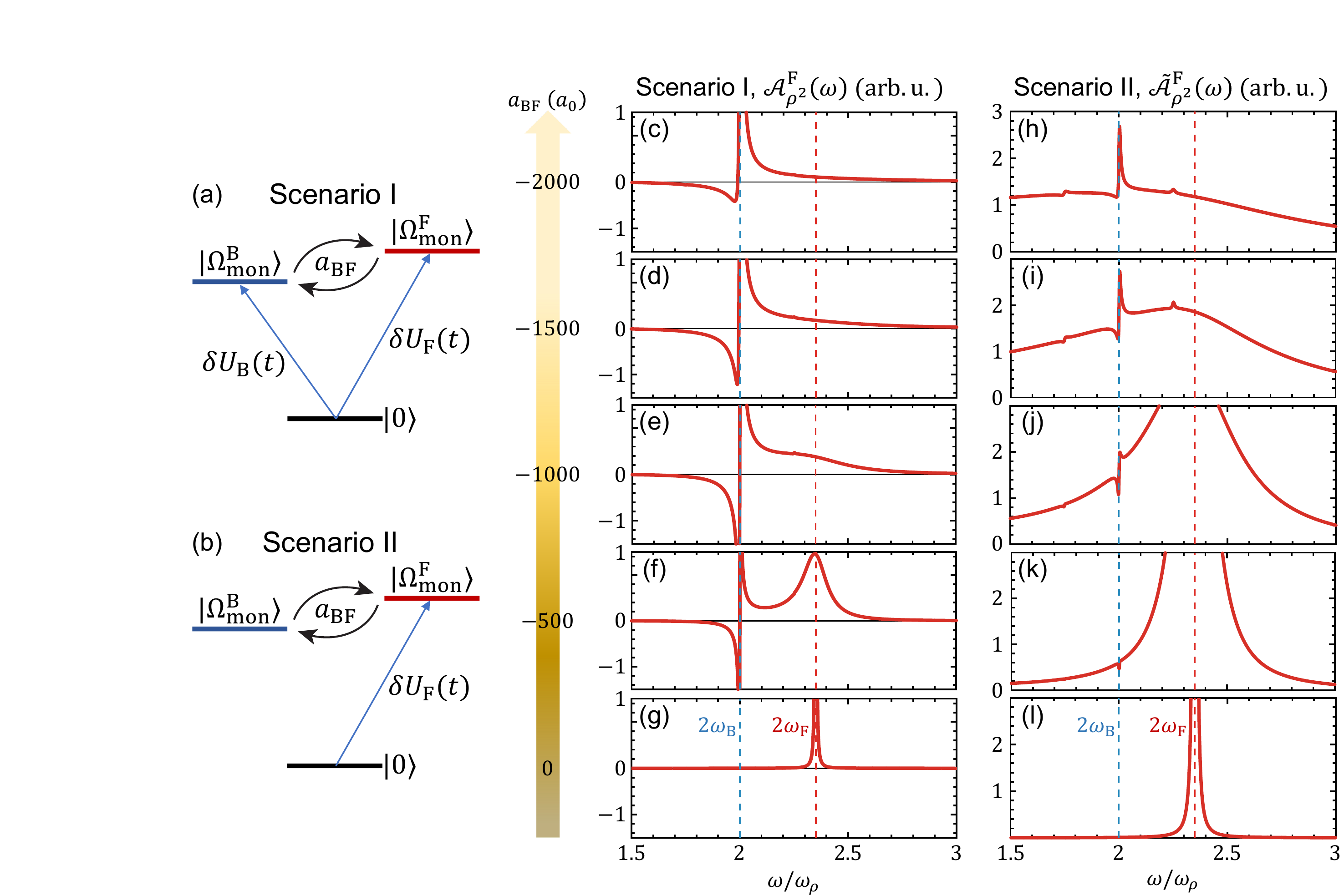} 
\caption{ Monopole modes in the impurity regime $N_{\rm F} = 10^{-2} N_{\rm B}\ll N_{\rm B}$. Effectively, the Bose-Fermi mixture represents a system of coupled harmonic oscillators, where the trap modulation launches both the bosonic and fermionic monopole modes, coupled to each other for $a_{\rm BF}\neq 0$. We consider two types of driving protocols in (a) and (b).
The experimental situation would correspond to the scenario I in (a), where the trap modulation perturbs both the bosonic and fermionic traps, in turn directly launching the two modes simultaneously. In the scenario II of (b), the trap modulation couples directly only to the fermionic monopole mode, which then can excite the bosonic one.
The bosons remain largely unaffected by the dilute fermions, and, for this reason, we focus here on the fermionic spectral responses.
The coherent mixing between the two types of monopole modes competes with the broadening of the fermionic resonance, as the dilute fermions can easily scatter off the bosons -- this is illustrated in panels (l) and (k) [corresponding to the scenario II], where the sharp in frequency fermionic peak for $a_{\rm BF} = 0$ (l) becomes much broader for $a_{\rm BF} = -500a_0$ (k), yet the coupling to the bosonic resonance remains barely notable in (k).
Upon further increasing the interaction strength $|a_{\rm BF}|$, panels (l)-to-(h), the fermionic resonance has largely disappeared, while the bosonic contribution from the back-action of the bosonic cloud becomes notable. In addition to the bosonic resonance, we observe weak sidebands that originate from the longitudinal monopole modes and their mixing with the transverse ones (see text).
In the scenario I (a), the optical modulation directly launches the bosonic monopole mode, leading to an interplay between the aforementioned fermionic broadening and coherent driving from the bosonic cloud. This is illustrated in panels (g)-to-(c), as the fermionic resonance becomes weaker and broader until it  is completely gone, while the mixing with the bosonic monopole mode quickly becomes strong and completely dominates the fermionic response for large $|a_{\rm BF}|$.
We note characteristic Fano interference profiles that come from the coherent mixing between the two types of monopole modes. For concreteness, here we showed the results for $a_{\rm B} = 0$, meaning that bosons are collisionless, but we emphasize that similar conclusions hold for other regimes.
}
\label{fig::figure-polaronic}
\end{figure*}

\subsection{Impurity regime $N_{\rm F} \ll N_{\rm B}$}
\label{subsec:Impurity regime}

Our analysis of the Bose-Fermi mixture starts from the limit of dilute fermions $N_{\rm F}\ll N_{\rm B}$. In this case, the bosonic response is barely affected by the presence of fermions, allowing us to focus on the fermionic spectral function. 

For $a_{\rm BF} = 0$, i.e., when the two fluids are not coupled to each other and the fermionic gas is non-interacting, fermions behave similar to bosons in the collisionless regime studied in the preceding subsection, except now the fermionic response picks up a resonance at $\Omega_{\rm mon}^{\rm F} = 2\omega_{\rho}^{\rm F}$ [Fig.~\ref{fig::figure-polaronic}(g),(l)]. This fermionic monopole frequency is generically different from the bosonic one because of the difference in
masses $m_{\rm F}/m_{\rm B}\approx 40/23$ and also because we choose $\lambda \approx 2.4$, encoding
the fact that in the MIT experiment~\cite{Yan2023}, the trapping potentials for bosons and fermions are different. As such, the Bose-Fermi mixture effectively represents a three-level system shown in Fig.~\ref{fig::figure-polaronic}(a),(b). In this sense, trap perturbations  act as if they excite the two monopole modes, which are coupled to each other for $a_{\rm BF} \neq 0$.

We consider two trap modulations shown in Fig.~\ref{fig::figure-polaronic}(a),(b). The scenario I in Fig.~\ref{fig::figure-polaronic}(a) corresponds to  an experiment where the trap perturbation simultaneously affects the bosonic and fermionic traps clouds and, as such, directly launches the two monopole modes. The respective fermionic response $A_{\rho^2}^{\rm F}(\omega)$ exhibits two key features. First, as the interaction strength $|a_{\rm BF}|$ is increased, the fermionic resonance at $2\omega_{\rm F}$ quickly becomes broad and weak, until it is no longer observable at large $|a_{\rm BF}|$. At the same time, $A_{\rho^2}^{\rm F}(\omega)$ acquires an additional contribution from the bosonic resonance, which eventually dominates the fermionic response. Second, for intermediate interactions, which are strong enough so that the bosonic spectral weight is appreciable but not too strong so that the fermionic peak is still sharp, we find coherent mixing between the bosonic and fermionic monopole modes. The spectral lines, even for large $|a_{\rm BF}|$, acquire characteristic Fano interference profiles typical of a system of coherently coupled harmonic oscillators~\cite{miroshnichenko2010fano}. A prerequisite for observing such profiles is that one of the excited states should be relatively long-lived. This suggests that the bosonic fluid should be either collisionless or hydrodynamic -- this aspect will become clearer when we discuss the mixture regime below.

The fermionic behavior at large $|a_{\rm BF}|$ is understood as the interplay between coherent driving from the bosons and incoherent broadening of the fermionic resonance. Indeed, as $|a_{\rm BF}|$ is increased, the dilute fermions can easily scatter off the relatively dense bosons, explaining the eventual disappearance of the fermionic resonance. Simultaneously, the bosonic monopole mode, which has been directly excited by the trap perturbation, acts as a coherent drive to the fermions and, as such, dominates the fermionic response.

This physical picture is further supported by the scenario II in Fig.~\ref{fig::figure-polaronic}(b), where the trap modulation directly launches only the fermionic monopole mode, which then can excite the bosonic one. The corresponding spectral function $\tilde{A}_{\rho^2}^{\rm F}(\omega)$ shows that by increasing $|a_{\rm BF}|$ from zero to $500 a_0$ [Fig.~\ref{fig::figure-polaronic}(l),(k)], the fermionic resonance becomes 
 dramatically broader, whereas the mixing with the bosonic resonance is barely observable. This is because the collision integral, which determines the linewidth of the fermionic resonance, scales as $a_{\rm BF}^2$, whereas the (mean-field) coherent coupling to bosons is only linear.  
Remarkably, upon further increasing $|a_{\rm BF}|$, the fermionic response becomes so broad in frequency that it has negligible spectral overlap with the initial sharp fermionic resonance. One could even be tempted to argue that as evidenced by this density-density-like response function, the fermion is essentially gone. Such a statement, however, is not entirely correct because the fermion still remains a well-defined quasiparticle, describable within the Landau Fermi-liquid theory that we employ here.

In principle, the fermionic response can become sharp again for larger values of $|a_{\rm BF}|$, i.e., upon entering the hydrodynamic regime. For dilute fermions, the values required for this are so large that we do not even consider this possibility. Such a situation can occur for reasonable parameters if the fermionic density is appreciable, as we discuss in the following subsection.

Besides this dramatic broadening of the fermionic resonance at large $|a_{\rm BF}|$, the spectral function $\tilde{A}_{\rho^2}^{\rm F}(\omega)$ also acquires a notable contribution from the bosonic resonance. To understand the physical picture, we note that the optical modulation launches the fermionic monopole mode, that is essentially featureless in frequency. Subsequently, this broadband fermionic mode drives the bosonic one, and the back-action from this long-lived excitation in turn gives rise to the relatively sharp bosonic feature in $\tilde{A}_{\rho^2}^{\rm F}(\omega)$. Notably, we find that $\tilde{A}_{\rho^2}^{\rm F}(\omega)$ also displays weak Stokes and anti-Stokes sidebands near the bosonic resonance. 
The higher energy sideband has frequency close to $\Omega_{\rm mon}^{\rm B}+ 2\omega_z^{\rm F}$, which is due to a process where one quantum of the transverse bosonic mode and one quantum of the longitudinal fermionic mode are both being excited. The lower sideband has frequency close to $\Omega_{\rm mon}^{\rm B}- 2\omega_z^{\rm F}$ and corresponds to a process where one quantum of the transverse bosonic mode is being excited and one quantum of the longitudinal fermionic mode is being depopulated. Both types of processes are possible for the thermal equilibrium state, the longitudinal and transverse monopole modes can mix for $a_{\rm BF}\neq 0$, as it was in Fig.~\ref{fig::bosons_monopole}, and, as such, the system seems to exhibit two additional well-defined collective modes, which manifest in the fermionic response through the same back-action mechanism as for the bosonic monopole mode. These sidebands, together with the Fano interference profiles discussed above, indicate that the Bose-Fermi mixture is expected to give rise to prominent and tunable nonlinearities -- this exciting research direction is left for future work.

\begin{figure}[t!]
\centering
    \includegraphics[width=1\linewidth]{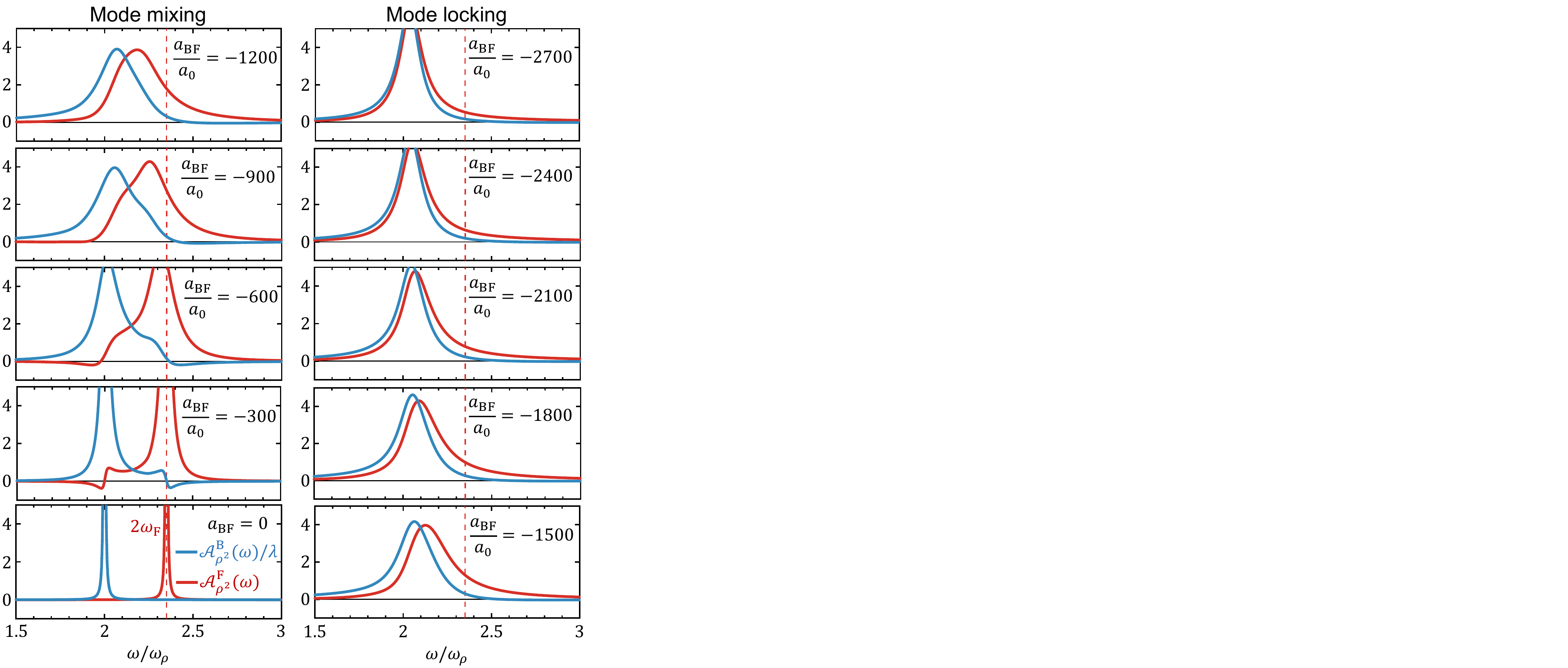} 
\caption{ Monopole modes in the mixture regime $N_{\rm F} = N_{\rm B}$, with bosons being nominally collisionless $a_{\rm B} = 50 a_0$.
When the Bose-Fermi coupling is not too strong, $|a_{\rm BF}|\lesssim 10^3 a_0$, we find that the bosonic and fermionic monopole modes exhibit both coherent mixing with each other and incoherent broadening of the respective resonances. For stronger interactions, $10^3 a_0\lesssim |a_{\rm BF}|\lesssim 2\times 10^3 a_0$, each of the spectral functions displays a single Lorentzian behavior, meaning that the coherent mixing between the two monopole modes is no longer observable -- this regime can be associated with the crossover region. Notably, as $|a_{\rm BF}|$ is further increased, the system turns into the hydrodynamic regime, where not only the two Lorentzians become sharper, as it was in Fig.~\ref{fig::bosons_monopole}, they also merge to have the same frequency -- an effect referred to as mode locking.
}
\label{fig::figure_mixture_CL}
\end{figure}

\begin{figure}[t!]
\centering
    \includegraphics[width=1\linewidth]{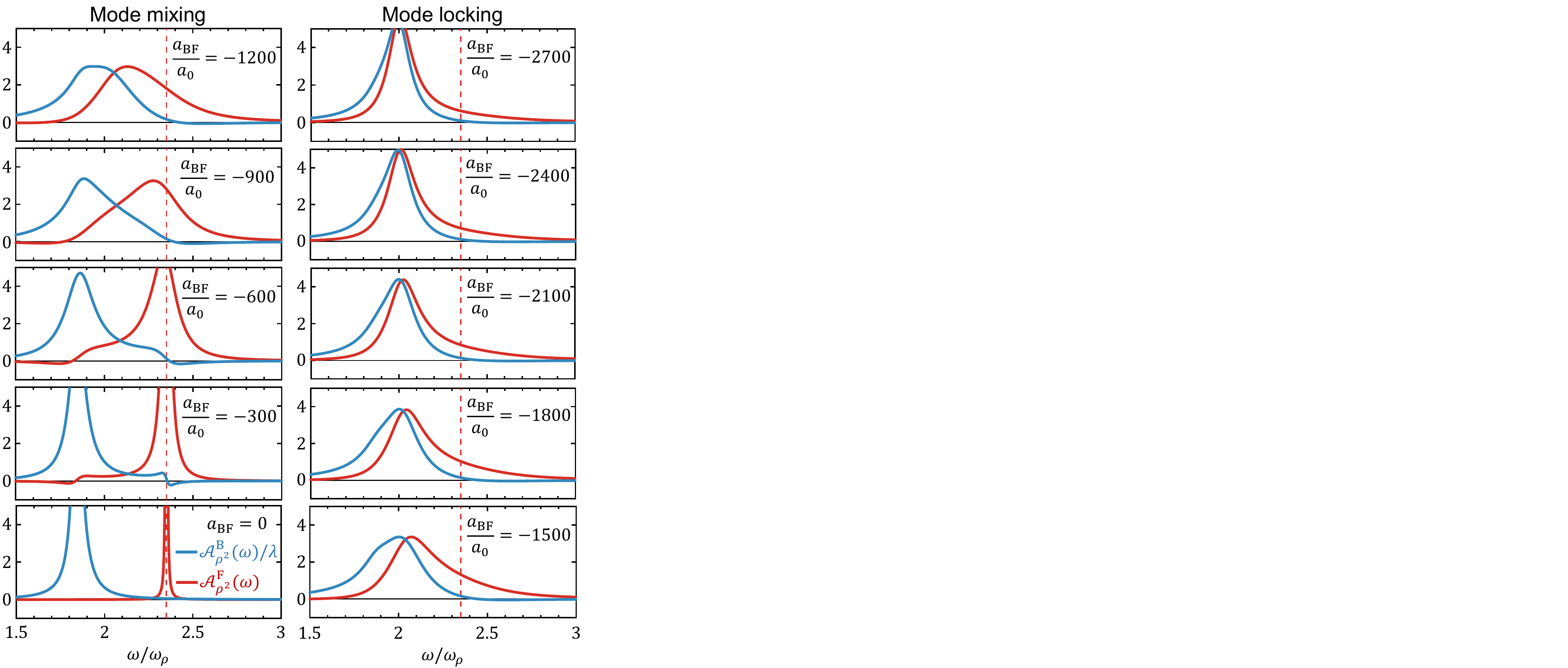} 
\caption{ The same analysis for the mixture regime $N_{\rm F} = N_{\rm B}$ as in Fig.~\ref{fig::figure_mixture_CL} but now bosons are nominally hydrodynamic $a_{\rm B} = 5000 a_0$. Quite remarkably, for $|a_{\rm BF}|\lesssim 600 a_0$, we find that the collisionless fermions display coherent mixing with the hydrodynamic bosonic mode. A bit unusual is that this latter mode also exhibits initial broadening as the interaction strength $|a_{\rm BF}|$ is increased. The rest of the results qualitatively follow those in Fig.~\ref{fig::figure_mixture_CL}, including eventual mode locking characteristic of the hydrodynamic regime of the mixture as a whole.
}
\label{fig::figure_mixture_hydro}
\end{figure}

\subsection{Mixture regime $N_{\rm F} \simeq N_{\rm B}$}
\label{subsec:Mixture regime}

We finally turn to investigate the mixture regime where the fermionic density is comparable to the bosonic one ($N_{\rm F} \simeq N_{\rm B}$).

Figure~\ref{fig::figure_mixture_CL} summarizes our results for the case with the bosons being nominally collisionless $a_{\rm B} = 50 a_0$. We find that a small interaction strength $|a_{\rm BF}| \lesssim 10^3 a_0$ gives rise to both incoherent broadenings of otherwise sharp bosonic and fermionic resonances and coherent mixing between them. For larger interaction strengths, $10^3 a_0\lesssim |a_{\rm BF}|\lesssim 2\times 10^3 a_0$, the incoherent part starts to dominate, and the spectral functions appear as rather featureless distinct Lorentzians. At even stronger interactions and in contrast to the impurity regime discussed above, the system becomes hydrodynamic. Both Lorentzians become narrower and they merge into a single, rather than two distinct, resonance -- an effect we refer to as mode locking. This synchronization effect represents a hallmark of the hydrodynamic regime of the Bose-Fermi mixture as a whole.

Similar results hold for the case where bosons are nominally hydrodynamic $a_{\rm B} = 5\times 10^3 a_0$, as summarized in Fig.~\ref{fig::figure_mixture_hydro}. Remarkably, when the interaction $|a_{\rm BF}|\lesssim 600 a_0$ is not too strong, we now observe that the coherent mixing is between the collisionless fermionic monopole mode and truly many-body hydrodynamic bosonic mode. We also find that the incoherent scatterings play a role and become dominant in the crossover region, where the two response function appear as two broad Lorentzians. Even the bosonic hydrodynamic mode becomes broader with increasing $|a_{\rm BF}|$ in this region. Finally, as $|a_{\rm BF}|$ is further increased, the Bose-Fermi mixture enters the hydrodynamic regime as a whole, characterized by the aforementioned mode locking and sharpening of the response functions.

\section{Conclusion and Outlook}
\label{sec:conclusion}

Our work elucidates the collective near-equilibrium dynamics of quantum Bose-Fermi mixtures in trapped cold atom systems.
We analyze linear responses of the system to changes in the confining potential and find that they differ dramatically depending on the parameter regimes. We observe several non-trivial phenomena, including coupling and interference between bosonic and fermionic collective modes, strong damping of these modes, and the emergence of hydrodynamics in the strongly interacting regime.
For intermediate and strong interactions, we observe appreciable mode mixing, manifesting, for instance, as the appearance of a longitudinal monopole resonance even when the optical perturbation is transverse. In the dilute fermion limit, the bare fermionic resonances quickly become featureless in frequency upon increasing the boson-fermion coupling strengths, which is reminiscent of the physics of Bose polarons. When bosonic and fermionic densities are comparable and interactions are strong enough to make the mixture hydrodynamic, we find mode locking between the bosonic and fermionic responses. Remarkably,  we also observe coherent mode mixing even when bosons are hydrodynamic and fermions are collisionless. The coherent mode mixing effects we uncover between longitudinal and transverse monopole modes, manifesting as Fano interference profiles and the emergence of the Stokes and anti-Stokes sidebands, indicate that Bose-Fermi mixtures should exhibit strong and tunable nonlinearities. While nonlinear effects are beyond the linearized Boltzmann framework analyzed here, they are a promising direction for future inquiry. Such non-linearities can potentially be used to explore analogues of non-linear optical phenomena using collective excitations rather than $\Lambda$-like atomic gases~\cite{fleischhauer2005electromagnetically,miroshnichenko2010fano}. Potential applications include generating single and two mode squeezing~\cite{braunstein2005squeezing}, entangling different collective modes~\cite{lukin2000entanglement}, and achieving reflections with phase conjugation~\cite{zel2013principles}.

Analysis of the dynamics of the cold atomic Bose-Fermi mixture was made possible by the GMM-MoM developed here. 
This method allows one to efficiently compute the linear-response properties, including the collective modes and transport coefficients,
of 
a large range of quantum many-body systems that are describable within the framework of the  Boltzmann equation. While this work focuses on a spatially-inhomogeneous 3D system, our approach will also accelerate investigations of salient 2D systems. Examples of phenomena in such systems that require further elucidation include electron hydrodynamics in vdW semiconductors, spin transport in bernal graphene, and the collisionless-to-hydrodynamic crossover in pancake-like dipolar gases~\cite{zhou2022isospin,moll2016evidence,bandurin2016negative,crossno2016observation,babadi2012collective}.
The tables of GMM fits to equilibrium distribution functions we provide are generic and can be directly applied to a broad class of interesting many-body systems~\cite{GMM-MoM_codebase}.

While the trapped cold-atom mixture we analyzed in this work has particle-conserving scattering processes, particle non-conserving collisions are also ubiquitous in solid-state and cold-atom platforms. Such processes are natural for systems containing phonons, magnons, and photons. 
Particle non-conserving collision integrals, however, are usually much simpler compared to the particle-conserving ones considered in this work and therefore are expected to be straightforwardly tractable 
using the same GMM-MoM approach. An 
additional step would be required to treat systems that have explicitly non-local (but spherically symmetric) interactions such as those generated by screened Coulomb or dipolar 
forces. Systems with pure Coulomb interactions may need to be modeled with Boltzmann-Vlasov equations; we expect that our approach facilitates analysis of such models as well. The GMM representaion of the distribution functions may not reduce the collision integrals to fully analytical expressions in these cases, unlike the contact $s$-wave interaction in the cold atom mixture considered here. This representation would still, however, dramatically simplify the collision integral, reducing it to a low-dimensional one that can easily be computed numerically using limited computational resources.

There are two generalizations of the GMM-MoM approach that may increase its efficiency. For systems that are spatially inhomogeneous, such as optically-trapped cold atomic gases similar to those considered in this work,
one may extend the GMM to an ansatz that encodes the full dependence of the equilibrium distribution functions on both real-space and momentum coordinates. This will further facilitate calculations of the collision integrals. Additionally, one may try mixture models composed of functions other than Gaussians that still allow simplification of the collision integral via Wick's theorem. Possible options include poly-Gaussian functions, composed of a sum of polynomials multiplied by a Gaussian, and error functions. Both of these ansatzes still enable semi-analytical computation of the collision integrals and may prove advantageous, for example, in the analysis of dense fermionic systems.

Aside from these simple generalizations, there are three non-trivial promising extensions of our GMM-MoM framework. 
Firstly, one can try to use a GMM with time-dependent Gaussians to investigate the far-from-equilibrium dynamics of quantum many-body systems. This approach would apply to systems which can be modeled using a Boltzmann equation with time-dependent distribution functions.
If such a time-dependent GMM-MoM proves accurate, it would enable significant progress in several open topics in modern physics. Examples include the behavior of non-equilibrium many-body steady-states, with applications to spintronics devices and field-effect transistors essential for classical computer hardware, fundamental questions about turbulence and hydrodynamics, and the phenomenology of light-induced phases of matter such as light-induced superconductivity and magnetism. 
Secondly, the Fano interference profiles and Stokes and anti-Stokes sidebands we find in the Bose-Fermi mixture studied here motivates the extension of the GMM-MoM framework to include nonlinear effects.  
Such an extension would facilitate analysis of the non-linear electrodynamics of correlated materials, allowing, for example, study of the temperature dependence of photocurrents in photovoltaic semiconductors or analysis of second or third harmonic generation in systems such as striped cuprate semiconductors which exhibit intertwined order parameters~\cite{zhao2017global,rajasekaran2018probing,torre2021mirror,nicoletti2022coherent,dolgirev2021theory,dolgirev2022optically}. Finally, we can try to extend the GMM-MoM framework to strongly-correlated systems beyond Landau Fermi liquids which are describable using a Boltzmann equation. Such an extension would necessitate a Gaussian mixture ansatz on the level of Green's functions rather than on the single-particle distribution functions at the heart of the Boltzmann model. 
Generalizing the GMM-MoM in this way would enable the analysis of non-Fermi liquids, which are believed to be the key to understanding strongly-correlated phases of matter.

Our work opens the door to rigorous theoretical investigation of many fundamental and technologically-important quantum many-body systems describable by Boltzmann equations which have previously evaded analysis. Furthermore, extensions of the method may provide an additional tool to analyze quantum many-body systems that are out-of-equilibrium, manifest non-linear behavior, or are strongly-correlated.

\section*{ACKNOWLEDGEMENTS}
The authors would like to thank N. Leitao, M. Babadi, M. Lukin, A. Rubio, S. Sachdev, P. Narang, A. Tarkhov, A. Atanasov, P. Volkov, I. Esterlis and M. Hafezi  for fruitful discussions. The work of P.E.D. was sponsored by the Army Research Office and was accomplished under Grant Number W911NF-21-1-0184. K.S. acknowledges funding from NSF EAGER-QAC-QCH award No. 2037687. E.D.
acknowledges support from ARO grant number W911NF-20-1-0163, the SNSF
project 200021\_212899. M.~Z. acknowledges support from NSF, AFOSR MURI on Ultracold Molecules and the Vannevar Bush Faculty Fellowship.

\bibliography{CM}

\clearpage
\newpage

\onecolumngrid
\appendix

\section{Evaluation of the matrix elements associated with the monopole mode}
\label{Appendix_monopole}

When working with the monopole basis functions, for future convenience, we redefine them as:
\begin{align}
    \phi_\alpha(\bm r,\bm p) = \beta^{N_\alpha} \kappa^{p_\alpha + r_\alpha/2} \times \rho^{2m_\alpha} p_\rho^{2n_\alpha} (\bm \rho \cdot\bm p_\rho )^{k_
    \alpha} \times z^{2p_\alpha} p_z^{2q_\alpha} (z p_z)^{r_\alpha},\quad  N_\alpha \equiv m_\alpha + n_\alpha + k_\alpha + p_\alpha + q_\alpha + r_\alpha. \label{eqn:mon_basis_full}
\end{align}
The GMM fits in the dimensionless variables read:
\begin{gather}
    n_{\rm B,eq}(\bm r,\bm p) = \frac{1}{\displaystyle\frac{1}{z_{\rm B}(\bm r)} \exp\Big(\displaystyle\frac{\beta p^2}{2}\Big) - 1} \approx \sum_s a_{{\rm B},s} \exp\Big( - \frac{\beta p^2}{2\gamma_{{\rm B},s}^2} \Big),\label{eqn:n_GMM_Bose_full}\\
    n_{\rm F,eq}(\bm r,\bm p) = \frac{1}{\displaystyle\frac{1}{z_{\rm F}(\bm r)} \exp\Big(\displaystyle\frac{\beta p^2 m_{\rm B}}{2m_{\rm F}}\Big) + 1} \approx \sum_s a_{{\rm F},s} \exp\Big( - \frac{\beta p^2}{2\gamma_{{\rm F},s}^2} \Big). \label{eqn:n_GMM_Fermi_full}
\end{gather}
We note that the coefficients $a_{{\rm B}, s}$ and $\gamma_{{\rm B},s}$ ($a_{{\rm F}, s}$ and $\gamma_{{\rm F},s}$) depend on the real-space coordinate $\bm r$ only through the local fugacity $z_{\rm B}(\bm r)$ ($z_{\rm F}(\bm r)$). A simplification that makes our numerical calculations faster and that we will use below is that the fugacities, in turn, depend only on the one-dimensional combination $\tilde{r}^2 = {\beta} ({\rho}^2 +  \kappa {z}^2)$. This useful but not crucial feature comes from the fact that the trapping potentials for bosons and fermions are similar and harmonic. If the potentials are anharmonic or have different from each other forms, one can easily extend the formalism below to account for such a situation. Besides, while the GMM fits are necessary for the collision matrices only, they also facilitate numerical evaluations of the rest of the matrices entering the linearized kinetic equations. This Appendix provides the expressions for all of the matrix elements associated with the monopole mode.

\subsection{Matrix elements of $\hat{\rm M}$}

By rescaling $\rho \sqrt{\beta } \to \rho$, $ z \sqrt{\beta \kappa} \to z $, and $ p\sqrt{\beta }  \to p $, the matrix elements of $\hat{\rm M}$ in Eq.~\eqref{eqn:M_def} can be written as:
\begin{align}
    ({\rm M}_{\rm B/F})_{\alpha\beta} = \frac{1}{\beta^3\sqrt{\kappa}} \int d^3\bm r\frac{d^3 \bm p}{(2\pi)^3} \rho^{2m_\alpha + 2 m_\beta} p_\rho^{2n_\alpha + 2n_\beta} (\bm \rho \cdot\bm p_\rho )^{k_
    \alpha + k_\beta} \times z^{2p_\alpha + 2p_\beta} p_z^{2q_\alpha + 2q_\beta} (z p_z)^{r_\alpha + r_\beta} \, n (1\pm n).
\end{align}
In these new variables, the equilibrium distribution functions $n = n_{\rm B/F,eq}$ depend on $r$ and $p$ only, which allows us to perform the angular integration first, leading to:
\begin{align}
    {\rm M}_{\alpha\beta}  = &
    \frac{1}{2\pi\beta^3\sqrt{\kappa}}h(k_\alpha + k_\beta) g\Big( p_\alpha + p_\beta + \frac{r_\alpha + r_\beta}{2}, m_\alpha + m_\beta + \frac{k_\alpha + k_\beta}{2} \Big)
    g\Big( q_\alpha + q_\beta + \frac{r_\alpha + r_\beta}{2}, n_\alpha + n_\beta + \frac{k_\alpha + k_\beta}{2} \Big) \notag\\
    & \times
    \int \limits_0^\infty dr \, r^{2 m_\alpha + 2 m_\beta + k_\alpha + k_\beta + 2p_\alpha + 2p_\beta + r_\alpha + r_\beta + 2}
    \int \limits_0^\infty dp \, p^{2 n_\alpha + 2 n_\beta + k_\alpha + k_\beta + 2q_\alpha + 2q_\beta + r_\alpha + r_\beta + 2} \times n(1\pm n),
\end{align}
where we defined:
\begin{gather}
    h(n) \equiv  \int_0^{2\pi} [\cos(\psi)]^n \frac{d\psi}{2\pi}  = 
    \displaystyle\frac{ E(n) n!}{2^n[(n/2)!]^2 }, \\
    g(m, n) \equiv  \int_{-1}^1 x^{2m} (1 -x^2)^n  dx =  \frac{\Gamma(m + 1/2)\Gamma(n + 1)}{\Gamma(m + n + 3/2)}, \text{ provided } m,n \in 0,1,2,\dots
\end{gather}
$E(n) = 1$ if $n$ is even and zero otherwise. We compute the momentum integral analytically using the GMM fits and $f(n,\gamma) \equiv \displaystyle\int\limits_0^\infty dp\, p^n \exp\Big( -\frac{p^2}{2\gamma^2}\Big) = 2^{(n - 1)/2} \gamma^{n + 1}\Gamma\Big( \frac{n + 1}{2} \Big)$. The remaining one-dimensional integration over $r$ is then done numerically.

\subsection{Matrix elements of $\hat{\rm H}$}

We first note that the relevant Poisson brackets in Eq.~\eqref{eqn:def_H_matr} can be written as:
\begin{align}
    \{ \phi_\beta, {\cal H}_{\rm B}\} & = \phi_\beta \Big[ 
    2m_\beta \frac{\bm \rho \cdot \bm p_\rho}{\rho^2} + k_\beta \frac{p_\rho^2}{\bm \rho \cdot \bm p_\rho} + (2p_\beta + r_\beta)\frac{p_z}{z} - \gamma_{\rm B} \Big( 2n_\beta \frac{\bm \rho \cdot \bm p_\rho}{p_\rho^2} + k_\beta \frac{\rho^2}{\bm \rho \cdot \bm p_\rho} + \kappa(2q_\beta + r_\beta)\frac{z}{p_z} \Big)
    \Big] \label{eqn:PB_1}
    ,\\
    \{ \phi_\beta, {\cal H}_{\rm F}\} & = \phi_\beta \Big[  \Big(
    2m_\beta \frac{\bm \rho \cdot \bm p_\rho}{\rho^2} + k_\beta \frac{p_\rho^2}{\bm \rho \cdot \bm p_\rho} + (2p_\beta + r_\beta)\frac{p_z}{z} \Big) \frac{m_{\rm B}}{m_{\rm F}} - \gamma_{\rm F} \Big( 2n_\beta \frac{\bm \rho \cdot \bm p_\rho}{p_\rho^2} + k_\beta \frac{\rho^2}{\bm \rho \cdot \bm p_\rho} + \kappa(2q_\beta + r_\beta)\frac{z}{p_z} \Big)
    \Big],\label{eqn:PB_2}
\end{align}
where we defined $\gamma_{\rm B} \equiv 1 + \displaystyle \frac{\beta}{\tilde{r}}  \partial_{\tilde{r}} \Sigma_{\rm HF, B}(\tilde{r})$ and $\gamma_{\rm F} \equiv \lambda + \displaystyle \frac{\beta}{\tilde{r}}  \partial_{\tilde{r}} \Sigma_{\rm HF, F}(\tilde{r})$ so that $\partial_{\bm \rho} {\cal H}_{\rm B/F} = \gamma_{\rm B/F} \bm \rho$ and $\partial_{z} {\cal H}_{\rm B/F} = \gamma_{\rm B/F} \kappa z$. Plugging this into Eq.~\eqref{eqn:def_H_matr} and following the preceding subsection, for the fermionic sector, we arrive at:
\begin{align}
    ({\rm H}_{\rm F} & )_{\alpha\beta}   = \frac{1}{2\pi\beta^3\sqrt{\kappa}}
    \Big\{\\
    &
    \frac{m_{\rm B}}{m_{\rm F}}\Big[ (2m_\beta h(k_\alpha + k_\beta + 1) +k _\beta h(k_\alpha + k_\beta - 1)) g\Big(p_\alpha + p_\beta + \frac{r_\alpha + r_\beta }{2},m_\alpha + m_\beta + \frac{k_\alpha + k_\beta - 1 }{2} \Big)   \notag\\
    &
    \times
    g\Big(q_\alpha + q_\beta + \frac{r_\alpha + r_\beta }{2},n_\alpha + n_\beta + \frac{k_\alpha + k_\beta + 1 }{2} \Big)  +
    \sqrt{\kappa}(2p_\beta + r_\beta) h(k_\alpha + k_\beta)
    \notag\\
    &
    \times
    g\Big(p_\alpha + p_\beta + \frac{r_\alpha + r_\beta - 1}{2},m_\alpha + m_\beta + \frac{k_\alpha + k_\beta  }{2} \Big)
    g\Big(q_\alpha + q_\beta + \frac{r_\alpha + r_\beta +1 }{2},n_\alpha + n_\beta + \frac{k_\alpha + k_\beta  }{2} \Big) \Big]
    \notag\\
    &
    \times
    \int \limits_0^\infty dr \, r^{2 m_\alpha + 2 m_\beta + k_\alpha + k_\beta + 2p_\alpha + 2p_\beta + r_\alpha + r_\beta + 1}
    \int \limits_0^\infty dp \, p^{2 n_\alpha + 2 n_\beta + k_\alpha + k_\beta + 2q_\alpha + 2q_\beta + r_\alpha + r_\beta + 3} \, n_{\rm F,eq}(1 - n_{\rm F,eq}) \notag\\
    &
    - \Big[(2n_\beta h(k_\alpha + k_\beta + 1) + k_\beta h(k_\alpha + k_\beta - 1)) g\Big(p_\alpha + p_\beta + \frac{r_\alpha + r_\beta }{2},m_\alpha + m_\beta + \frac{k_\alpha + k_\beta + 1 }{2} \Big) \notag\\
    &
    \times
    g\Big(q_\alpha + q_\beta + \frac{r_\alpha + r_\beta }{2},n_\alpha + n_\beta + \frac{k_\alpha + k_\beta - 1 }{2} \Big) + \sqrt{\kappa}(2q_\beta + r_\beta) h(k_\alpha + k_\beta)\notag
    \notag\\
    &
    \times
    g\Big(p_\alpha + p_\beta + \frac{r_\alpha + r_\beta  + 1}{2},m_\alpha + m_\beta + \frac{k_\alpha + k_\beta  }{2} \Big)
    g\Big(q_\alpha + q_\beta + \frac{r_\alpha + r_\beta - 1}{2},n_\alpha + n_\beta + \frac{k_\alpha + k_\beta }{2} \Big) \Big]
    \notag\\
    &
    \times
    \int \limits_0^\infty dr \, r^{2 m_\alpha + 2 m_\beta + k_\alpha + k_\beta + 2p_\alpha + 2p_\beta + r_\alpha + r_\beta + 3} \gamma_{\rm F}(r)
    \int \limits_0^\infty dp \, p^{2 n_\alpha + 2 n_\beta + k_\alpha + k_\beta + 2q_\alpha + 2q_\beta + r_\alpha + r_\beta + 1} \, n_{\rm F,eq}(1 - n_{\rm F,eq}) \notag \Big\}.
\end{align}
The expression for the bosonic sector is obtained from this one by substituting $n_{\rm F,eq}(1 - n_{\rm F,eq}) \to n_{\rm B,eq}(1 + n_{\rm B,eq})$,  putting the mass ratio to one $\displaystyle \frac{m_{\rm B}}{m_{\rm F}} \to 1$, and by replacing $\gamma_{\rm F}\to \gamma_{\rm B}$. As in the preceding subsection, we evaluate all the momentum integrals analytically and all the real-space integrals numerically.

\subsection{Matrix elements of $\hat{\rm S}$}

The self-energies entering the matrix $\hat{\rm \Sigma}$, cf. Eq.~\eqref{eqn:Sigma_gen}, take the following form:
\begin{align}
    \Sigma_{\rm B}[\Delta_{\rm B} \phi_\beta](\rho, z) = 2g_{\rm B} \int \frac{d^3 \bm p}{(2\pi)^3} n_{\rm B,eq}(1 + n_{\rm B,eq}) \phi_\beta \to \frac{2g_{\rm B}}{\beta^{3/2}} \rho^{2m_\beta + k_\beta} z^{2p_\beta + r_\beta} v_{{\rm B},\beta}(r),
\end{align}
where we rescaled the variables in the usual way, i.e., $\rho \sqrt{\beta } \to \rho$, $ z \sqrt{\beta \kappa} \to z $, and $ p\sqrt{\beta }  \to p$, and introduced:
\begin{align}
    v_{{\rm B/F},\beta}(r) \equiv \frac{1}{(2\pi)^2} h(k_\beta) g\Big( q_\beta + \frac{r_\beta}{2}, n_\beta + \frac{k_\beta}{2} \Big) \int\limits_0^\infty dp\, p^{2n_\beta + k_\beta + 2 q_\beta + r_\beta + 2} n(r,p)(1 \pm n(r,p)).
\end{align}
Plugging this into Eq.~\eqref{eqn:SB_gen}, we obtain:
\begin{align}
    ({\rm S}_{\rm B} & )_{\alpha\beta}  = \frac{2g_{\rm B}}{2\pi \beta^{7/2}\sqrt{\kappa}} \Big\{\\
    &
    \Big[ (2m_\alpha h(k_\alpha + 1) + k_\alpha h(k_\alpha - 1)) g\Big(q_\alpha + \frac{r_\alpha}{2}, n_\alpha + \frac{k_\alpha + 1}{2} \Big)
    g\Big( p_\alpha + p_\beta + \frac{r_\alpha + r_\beta}{2}, m_\alpha + m_\beta + \frac{k_\alpha + k_\beta - 1}{2} \Big)\notag\\
    &
    +\sqrt{\kappa} (2p_\alpha + r_\alpha)h(k_\alpha)
    g\Big( q_\alpha + \frac{r_\alpha + 1}{2}, n_\alpha + \frac{k_\alpha}{2} \Big) g\Big( p_\alpha + p_\beta + \frac{r_\alpha + r_\beta - 1}{2}, m_\alpha + m_\beta + \frac{k_\alpha + k_\beta}{2} \Big)
    \Big]\notag\\
    &\times
    \int \limits_0^\infty dr \, r^{2 m_\alpha + 2 m_\beta + k_\alpha + k_\beta + 2p_\alpha + 2p_\beta + r_\alpha + r_\beta + 1} v_{\rm B,\beta}
    \int \limits_0^\infty dp \, p^{2 n_\alpha  + k_\alpha + 2q_\alpha + r_\alpha + 3} \, n_{\rm B,eq}(1 + n_{\rm B,eq}) \notag\\
    & 
    - \Big[ (2n_\alpha h(k_\alpha + 1) + k_\alpha h(k_\alpha - 1)) g\Big(q_\alpha + \frac{r_\alpha}{2}, n_\alpha + \frac{k_\alpha - 1}{2} \Big)
    g\Big( p_\alpha + p_\beta + \frac{r_\alpha + r_\beta}{2}, m_\alpha + m_\beta + \frac{k_\alpha + k_\beta + 1}{2} \Big) \notag\\
    &
    +\sqrt{\kappa} (2q_\alpha + r_\alpha)h(k_\alpha)
    g\Big( q_\alpha + \frac{r_\alpha - 1}{2}, n_\alpha + \frac{k_\alpha}{2} \Big) g\Big( p_\alpha + p_\beta + \frac{r_\alpha + r_\beta + 1}{2}, m_\alpha + m_\beta + \frac{k_\alpha + k_\beta}{2} \Big)
    \Big]\notag\\
    &\times
    \int \limits_0^\infty dr \, r^{2 m_\alpha + 2 m_\beta + k_\alpha + k_\beta + 2p_\alpha + 2p_\beta + r_\alpha + r_\beta + 3} v_{\rm B,\beta}\gamma_{\rm B}
    \int \limits_0^\infty dp \, p^{2 n_\alpha  + k_\alpha + 2q_\alpha + r_\alpha + 1} \, n_{\rm B,eq}(1 + n_{\rm B,eq})
    \Big\}.\notag
\end{align}
Similar expressions hold for $({\rm S}_{\rm BF})_{\alpha\beta}$ and $({\rm S}_{\rm FB})_{\alpha\beta}$.

\subsection{Matrix elements ${\rm I}_{\alpha\beta}$}
\label{appendix:coll_int}

In this subsection, we show how to evaluate the collision matrix ${\rm I}_{\alpha\beta}$ in Eq.~\eqref{eqn_coll_matr_in} associated with incoherent boson-boson scatterings. The main idea of the construction below is that the GMM fits allow one to represent the complicated matrix elements as sums of Gaussian integrals that can be evaluated analytically over the 12-dimensional momentum space (the integration over the remaining 3-dimensional real space is then done numerically). Some difficulties still exist, such as taking into account the momentum and energy conservation laws -- we discuss this below.

As a first step, we change variables to the center-of-mass frame: $\bm p = \frac{1}{2}\bm P + \bm q$, $\bm p' = \frac{1}{2}\bm P - \bm q$, $\bm p_1 = \frac{1}{2}\bm P' + \bm q'$, and $\bm p_1' = \frac{1}{2}\bm P' - \bm q'$. The momentum conservation leads to $\bm P' = \bm P$, while the energy conservation enforces $q' = q$:
\begin{align}
    \int d^3\bm r\frac{d^3 \bm p}{(2\pi)^3} \frac{d^3 \bm p'}{(2\pi)^3}\frac{d^3 \bm p_1}{(2\pi)^3}\frac{d^3 \bm p'_1}{(2\pi)^3}  (2\pi)^3\delta(\bm p + 
    \bm p' - \bm p_1 - \bm p_1') \times (2\pi)\delta(\varepsilon_{\rm B}(\bm p) & +  \varepsilon_{\rm B}(\bm p')   - \varepsilon_{\rm B}(\bm p_1) -\varepsilon_{\rm B}(\bm p_1')) \notag\\
    &
    \to
     \int d^3\bm r\frac{d^3 \bm P}{(2\pi)^3} \frac{d^3 \bm q}{(2\pi)^3}\frac{d^3 \bm q'}{(2\pi)^3}  (2\pi)\delta(q^2 - q'^2)\notag.
\end{align}
It will be convenient to write $S[\phi_\alpha] = \beta^{N_\alpha} \kappa^{p_\alpha + r_\alpha/2} \rho^{2m_\alpha+ k_\alpha} z^{2p_\alpha + r_\alpha} S_\alpha[\bm P,\bm q,\bm q']$, where $S_\alpha[\bm P,\bm q,\bm q']$ is a known polynomial of its arguments. Here, it is implicit that we consider all the in-plane momenta relative to $\bm \rho$ -- this is allowed to do since, for the monopole mode, the integrals do not depend on the angle $\phi$ between $\bm \rho$ and the $x$-axis. Rescaling the variables in the usual manner, i.e., $\rho \sqrt{\beta } \to \rho$, $ z \sqrt{\beta \kappa} \to z $, and $ p\sqrt{\beta }  \to p$, we arrive at: 
\begin{align}
({\rm I}_{\rm B})_{\alpha\beta} = & - \frac{{g}^2_{\rm B} }{2\beta^5 \sqrt{\kappa}}  \int d^3\bm r \, \rho^{2m_\alpha + 2m_\beta + k_\alpha + k_\beta} z^{2p_\alpha + 2p_\beta + r_\alpha + r_\beta}\int \frac{d^3 \bm P}{(2\pi)^3} \frac{d^3 \bm q}{(2\pi)^3}\frac{d^3 \bm q'}{(2\pi)^3}  (2\pi)\delta(q^2 - q'^2)\notag\\
    &
    \times (1 + n_{\rm B, eq}(\bm r,\bm p))(1 + n_{\rm B, eq}(\bm r,\bm p'))n_{\rm B, eq}(\bm r,\bm p_1) n_{\rm B, eq}(\bm r,\bm p_1') \times {\rm Poly}[\bm P, \bm q, \bm q']. \label{eqn:I_B_v0}
\end{align}
We note that the real-space angular integration can be readily done:
\begin{align}
     \int d^3\bm r \, \rho^{2m_\alpha + 2m_\beta + k_\alpha + k_\beta} z^{2p_\alpha + 2p_\beta + r_\alpha + r_\beta} \to 2\pi g\Big( p_\alpha + p_\beta +  \frac{r_\alpha + r_\beta}{2}, & m_\alpha + m_\beta + \frac{k_\alpha + k_\beta}{2} \Big)\notag\\
     &\times \int\limits_0^\infty dr\, r^{2m_\alpha + 2m_\beta + k_\alpha + k_\beta + 2p_\alpha + 2p_\beta + r_\alpha + r_\beta +2}\notag.
\end{align}
Here ${\rm Poly}[\bm P, \bm q, \bm q'] = S_\alpha[\bm P,\bm q,\bm q'] \times S_\beta[\bm P,\bm q,\bm q']$ is a polynomial of its arguments. It is worth pointing out that polynomials are particularly suitable for our construction of the GMM fits because, as we mentioned, after all manipulations, we will end up having a Gaussian integral. As such, when the integrand contains a polynomial, one can simply apply Wick's theorem to get analytical results.

To proceed, we expand the product over the distribution functions in Eq.~\eqref{eqn:I_B_v0} as $(1 + n)(1 + n')n_1 n_1' = n_1 n_1' + n_1 n_1' n + n_1 n_1'n' + n_1 n_1'nn'$ so that the matrix ${\rm I}_{\rm B}$ is a sum of four terms. Substituting the GMM fits,  we get for the first (classical) term~\footnote{ We call the first term $\sim n_1 n_1'$ classical because it is the only one (among the four terms contributing to ${\rm I}_{\rm B}$) that survives in the classical limit $1 + n \approx 1$.} the following:
\begin{align*}
    n_1 n_1' = \sum_{s_1 s_2} a_{s_1} a_{s_2} \exp\Big\{ - \frac{p_1^2}{2 \gamma_{s_1}^2} - \frac{p_1'^2}{2 \gamma_{s_2}^2}\Big\}
    = \sum_{s_1 s_2} a_{s_1} a_{s_2} \exp\Big\{  - \Big(\frac{1}{2\gamma_{s_1}^2} + \frac{1}{2\gamma_{s_2}^2} \Big) \Big(\frac{P^2}{4} + q^2\Big) - \Big(\frac{1}{2\gamma_{s_1}^2} - \frac{1}{2\gamma_{s_2}^2} \Big)\bm P\cdot \bm q' 
    \Big\}.
\end{align*}
Similar expressions hold for the other three (quantum) terms. We note that the prefactor in front of the exponent depends only on the real-space coordinate $\bm r$, while the exponent itself contains the full information about the momentum variables. Since the real-space integration will be evaluated numerically, from now on, we focus on the exponent and analytically evaluate the momentum integrals. To capture each of the four terms contributing to ${\rm I}_{\rm B}$, we write the exponent as:
\begin{align}
    \exp \Big\{ - \frac{P^2}{2\gamma_{P}^2} - \frac{1}{2\gamma_q^2}(q^2 + q'^2) + \frac{1}{\gamma_P} (A \bm q + B \bm q') \cdot \bm P \Big\}. \label{eqn:exp_gen}
\end{align}
The coefficients $\gamma_P$, $\gamma_q$, $A$, and $B$ can be simply written in terms of the parameters $\gamma_{s_i}$ entering the GMM fits. Explicitly, for the first classical term we have:
\begin{align*}
    \frac{1}{\gamma_P^2} = \frac{1}{4\gamma_{s_1}^2} +\frac{1}{4\gamma_{s_2}^2},\quad 
    \frac{1}{\gamma_q^2} = \frac{1}{2\gamma_{s_1}^2} +\frac{1}{2\gamma_{s_2}^2} ,\quad
    A = 0,\quad B = \gamma_P \Big(\frac{1}{2\gamma_{s_2}^2} - \frac{1}{2\gamma_{s_1}^2} \Big).
\end{align*}
Similar expressions can be easily obtained for the remaining three terms (in general, we have nonzero values for $A$ and $B$).

With the form in Eq.~\eqref{eqn:exp_gen}, the three-dimensional Gaussian integral over $\bm P$ can be readily evaluated:
\begin{align}
    \int \frac{d^3 \bm P}{(2\pi)^3} \exp \Big\{ - \frac{P^2}{2\gamma_{P}^2} - \frac{q^2}{\gamma_q^2} + \frac{1}{\gamma_P} (A \bm q + B \bm q') \cdot \bm P \Big\} {\rm Poly}[\bm P, \bm q, \bm q'] & = \exp \Big\{  - \frac{q^2}{\gamma_q^2} + \frac{1}{2} (A \bm q + B \bm q')^2 \Big\} {\rm Poly}[\bm q, \bm q'] \notag\\
    & = \exp \Big\{  - C_1 q^2 + C_2 \bm q \cdot \bm q' \Big\} {\rm Poly}[\bm q, \bm q'].\label{eqn:coll_int_P}
\end{align}
In obtaining this result, we used the energy conservation constraint $q^2 = q'^2$. In the first equality, the new polynomial ${\rm Poly}[\bm q, \bm q']$ no longer depends on $\bm P$ but it does depend on $\bm r$ through the coefficients $\gamma_P$, $A$, and $B$ (we will not explicitly indicate the dependence on $\bm r$). In practice, to get this polynomial, we write a simple {\it Mathematica} subroutine that computes the Gaussian integral over $\bm P$. Here we defined $C_1 = \displaystyle\frac{1}{\gamma_q^2} - \frac{1}{2} A^2 - \frac{1}{2}B^2$ and $C_2 = A B$.

We are left with a Gaussian integral over $\bm q$ and $\bm q'$. However, what makes this integral hard is the energy constraint $\delta(q^2 - q'^2)$ and the appearance of the terms $\sim \bm q\cdot \bm q'$ in the exponent Eq.~\eqref{eqn:coll_int_P}. We now note that the expression in the exponent can be written as a square of a vector (assuming $q = q'$)
\begin{align}
    C_1 q^2 - C_2 \bm q\cdot \bm q'   \xlongequal{q = q'} (\tilde{A} \bm q + \tilde{B}\bm q')^2. 
\end{align}
Indeed, simple algebra gives:
\begin{align}
    \tilde{A} = \sqrt{\frac{1}{2}\Big(C_1 + \sqrt{C_1^2 - C_2^2 } \Big)},\quad \tilde{B} = -\text{sign}(C_2) \sqrt{\frac{1}{2}\Big(C_1 - \sqrt{C_1^2 - C_2^2 } \Big)}.
\end{align}
This allows one to change variables to $\bm k = \tilde{A} \bm q + \tilde{B}\bm q'$ and $\bm k' = \tilde{A} \bm q' + \tilde{B}\bm q$; under such a linear transformation, we get ($\tilde{A}\neq \tilde{B}$):
\begin{align*}
    \delta(q^2 - q'^2) = (\tilde{A}^2 - \tilde{B}^2)  \delta(k^2 - k'^2),\quad \int \frac{d^3 \bm q}{(2\pi)^3}\frac{d^3 \bm q'}{(2\pi)^3} \to \frac{1}{(\tilde{A}^2 - \tilde{B}^2)^3} \int \frac{d^3 \bm k}{(2\pi)^3}\frac{d^3 \bm k'}{(2\pi)^3}. \label{eqn:coll_int_v0}
\end{align*}
We, therefore, obtain:
\begin{align}
    \int \frac{d^3 \bm q}{(2\pi)^3}\frac{d^3 \bm q'}{(2\pi)^3} (2\pi) \delta(q^2 - q'^2) \exp \Big\{  - C_1 q^2 & + C_2 \bm q \cdot \bm q' \Big\} {\rm Poly}[\bm q, \bm q'] \notag\\
    &
    = \frac{1}{(\tilde{A}^2 - \tilde{B}^2)^2} \int \frac{d^3 \bm k}{(2\pi)^3}\frac{d^3 \bm k'}{(2\pi)^3} (2\pi) \delta(k^2 - k'^2) e^{-k^2} {\rm Poly}[\bm k, \bm k'].
\end{align}
Here, the new polynomial ${\rm Poly}[\bm k, \bm k']$ is obtained from ${\rm Poly}[\bm q, \bm q']$ by simply substituting $\bm q = (\tilde{A}\bm k - \tilde{B}\bm k')/(\tilde{A}^2 - \tilde{B}^2)$ and $\bm q' = (\tilde{A}\bm k' - \tilde{B}\bm k)/(\tilde{A}^2 - \tilde{B}^2)$, i.e., through the inverse linear transformation. The integral in Eq.~\eqref{eqn:coll_int_v0} is already feasible for further analytical/numerical evaluations. Indeed, in the spherical coordinates, it becomes:
\begin{align}
Q_{\alpha\beta}(r) = \frac{\pi}{(2\pi)^6(\tilde{A}^2 - \tilde{B}^2)^2} \int\limits_0^\infty dk\, k^3  e^{-k^2} \int\limits_0^{2\pi} d\varphi\int\limits_0^{2\pi} d\varphi' \int\limits_{-1}^1 d( \cos\theta) \int\limits_{-1}^1 d ( \cos\theta') \,
{S}_{\alpha\beta}[k,\phi,\phi', \theta,\theta'],
\end{align}
where $S_{\alpha\beta}[k,\phi,\phi', \theta,\theta']$ is obtained from ${\rm Poly}[\bm k, \bm k']$ by substituting the spherical coordinates and $k = k'$. Importantly, each function $S_{\alpha\beta}$ is some known polynomial of $\cos\theta$, $\sin\theta$, $\cos\theta'$, $\sin\theta'$, $\cos\phi$, $\sin\phi$, $\cos\phi'$, and $\sin\phi'$ -- as such, the angular integration can be done analytically. In our numerics, we write a simple {\it Mathematica} code that symbolically evaluates this four-dimensional angular integration. The remaining Gaussian integral over $k$ is then also evaluated analytically (we write a separate {\it Mathematica} subroutine for this step as well).

This completes our protocol for evaluating the momentum integrals. At this stage, we are left with the integral over $\bm r$, which is reduced to a one-dimensional integral over $r$ and then evaluated numerically. Explicitly, the first classical contribution to $({\rm I}_{\rm B})_{\alpha\beta}$ reads:
\begin{align}
    ({\rm I}_{\rm B})_{\alpha\beta,1}   =  - \frac{{g}^2_{\rm B} \pi }{\beta^5 \sqrt{\kappa}}  \sum_{s_1,s_2} g\Big( p_\alpha + p_\beta +  \frac{r_\alpha + r_\beta}{2}, & m_\alpha + m_\beta + \frac{k_\alpha + k_\beta}{2} \Big) \\
    &
    \times \int\limits_0^\infty dr\, r^{2m_\alpha + 2m_\beta + k_\alpha + k_\beta + 2p_\alpha + 2p_\beta + r_\alpha + r_\beta +2} a_{s_1}(r)a_{s_2}(r) Q_{\alpha\beta,1}(r),
    \notag
\end{align}
and similar expressions hold for the remaining three contributions.

\subsection{Matrix elements of $\hat{\rm J}$}

We turn to discuss the matrix $\hat{\rm J}$ associated with collisions between bosons and fermions, Eqs.~\eqref{eqn_coll_matr_in_v2}-\eqref{eqn_coll_matr_fin}. Now, the corresponding matrix elements are evaluated in exactly the same manner as in the preceding subsection. The only new feature we need to address in this subsection is that the bosonic and fermionic masses might differ. In this case, the transformation to the center-of-mass frame now reads:
\begin{align}
    \bm p = \frac{m_{\rm B}}{m_{\rm B} + m_{\rm F}}\bm P + \bm q,\, \bm p' = \frac{m_{\rm F}}{m_{\rm B} + m_{\rm F}} \bm P - \bm q,\, 
    \bm p_1 =  \frac{m_{\rm B}}{m_{\rm B} + m_{\rm F}} \bm P' + \bm q',\, \bm p_1' = \frac{m_{\rm F}}{m_{\rm B} + m_{\rm F}}\bm P' - \bm q'.
\end{align}
As above, the momentum conservation enforces $\bm P = \bm P'$, while the energy conservation gives $q = q'$:
\begin{align}
    \int d^3\bm r\frac{d^3 \bm p}{(2\pi)^3} \frac{d^3 \bm p'}{(2\pi)^3}\frac{d^3 \bm p_1}{(2\pi)^3}\frac{d^3 \bm p'_1}{(2\pi)^3}  (2\pi)^3\delta(\bm p +
    \bm p' -  \bm p_1 - \bm p_1') & \times (2\pi)\delta(\varepsilon_{\rm B}(\bm p)  +  \varepsilon_{\rm F}(\bm p')   - \varepsilon_{\rm B}(\bm p_1) -\varepsilon_{\rm F}(\bm p_1')) \notag\\
    &
    \to \frac{2 m_{\rm B}m_{\rm F}}{m_{\rm B} + m_{\rm F}}
     \int d^3\bm r\frac{d^3 \bm P}{(2\pi)^3} \frac{d^3 \bm q}{(2\pi)^3}\frac{d^3 \bm q'}{(2\pi)^3}  (2\pi)\delta(q^2 - q'^2)\notag.
\end{align}
The rest of the computation follows the preceding subsection step-by-step.

\section{Evaluation of the matrix elements associated with the quadrupole mode}
\label{Appendix_quad}

When working with the quadrupole basis functions, we also use the definitions in Eqs.~\eqref{eqn:mon_basis_full}-\eqref{eqn:n_GMM_Fermi_full}. We rescale the functions $\xi_i(\bm r,\bm p)$ by an overall factor of $\beta$ and rewrite them as: $\xi_{i}(\bm r,\bm p) = \beta  \rho^{\mu_i} p_\rho^{\nu_i} \cos(2\phi + \nu_i \phi_p)$, where $\phi$ is the angle between $\bm \rho$ and the $x$-axis and $\phi_p$ is the angle between $\bm \rho$ and $\bm p_\rho$; $(\mu_1,\nu_1) = (2,0)$, $(\mu_2,\nu_2) = (1,1)$, and $(\mu_3,\nu_3) = (0,2)$. Below we provide all the matrix elements associated with the quadrupole mode.

\subsection{Matrix elements of $\hat{\rm M}$}

Similar algebra as in the preceding Appendix gives:
\begin{align}
    {\rm M}^{ij}_{\alpha\beta}  = &
    \frac{1}{4\pi\beta^3\sqrt{\kappa}}\tilde{h}(|\nu_i - \nu_j|, k_\alpha + k_\beta) g\Big( p_\alpha + p_\beta + \frac{r_\alpha + r_\beta}{2}, m_\alpha + m_\beta + \frac{k_\alpha + k_\beta + \mu_i + \mu_j}{2} \Big) \notag\\
    &\times
    g\Big( q_\alpha + q_\beta + \frac{r_\alpha + r_\beta}{2}, n_\alpha + n_\beta + \frac{k_\alpha + k_\beta + \nu_i + \nu_j}{2} \Big)
    \int \limits_0^\infty dr \, r^{2 m_\alpha + 2 m_\beta + k_\alpha + k_\beta + 2p_\alpha + 2p_\beta + r_\alpha + r_\beta + \mu_i + \mu_j + 2}
    \notag\\
    & \times
    \int \limits_0^\infty dp \, p^{2 n_\alpha + 2 n_\beta + k_\alpha + k_\beta + 2q_\alpha + 2q_\beta + r_\alpha + r_\beta + \nu_i + \nu_j + 2} \, n(1\pm n),
\end{align}
where we defined:
\begin{align}
    \tilde{h}(n,k) \equiv \int_0^{2\pi} \cos(n \psi) [\cos{\psi}]^k \frac{d\psi }{2\pi} = \frac{k!}{2^k} \frac{\theta(k -|n|) E(k+n) }{[(\frac{k-n}{2})!][(\frac{k+n}{2})!]}.
\end{align}
In the derivations here, we used the same rescaling as above, namely: $ \rho \sqrt{\beta } \to \rho$, $ z \sqrt{\beta \kappa} \to z $, and $ p\sqrt{\beta }  \to p$.

\subsection{Matrix elements of $\hat{\rm H}$}

We note that $\{\xi_j \phi_\beta, {\cal H}_{\rm B/F} \} = \xi_j\{ \phi_\beta, {\cal H}_{\rm B/F} \} + \phi_\beta \{\xi_j, {\cal H}_{\rm B/F} \}$. The first term was evaluated above, in Eqs.~\eqref{eqn:PB_1}-\eqref{eqn:PB_2}, while the second term can be written as $\{\xi_j, {\cal H}_{\rm B/F} \} = \sum_k X^{\rm B/F}_{jk}\xi_k$, where
\begin{align}
    X_{jk}^{\rm B} = \begin{bmatrix}
    0 & 2 & 0\\
    -\gamma_{\rm B} & 0 & 1\\
    0 & -2\gamma_{\rm B} & 0
    \end{bmatrix},
    \qquad
    X_{jk}^{\rm F} = \displaystyle \begin{bmatrix}
    0 & 2m_{\rm B}/m_{\rm F} & 0\\
    -\gamma_{\rm F} & 0 & m_{\rm B}/m_{\rm F} \\
    0 & -2\gamma_{\rm F} & 0
    \end{bmatrix}.
\end{align}
Following the above form for the Poisson brackets, we write $({\rm H}_{\rm F})_{\alpha\beta}^{ij} = ({\rm H}_{\rm F})_{\alpha\beta,1}^{ij} +  ({\rm H}_{\rm F})_{\alpha\beta,2}^{ij}$, and a straightforward algebra in the spirit of preceding calculations gives:
\begin{align}
   & ({\rm H}_{\rm F}  )_{\alpha\beta,1}^{ij}   = \int d^3\bm r \frac{d^3 \bm p}{(2\pi)^3} \xi_i\xi_j \phi_\alpha \{ \phi_\beta, {\cal H}_{\rm F} \} \times n_{\rm eq, F} (1 - n_{\rm eq, F}) 
     = \frac{1}{4\pi\beta^3\sqrt{\kappa}}
    \Big\{\\
    &
    \frac{m_{\rm B}}{m_{\rm F}}\Big[ (2m_\beta \tilde{h}(|\nu_i - \nu_j|,k_\alpha + k_\beta + 1) +k _\beta \tilde{h}(|\nu_i - \nu_j|,k_\alpha + k_\beta - 1))\notag\\
    &\times 
    g\Big(p_\alpha + p_\beta + \frac{r_\alpha + r_\beta }{2},m_\alpha + m_\beta + \frac{k_\alpha + k_\beta + \mu_i + \mu_j - 1 }{2} \Big)   
    g\Big(q_\alpha + q_\beta + \frac{r_\alpha + r_\beta }{2},n_\alpha + n_\beta + \frac{k_\alpha + k_\beta + \nu_i + \nu_j + 1 }{2} \Big)\notag\\
    & 
    +
    \sqrt{\kappa}(2p_\beta + r_\beta) \tilde{h}(|\nu_i - \nu_j|,k_\alpha + k_\beta)
    \notag\\
    &
    \times
    g\Big(p_\alpha + p_\beta + \frac{r_\alpha + r_\beta - 1}{2},m_\alpha + m_\beta + \frac{k_\alpha + k_\beta  +\mu_i + \mu_j}{2} \Big)
    g\Big(q_\alpha + q_\beta + \frac{r_\alpha + r_\beta +1 }{2},n_\alpha + n_\beta + \frac{k_\alpha + k_\beta + \nu_i + \nu_j  }{2} \Big) \Big]
    \notag\\
    &
    \times
    \int \limits_0^\infty dr \, r^{2 m_\alpha + 2 m_\beta + k_\alpha + k_\beta + 2p_\alpha + 2p_\beta + r_\alpha + r_\beta + \mu_i + \mu_j + 1}
    \int \limits_0^\infty dp \, p^{2 n_\alpha + 2 n_\beta + k_\alpha + k_\beta + 2q_\alpha + 2q_\beta + r_\alpha + r_\beta + \nu_i + \nu_j + 3} \, n_{\rm F,eq}(1 - n_{\rm F,eq}) \notag\\
    &
    - \Big[(2n_\beta \tilde{h}(|\nu_i - \nu_j|,k_\alpha + k_\beta + 1) + k_\beta \tilde{h}(|\nu_i - \nu_j|,k_\alpha + k_\beta - 1))\notag\\
    &\times
    g\Big(p_\alpha + p_\beta + \frac{r_\alpha + r_\beta }{2},m_\alpha + m_\beta + \frac{k_\alpha + k_\beta + \mu_i + \mu_j + 1 }{2} \Big)
    g\Big(q_\alpha + q_\beta + \frac{r_\alpha + r_\beta }{2},n_\alpha + n_\beta + \frac{k_\alpha + k_\beta + \nu_i + \nu_j - 1 }{2} \Big) \notag\\
    &
    + \sqrt{\kappa}(2q_\beta + r_\beta) \tilde{h}(|\nu_i - \nu_j|,k_\alpha + k_\beta)\notag
    \notag\\
    &
    \times
    g\Big(p_\alpha + p_\beta + \frac{r_\alpha + r_\beta  + 1}{2},m_\alpha + m_\beta + \frac{k_\alpha + k_\beta + \mu_i + \mu_j }{2} \Big)
    g\Big(q_\alpha + q_\beta + \frac{r_\alpha + r_\beta - 1}{2},n_\alpha + n_\beta + \frac{k_\alpha + k_\beta + \nu_i + \nu_j }{2} \Big) \Big]
    \notag\\
    & 
    \times
    \int \limits_0^\infty dr \, r^{2 m_\alpha + 2 m_\beta + k_\alpha + k_\beta + 2p_\alpha + 2p_\beta + r_\alpha + r_\beta + \mu_i + \mu_j + 3} \gamma_{\rm F}(r)
    \int \limits_0^\infty dp \, p^{2 n_\alpha + 2 n_\beta + k_\alpha + k_\beta + 2q_\alpha + 2q_\beta + r_\alpha + r_\beta + \nu_i + \nu_j + 1} \, n_{\rm F,eq}(1 - n_{\rm F,eq}) \notag \Big\}.
\end{align}
We also get:
\begin{align}
     ({\rm H}_{\rm F}  & )_{\alpha\beta,2}^{ij}  = \int d^3\bm r \frac{d^3 \bm p}{(2\pi)^3} \xi_i \phi_\alpha \phi_\beta \{ \xi_j, {\cal H}_{\rm F} \} \times n_{\rm eq, F} (1 - n_{\rm eq, F})=  \frac{1}{4\pi\beta^3\sqrt{\kappa}}\sum_k 
     \tilde{h}(|\nu_i - \nu_k|, k_\alpha + k_\beta) \\
    &
     \times
      g\Big( p_\alpha + p_\beta + \frac{r_\alpha + r_\beta}{2}, m_\alpha + m_\beta + \frac{k_\alpha + k_\beta + \mu_i + \mu_k}{2} \Big) 
    g\Big( q_\alpha + q_\beta + \frac{r_\alpha + r_\beta}{2}, n_\alpha + n_\beta + \frac{k_\alpha + k_\beta + \nu_i + \nu_k}{2} \Big)\notag\\
    &\times
    \int \limits_0^\infty dr \, r^{2 m_\alpha + 2 m_\beta + k_\alpha + k_\beta + 2p_\alpha + 2p_\beta + r_\alpha + r_\beta + \mu_i + \mu_k + 2} X_{jk}^{\rm F}(r)\notag\\
    &
    \times
    \int \limits_0^\infty dp \, p^{2 n_\alpha + 2 n_\beta + k_\alpha + k_\beta + 2q_\alpha + 2q_\beta + r_\alpha + r_\beta + \nu_i + \nu_k + 2} \, n_{\rm F,eq}(1 - n_{\rm F,eq}).\notag
\end{align}
Similar expressions can be obtained for the bosonic sector (or use the same expressions but substitute: $n_{\rm F,eq}(1 - n_{\rm F,eq}) \to n_{\rm B,eq}(1 + n_{\rm B,eq})$, $\displaystyle\frac{m_{\rm B}}{m_{\rm F}} \to 1$, $\gamma_{\rm F}\to \gamma_{\rm B}$, and $X^{\rm F}_{jk} \to X^{\rm B}_{jk}$).

\subsection{Matrix elements of $\hat{\rm S}$}

We note that the self-energies entering the matrix $\hat{\rm S}$ in Eq.~\eqref{eqn:Sigma_gen} now depend not only on $\rho$ but also on the angle $\phi$ between $\bm \rho$ and the $x$-axis:
\begin{align}
    {\rm \Sigma}_{\rm B}[\Delta_{\rm B}\xi_j\phi_\beta] (\rho,z,\phi) = 2g_{\rm B} \int \frac{d^3 \bm p}{(2\pi)^3} n_{\rm B,eq}(1 + n_{\rm B,eq}) \xi_j \phi_\beta \to \frac{2g_{\rm B}}{\beta^{3/2}} \cos{(2\phi)} \rho^{2m_\beta + k_\beta + \mu_j} z^{2p_\beta + r_\beta} v_{{\rm B},\beta j}(r),
\end{align}
where we rescaled the variables as $\rho \sqrt{\beta } \to \rho$, $ z \sqrt{\beta \kappa} \to z $, and $ p\sqrt{\beta }  \to p$ and introduced:
\begin{align}
    v_{{\rm B/F},\beta j}(r) \equiv \frac{1}{(2\pi)^2} \tilde{h}(\nu_j, k_\beta) g\Big( q_\beta + \frac{r_\beta}{2}, n_\beta + \frac{k_\beta + \nu_j}{2} \Big) \int\limits_0^\infty dp\, p^{2n_\beta + k_\beta + 2 q_\beta + r_\beta + \nu_j + 2} \, n(r,p)(1 \pm n(r,p)).
\end{align}
Following the Poisson bracket expansion (see the preceding subsection), we write $({\rm S}_{\rm B})_{\alpha\beta}^{ij} = ({\rm S}_{\rm B})_{\alpha\beta,1}^{ij} + ({\rm S}_{\rm B})_{\alpha\beta,2}^{ij}$, where:
\begin{align}
    ( & {\rm S}_{\rm B})_{\alpha\beta,1}^{ij}     = \frac{2g_{\rm B}}{4\pi \beta^{7/2}\sqrt{\kappa}} \Big\{
    \Big[ (2m_\alpha \tilde{h}(\nu_i,k_\alpha + 1) + k_\alpha \tilde{h}(\nu_i,k_\alpha - 1))\\
    &
    \times 
    g\Big(q_\alpha + \frac{r_\alpha}{2}, n_\alpha + \frac{k_\alpha + \nu_i + 1}{2} \Big)
    g\Big( p_\alpha + p_\beta + \frac{r_\alpha + r_\beta}{2}, m_\alpha + m_\beta + \frac{k_\alpha + k_\beta + \mu_i + \mu_j - 1}{2} \Big)\notag\\
    &
    +\sqrt{\kappa} (2p_\alpha + r_\alpha)\tilde{h}(\nu_i,k_\alpha)
    g\Big( q_\alpha + \frac{r_\alpha + 1}{2}, n_\alpha + \frac{k_\alpha + \nu_i}{2} \Big) g\Big( p_\alpha + p_\beta + \frac{r_\alpha + r_\beta - 1}{2}, m_\alpha + m_\beta + \frac{k_\alpha + k_\beta + \mu_i + \mu_j}{2} \Big)
    \Big]\notag\\
    &\times
    \int \limits_0^\infty dr \, r^{2 m_\alpha + 2 m_\beta + k_\alpha + k_\beta + 2p_\alpha + 2p_\beta + r_\alpha + r_\beta + \mu_i + \mu_j + 1} v_{\rm B,\beta}
    \int \limits_0^\infty dp \, p^{2 n_\alpha  + k_\alpha + 2q_\alpha + r_\alpha + \nu_i + 3} \, n_{\rm B,eq}(1 + n_{\rm B,eq}) \notag\\
    & 
    - \Big[ (2n_\alpha \tilde{h}(\nu_i,k_\alpha + 1) + k_\alpha \tilde{h}(\nu_i,k_\alpha - 1))\notag\\
    &
    \times
    g\Big(q_\alpha + \frac{r_\alpha}{2}, n_\alpha + \frac{k_\alpha + \nu_i - 1}{2} \Big)
    g\Big( p_\alpha + p_\beta + \frac{r_\alpha + r_\beta}{2}, m_\alpha + m_\beta + \frac{k_\alpha + k_\beta + \mu_i + \mu_j + 1}{2} \Big) \notag\\
    &
    +\sqrt{\kappa} (2q_\alpha + r_\alpha)\tilde{h}(\nu_i,k_\alpha)
    g\Big( q_\alpha + \frac{r_\alpha - 1}{2}, n_\alpha + \frac{k_\alpha + \nu_i}{2} \Big) g\Big( p_\alpha + p_\beta + \frac{r_\alpha + r_\beta + 1}{2}, m_\alpha + m_\beta + \frac{k_\alpha + k_\beta + \mu_i + \mu_j}{2} \Big)
    \Big]\notag\\
    &\times
    \int \limits_0^\infty dr \, r^{2 m_\alpha + 2 m_\beta + k_\alpha + k_\beta + 2p_\alpha + 2p_\beta + r_\alpha + r_\beta + \mu_i + \mu_j + 3} v_{\rm B,\beta}\gamma_{\rm B}
    \int \limits_0^\infty dp \, p^{2 n_\alpha  + k_\alpha + 2q_\alpha + r_\alpha + \nu_i + 1} \, n_{\rm B,eq}(1 + n_{\rm B,eq})
    \Big\}\notag
\end{align}
and
\begin{align}
    ( & {\rm S}_{\rm B})_{\alpha\beta,2}^{ij} =   \frac{2 g_{\rm B}}{4\pi\beta^{7/2}\sqrt{\kappa}}\sum_k 
     \tilde{h}( \nu_k, k_\alpha)
      g\Big( p_\alpha + p_\beta + \frac{r_\alpha + r_\beta}{2}, m_\alpha + m_\beta + \frac{k_\alpha + k_\beta + \mu_k + \mu_j}{2} \Big) 
    g\Big( q_\alpha + \frac{r_\alpha}{2}, n_\alpha + \frac{k_\alpha + \nu_k}{2} \Big)\notag\\
    &\times
    \int \limits_0^\infty dr \, r^{2 m_\alpha + 2 m_\beta + k_\alpha + k_\beta + 2p_\alpha + 2p_\beta + r_\alpha + r_\beta + \mu_k + \mu_j + 2} X_{ik}^{\rm B} v_{{\rm B},\beta j}
    \int \limits_0^\infty dp \, p^{2 n_\alpha + k_\alpha + 2q_\alpha + r_\alpha + \nu_k + 2} \, n_{\rm B,eq}(1 + n_{\rm B,eq}).
\end{align}
Similar expressions can be obtained for $({\rm S}_{\rm BF})_{\alpha\beta}$ and $({\rm S}_{\rm FB})_{\alpha\beta}$.

\subsection{Matrix elements of $\hat{\rm I}$}

We compute the matrix $\hat{\rm I}$ in the same manner as in the preceding Appendix. The only new element we encounter here is that the polynomial integrands entering collision matrices now depend on the angle $\phi$ between $\bm \rho$ and the $x$-axis. To overcome this issue, we write (in the rescaled variables) expressions of the type: $S[\phi_\alpha\xi_i] \to S_{\alpha,i}[\bm P,\bm q,\bm q',\phi] \rho^{2m_\alpha + k_\alpha +\mu_i}z^{2p_\alpha + r_\alpha}$, where we also switched to the center-of-mass frame introduced in Appendix~\ref{appendix:coll_int}. These new functions $S_{\alpha,i}[\bm P,\bm q,\bm q',\phi]$ are polynomials of $\cos\phi$ and $\sin\phi$ allowing for efficient analytical/numerical computations of
\begin{align}
    S_{\alpha\beta}^{ij}[\bm P,\bm q,\bm q'] \equiv \int \frac{d\phi}{2\pi} S_{\alpha,i}[\bm P,\bm q,\bm q',\phi] S_{\beta,j}[\bm P,\bm q,\bm q',\phi].
\end{align}
In our numerics, we write a {\it Mathematica} subroutine that evaluates the functions $S_{\alpha\beta}^{ij}[\bm P,\bm q,\bm q']$, which are now polynomials of $\bm P$, $\bm q$, and $\bm q'$. Plugging these polynomials back into, for instance, the matrix ${\rm I}_{\rm B}$, we arrive at (compare to Eq.~\eqref{eqn:I_B_v0}):
\begin{align}
({\rm I}_{\rm B})_{\alpha\beta}^{ij} = & - \frac{{g}^2_{\rm B} }{2\beta^5 \sqrt{\kappa}}  \int d^3\bm r \, \rho^{2m_\alpha + 2m_\beta + k_\alpha + k_\beta + \mu_i + \mu_j} z^{2p_\alpha + 2p_\beta + r_\alpha + r_\beta }\int \frac{d^3 \bm P}{(2\pi)^3} \frac{d^3 \bm q}{(2\pi)^3}\frac{d^3 \bm q'}{(2\pi)^3}  (2\pi)\delta(q^2 - q'^2)\notag\\
    &
    \times (1 + n_{\rm B, eq}(\bm r,\bm p))(1 + n_{\rm B, eq}(\bm r,\bm p'))n_{\rm B, eq}(\bm r,\bm p_1) n_{\rm B, eq}(\bm r,\bm p_1') \times S_{\alpha\beta}^{ij}[\bm P, \bm q, \bm q']. \label{eqn:I_B_v0_quad}
\end{align}
The rest of the computation, including the computation of the matrix ${\rm \hat{J}}$ which accounts for collisions between bosons and fermions, follows the procedure in Appendix~\ref{Appendix_monopole} step-by-step.

\section{Benchmarking and applications of collision integral calculations}
\label{appendix: benchmarking}

\begin{figure*}[b!]
\centering
    \includegraphics[width=0.75\linewidth]{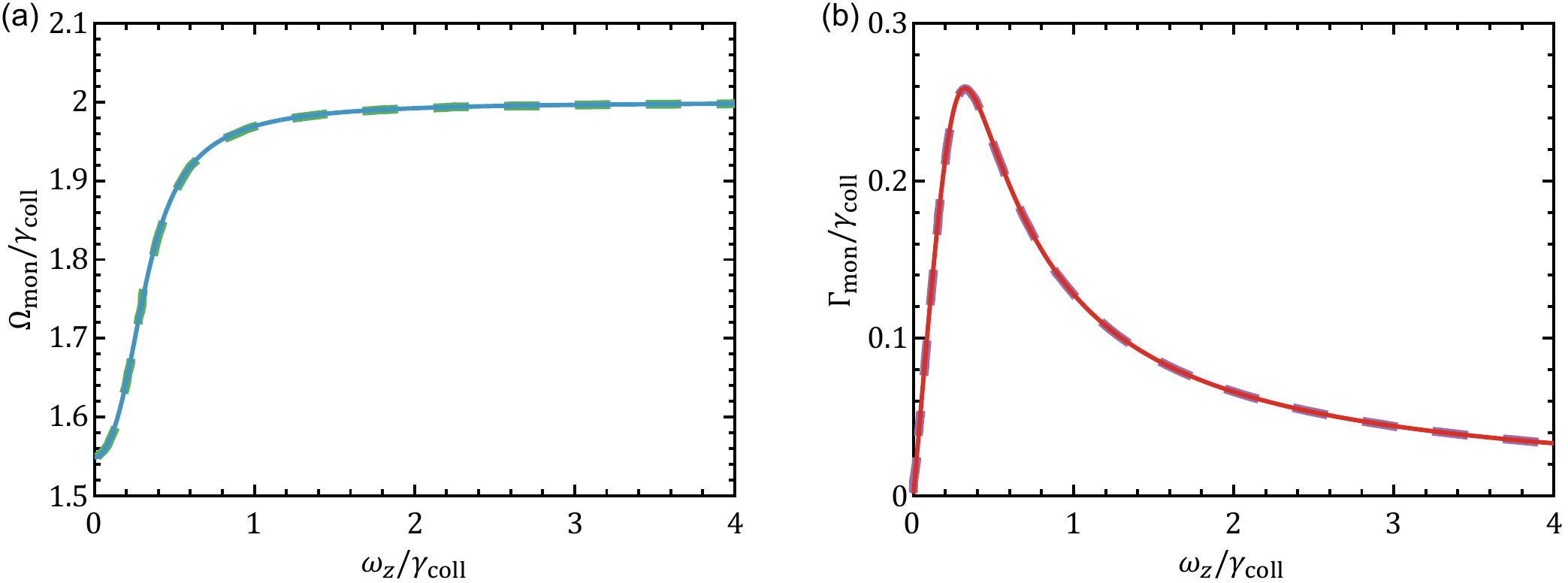} 
\caption{ Monopole mode of a classical gas, trapped into a harmonic potential with $\kappa = 10^{-2}$. Solid lines in (a) and (b) are the monopole mode frequency $\Omega_{\rm mon}$ and its dampening $\Gamma_{\rm mon}$, respectively, computed using the approach developed here. Dashed lines were extracted from Ref.~\cite{PhysRevA.60.4851}. Similarly to our results in the main text, cf. Fig.~\ref{fig::bosons_monopole}, the classical gas also exhibits a collisionless-to-hydrodynamic crossover, as evidenced, for instance, by the non-monotonic behavior of $\Gamma_{\rm mon}$.
}
\label{fig::benchmark_cl}
\end{figure*}


\subsection{Monopole mode of a classical gas}

A simple and straightforward way to benchmark the method developed in this work is to consider a classical gas trapped in a harmonic potential. Such a gas at equilibrium follows the Boltzmann-Maxwell distribution function, which is Gaussian. The Hartree-Fock self-energy of the gas can also be neglected. Correspondingly, in the GMM ansatz in Eq.~\eqref{eqn:GGM_gen}, we only need to keep track of a single harmonic ($M_0 = 1$). In case of a dilute gas, we can further approximate $1 + n_{\rm eq} \approx 1$, which simplifies the calculations below.

We focus on the monopole mode and consider the truncation order to be $M = 1$, so that in total we have 7 basis functions for the method of moments. All of the matrices entering the kinetic equation~\eqref{eqn:MoM} can  be computed analytically:
\begin{align}
    \frac{{\rm M}_{\alpha\beta}}{N_{\rm tot}} = \begin{bmatrix}
    1 &  \displaystyle\frac{2}{\beta} & \displaystyle\frac{2}{\beta} & 0 & \displaystyle\frac{1}{\kappa\beta} & \displaystyle\frac{1}{\beta} & 0\\
    \displaystyle\frac{2}{\beta} & \displaystyle\frac{8}{\beta^2} & \displaystyle\frac{4}{\beta^2} & 0 & \displaystyle\frac{2}{\kappa\beta^2} & \displaystyle\frac{2}{\beta^2} & 0\\
    \displaystyle\frac{2}{\beta} & \displaystyle\frac{4}{\beta^2} & \displaystyle\frac{8}{\beta^2} & 0 & \displaystyle\frac{2}{\kappa\beta^2}  & \displaystyle\frac{2}{\beta^2} & 0\\
    0 & 0 & 0 & \displaystyle\frac{2}{\beta^2} & 0 & 0  & 0 \\
    \displaystyle\frac{1}{\kappa\beta} & \displaystyle\frac{2}{\kappa\beta^2} & \displaystyle\frac{2}{\kappa\beta^2} & 0 & \displaystyle\frac{3}{\kappa^2\beta^2} & \displaystyle\frac{1}{\kappa\beta^2} & 0\\
    \displaystyle\frac{1}{\beta}  & \displaystyle\frac{2}{\beta^2} & \displaystyle\frac{2}{\beta^2} & 0 & \displaystyle\frac{1}{\kappa\beta^2} &  \displaystyle\frac{3}{\beta^2} & 0 \\
    0 & 0 & 0 & 0 & 0 & 0 & \displaystyle\frac{1}{\kappa\beta^2} 
    \end{bmatrix},\quad
    \frac{{\rm H}_{\alpha\beta}}{N_{\rm tot}} = \begin{bmatrix}
    0 & 0 & 0 & 0 & 0 & 0 & 0\\
    0 & 0 & 0 & -\displaystyle\frac{4}{\beta^2} & 0 & 0 & 0\\
    0 & 0 & 0 & \displaystyle\frac{4}{\beta^2}  & 0 & 0 & 0\\
    0 & \displaystyle \frac{4}{\beta^2} &  -\displaystyle \frac{4}{\beta^2}& 0 & 0 & 0 & 0\\
    0 & 0 & 0 & 0 & 0 & 0 & -\displaystyle\frac{2}{\kappa \beta^2}\\
    0 & 0 & 0 & 0 & 0 & 0 & \displaystyle\frac{2}{ \beta^2}\\
    0 & 0 & 0 & 0 & \displaystyle\frac{2}{\kappa\beta^2} & -\displaystyle\frac{2}{\beta^2} & 0
    \end{bmatrix},
\end{align}
where $N_{\rm tot} =  e^{\mu\beta}/(\beta^3\sqrt{\kappa})$ is the total number of particles and $\mu$ is the chemical potential. Note that following the main text notations, we set $m = 1$ in this subsection. The collision integral matrix ${\rm I}_{\alpha\beta}$ has only four nonzero matrix elements, namely: ${\rm I}_{33} = {\rm I}_{66} = N_{\rm tot} I$ and ${\rm I}_{36} = {\rm I}_{63} =-  N_{\rm tot} I$, where
\begin{align}
    I \approx & - \frac{{g}^2_{\rm B}}{2N_{\rm tot}}  \int d^3\bm r\frac{d^3 \bm p}{(2\pi)^3} \frac{d^3 \bm p'}{(2\pi)^3}\frac{d^3 \bm p_1}{(2\pi)^3}\frac{d^3 \bm p'_1}{(2\pi)^3} \times (2\pi)^3\delta(\bm p + 
    \bm p' - \bm p_1 - \bm p_1') \notag\\
    &\qquad\qquad\qquad\qquad\qquad\qquad\qquad
    \times (2\pi)\delta(\varepsilon_{p} + \varepsilon_{p'} - \varepsilon_{p_1} -\varepsilon_{p_1'})
    \times n_{\rm eq}(\bm r,\bm p_1) n_{\rm eq}(\bm r,\bm p_1') \times S[p_z^2]S[p_z^2].
\end{align}
By switching to the center-of-mass frame following Appendix~\ref{appendix:coll_int}, we get:
\begin{align}
    I \approx & - \frac{{g}^2_{\rm B}}{2N_{\rm tot}}  \int d^3\bm r \frac{d^3 \bm P}{(2\pi)^3} \frac{d^3 \bm q}{(2\pi)^3}\frac{d^3 \bm q'}{(2\pi)^3} \times \frac{\pi}{q} \delta(q - q') n_{\rm eq}(\bm r,\bm p_1) n_{\rm eq}(\bm r,\bm p_1') \times S[p_z^2]S[p_z^2].
\end{align}
Integration over the real-space gives: 
\begin{align}
    \int d^3\bm r \, n_{\rm eq}(\bm r,\bm p_1) n_{\rm eq}(\bm r,\bm p_1') = \frac{\pi^{\frac{3}{2}} e^{2\mu\beta}}{\beta^{\frac{3}{2}}\kappa^{\frac{1}{2}}} \exp
\left\{-\beta q^2 -\frac{1}{4}\beta P^2\right\}.
\end{align}
We note that $S[p_z^2] = 2(q_z^2 - q_z'^2)$, i.e., it does not depend on $\bm P$, which can now be easily integrated out:
\begin{align}
    \int \frac{d^3 \bm P}{(2\pi)^3} \exp\left\{-\frac{1}{4}\beta P^2\right\}  = \left(\frac{1}{\pi\beta}\right)^{\frac{3}{2}}.
\end{align}
The remaining integration is straightforward:
\begin{align}
    I \approx -\frac{4}{15\pi^3} \frac{{g}_{\rm B}^2 e^{\mu\beta}}{\beta^4} = -\frac{8}{15  \beta^2} \sigma n_0 v_T= -\frac{16}{15 \beta^2 } \gamma_{\rm coll},
\end{align}
where $v_T = \sqrt{8/(\pi \beta)}$ is the thermal velocity, $n_0$ is the density at the center of the trap, and $\gamma_{\rm coll} = \sigma n_0 v_T /2$.

Our numerical subroutines reproduce the above analytical matrices and reproduce the known results from Ref.~\cite{PhysRevA.60.4851}, as shown in Fig.~\ref{fig::benchmark_cl}. Specifically, Ref.~\cite{PhysRevA.60.4851} demonstrated that the method of moments with $M = 1$ applied to a classical thermal gas agrees remarkably well with more precise molecular dynamics simulations. From this result, one can also conclude that $M = 1$ is expected to be an excellent approximation. If one wants to study a quantum gas where the equilibrium distribution function is no longer Gaussian (as done in the main text), one may need to consider $M\geq 1$. In the main text, we fixed $M = 2$ and checked that further increasing $M$ does not result in an appreciable difference in the resulting spectral functions.

\subsection{Bose polaron decay rate}

\begin{figure*}[t!]
\centering
    \includegraphics[width=0.5\linewidth]{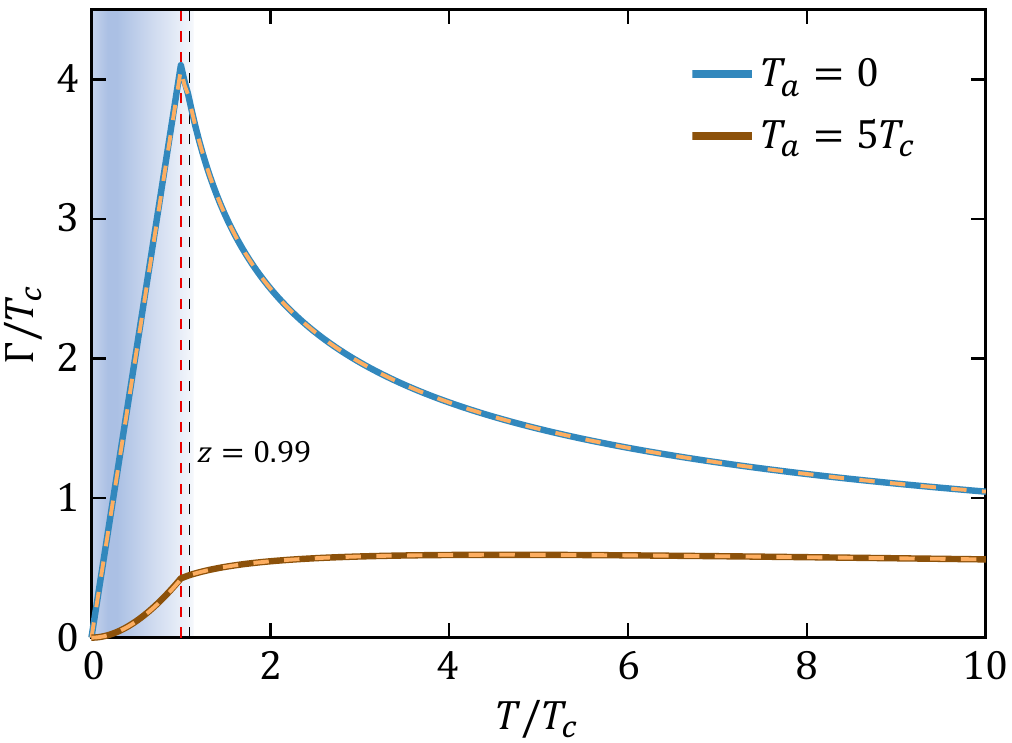} 
\caption{ The fermion (Bose polaron) decay rate as a function of temperature for unitary ($T_a = 0$) and non-unitary Bose-Fermi interactions ($T_a = 5 T_c$). Solid curves correspond to resuming the Taylor series expansion, Eq.~\eqref{eqn:TS_GAMMA} derived in Ref.~\cite{yan2020bose}. Dashed curves correspond to the GMM ansatz, Eq.~\eqref{eqn:GMM_GAMMA}, where the maximal bosonic fugacity is chosen to be $z_{\rm max} = 0.99$ (black dashed vertical line) so that we have $M_0 = 17$ Gaussians in total, cf. Fig.~\ref{fig::GMM fitting}(b). The domain of applicability of the GMM approach is $T\gtrsim 1.1T_c$, where the two types of calculations essentially coincide. Notably, the GMM predictions for $T \lesssim 1.1 T_c$ (shaded region) are also in remarkable agreement with the more accurate Taylor series resummation. 
}
\label{fig::Gamma_T}
\end{figure*}

In this subsection, we demonstrate that the GMM approach can be applied to experimentally relevant quantum many-body systems with strong interactions; in these systems, one cannot disregard the momentum dependence of the scattering amplitude. We follow  Ref.~\cite{yan2020bose} and consider the impurity limit where we have a single fermion (Bose polaron)  immersed into a homogeneous bosonic bath. The fermion decay rate $\Gamma$ in this case can be estimated as~\cite{yan2020bose}:
\begin{align}
    \Gamma = \frac{\lambda_{\rm F}^3}{m_r} \int \frac{d^3 \bm k_{\rm B}}{(2\pi)^3}\int \frac{d^3 \bm k_{\rm F}}{(2\pi)^3} e^{-\beta k_{\rm F}^2/(2m_{\rm F}) } n_{\rm B}(k_{\rm B}) \sigma(k) k,
\end{align}
where $ \bm v_{\rm rel} = \bm k/m_r = \bm k_{\rm F}/m_{\rm F} - \bm k_{\rm B}/m_{\rm B}$, $\bm k$ is the relative momentum between the two scattering particles, and $m_r = m_{\rm F}m_{\rm B}/(m_{\rm F} + m_{\rm B})$ is the reduced mass. We neglect the Bose-Bose interactions, $a_{\rm B} = 0$, corresponding to an ideal Bose gas. We will also consider  temperature $T$ to be both below and above $T_c$. Below $T_c$, the fermion decay rate comes from scatterings off the thermal bosons. For $T\leq T_c$, the bosonic chemical potential is zero $\mu_{\rm B} = 0$ ($z_{\rm B} = 1$). Since the Bose-Fermi coupling can be strong, we take into account the $k$-dependence of the scattering cross section:
\begin{align}
    \sigma(k) = \frac{4\pi a_{\rm BF}^2}{1 + k^2 a_{\rm BF}^2},
\end{align}
where $a_{\rm BF}$ is the Bose-Fermi scattering length.
At this stage, we employ the GMM expansion:
\begin{align}
    n_{\rm B}(p) \approx \sum_{s = 1}^{M_0} a_{{\rm B},s} \exp\Big( -\frac{\beta p^2}{2m_{\rm B} \gamma^2_{{\rm B},s}} \Big).
\end{align}
In what follows, we will avoid writing the subscript ${\rm B}$ to ease the notations, as it is clear that the GMM ansatz in this subsection refers solely to bosons. The scattering rate $\Gamma$ is then written as:
\begin{align}
    \Gamma \approx \frac{\lambda_{\rm F}^3}{m_r} \sum_{s = 1}^{M_0} a_{s} \int \frac{d^3 \bm k_{\rm B}}{(2\pi)^3}\int \frac{d^3 \bm k_{\rm F}}{(2\pi)^3} \frac{4\pi a_{\rm BF}^2 k}{1 + k^2 a_{\rm BF}^2} \exp{\Big[ -\frac{\beta k_{\rm F}^2 }{2m_{\rm F}}  -\frac{\beta {k}_{\rm B}^2 }{2m_{\rm B} \gamma^2_{s}} \Big]}.
\end{align}
To evaluate this integral, we introduce the following change of variables from $\bm k_{\rm F}$ and $\bm k_{\rm B}$ to $\bm K$ and $\bm k$:
\begin{align}
\begin{cases}
    \bm k_{\rm F} = \displaystyle\frac{\alpha}{1 + \alpha} \bm K + \frac{1 + \alpha}{1 + \alpha \gamma^2_{s}} \bm k\\
    \bm k_{\rm B} = \displaystyle\frac{1}{1 + \alpha} \bm K - \frac{1 + \alpha}{1 + \alpha \gamma^2_{s}} \gamma^2_{s}\bm k
\end{cases}
\end{align}
Here $\alpha = m_{\rm F}/m_{\rm B}$. This linear transformation (i) has unity Jacobian, (ii) respects that $\bm k$ is the relative momentum, and (iii) allows us to directly integrate out $\bm K$. After simple algebra, we obtain:
\begin{align}
    \Gamma = \frac{v_{\rm rel}}{\lambda_{\rm B}^3} \sum_{s = 1}^{M_0} a_{s} \gamma_{s}^3  \sqrt{\frac{1 + \alpha\gamma_{s}^2}{1 + \alpha}}\tilde{\sigma}\Big(\frac{T}{T_a}\frac{1 + \alpha\gamma_{s}^2}{1 + \alpha}\Big), \label{eqn:GMM_GAMMA}
\end{align}
where $v_{\rm rel} = \sqrt{8 T/(\pi m_r)}$, $T_a = 1/(2 m_r a_{\rm BF}^2)$, and
\begin{align}
    \tilde{\sigma}(y) = 8\pi a^2 \int_0^\infty d x \, x^3 \, [1 + y x^2]^{-1} \, e^{-x^2}.
\end{align}
If one were to use a Taylor series expansion instead, then one should substitute:
\begin{gather}
    a_{s} \to z^s,\, \gamma_{s} \to \frac{1}{\sqrt{s}},\, M_0\to \infty\Rightarrow
    \Gamma_{\rm TS} = \frac{v_{\rm rel}}{\lambda_{\rm B}^3} \sum_{s = 1}^{\infty} \frac{z^s}{s^{3/2}} \sqrt{\frac{1}{s} \frac{s + \alpha}{1 + \alpha} }\tilde{\sigma}\Big(\frac{T}{T_a}\frac{1}{s} \frac{s + \alpha}{1 + \alpha}\Big). \label{eqn:TS_GAMMA}
\end{gather}
The result in Eq.~\eqref{eqn:TS_GAMMA} exactly reproduces Eq.~(32) of Ref.~\cite{yan2020bose}.

The physics of the Bose polaron decay rate is quite fascinating, and the interested reader is referred to the thorough discussion in Ref.~\cite{yan2020bose}. Here, we focus on the validity of our GMM calculations. Figure~\ref{fig::Gamma_T} shows the comparison between the GMM result in Eq.~\eqref{eqn:GMM_GAMMA} and the Taylor series expression~\eqref{eqn:TS_GAMMA} for both unitary ($T_a = 0$) and non-unitary interactions ($T_a = 5T_c$). For the GMM ansatz, we set the maximal bosonic fugacity to be $z = 0.99$ so that we have $M_0 = 17$ Gaussians in total, cf. Fig.~\ref{fig::GMM fitting}(b). This choice of the maximal fugacity implies that the domain of applicability of the GMM calculations is $T\gtrsim 1.1T_c$, where the two types of calculations coincide.  Even for $T \lesssim 1.1 T_c$, the two approaches agree remarkably well (see Fig.~\ref{fig::Gamma_T}). 

To get a sense of how the two calculations differ from each other for $T\lesssim 1.1 T_c$, we consider the unitary limit with $T_a = 0$. In this case, both approaches predict linear in $T$ behavior for $T\leq T_c$ (where the bosonic fugacity is $z = 1$), meaning the maximal discrepancy between the two calculations can be estimated to be 
$|\Gamma(T_c) - \Gamma_{\rm TS}(T_c)|/\Gamma_{\rm TS}(T_c) \approx 0.7\%$. This remarkable agreement, in the most challenging regime at $T = T_c$ and with unitary interactions, stems from the fact that the bosonic GMM expansion can be accurately extrapolated outside its domain of applicability, cf. Fig.~\ref{fig::GMM fitting}(b), as discussed in Sec.~\ref{sec:GMM} of the main text.

\subsection{Shear viscosity  of a strongly-interacting two-component Fermi system}

\begin{figure*}[t!]
\centering
    \includegraphics[width=0.8\linewidth]{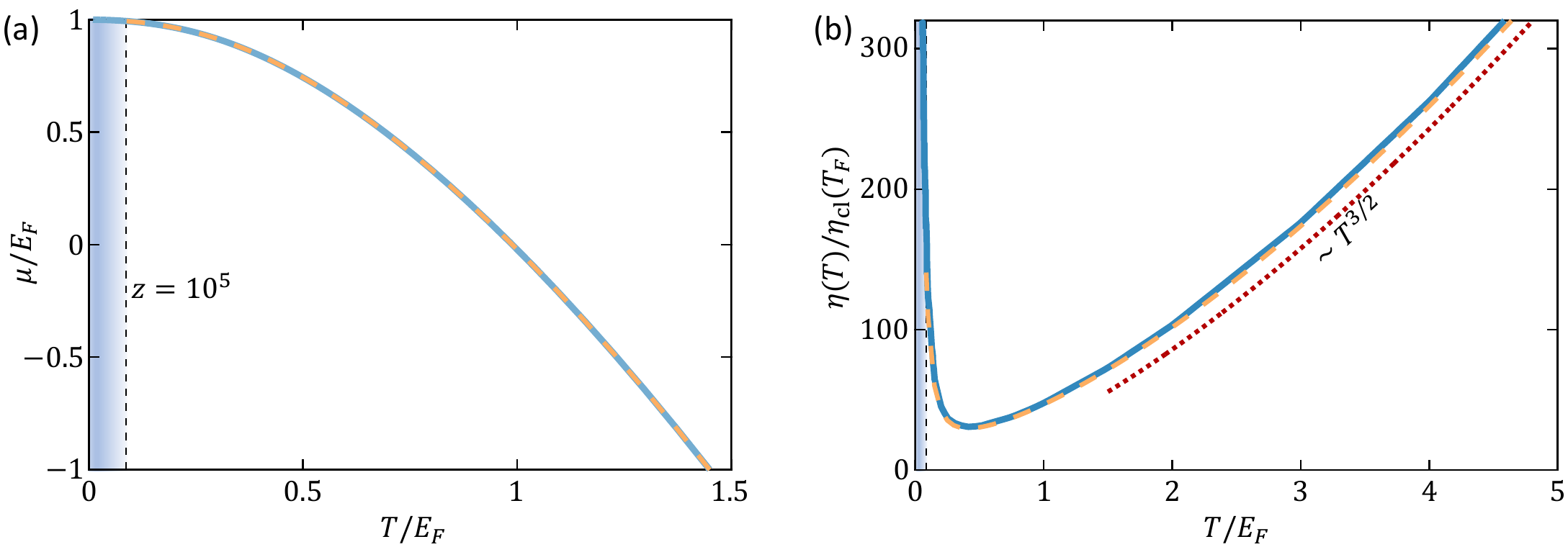} 
\caption{ Performance of the cGMM expansion on a strongly-interacting Fermi fluid. (a) The chemical potential $\mu$ as a function of $T$: the solid line encodes the exact dependence, while the dashed one corresponds to the cGMM. In our simulations, we chose $z_{\rm max} = 10^5$, which implies that the domain of applicability of the cGMM is $T\gtrsim0.1T_F$ (vertical dashed line), where the two calculation coincide. (b) The shear viscosity $\eta$ of the fluid as a function of $T$ at $k_F|a| = 4.5$ ($\eta$ is normalized by $\eta_{\rm cl}(T_F) = 5 \sqrt{mT_F}/(32a^2 \sqrt{\pi})$). The solid curve is taken from Ref.~\cite{massignan2005viscous}, while the dashed one is obtained using the cGMM, and the two results reasonably agree. The choice  $k_F|a| = 4.5$  implies that $T_a \approx 0.1 T_F$ and allows one to see  the $\sim T^{3/2}$ scaling (dotted line)  at high temperatures, $T_F,T_a \ll T$. The shaded regions in (a) and (b) are where the chosen cGMM is no longer applicable.
}
\label{fig::viscosity}
\end{figure*}

In this subsection, we demonstrate that the fermionic version of the GMM, cGMM, readily enables computation of interesting transport coefficients such as the shear viscosity. To this end, we follow Ref.~\cite{massignan2005viscous} and consider a homogeneous two-component fermionic fluid in its normal state where, within the method of moments, the shear viscosity is given by~\cite{massignan2005viscous}:
\begin{align}
    \eta = \frac{2}{T m^2} \frac{\Big[\displaystyle \int \frac{d^3 \bm p}{(2 \pi)^3} p_x^2 \,p_y^2 \, n_p (1 - n_p) \Big]^2 }{\displaystyle  \frac{1}{4}\int \frac{d^3 \bm p}{(2 \pi)^3} \int \frac{d^3 \bm p'}{(2 \pi)^3} \int d\Omega \frac{d\sigma(q)}{d\Omega} \frac{|\bm p - \bm p'|}{m} n_p n_{p'}(1 - n_{p_1})(1 - n_{p_1'}) \times (S[p_xp_y])^2}. \label{eqn:eta_def}
\end{align}
In an above expression, $n_p = [ \exp(\beta(\varepsilon_p - \mu))  + 1] ^{-1}$ is the Fermi-Dirac distribution function, $S[f] = f + f' - f_1 - f_1' \Rightarrow S[p_xp_y] = 2(q_x q_y - q'_x q'_y)$, and the differential cross section is given by ($\bm q = (\bm p - \bm p')/2$):
\begin{align}
    \frac{d\sigma(q)}{d\Omega } = \frac{a^2}{1 + q^2 a^2},
\end{align}
where $\Omega$ is the solid angle of the relative momentum $\bm q' = (\bm p_1 - \bm p_1')$ with respect to the direction of $\bm q$.

In the limit where the fermionic distribution function is approximately Gaussian ($E_F \ll T$), the shear viscosity can be computed analytically:
\begin{align}
    \eta = \frac{5}{8} \frac{\sqrt{\pi m T}}{\Bar{\sigma}(T/T_a)} \text{, where }
    {\Bar{\sigma}(y)} = \frac{4\pi a^2}{3} \int_0^\infty dx \, x^7 \, [1 + yx^2 ]^{-1} e^{-x^2}
\end{align}
where $T_a = 1/(m a^2)$. In the limit $T \ll T_a$, we have $\eta_{\rm cl} = 5 \sqrt{mT}/(32a^2 \sqrt{\pi}) \sim \sqrt{T}$. In the opposite limit, $T_a \ll T$, we get $\eta = 15 (m T)^{3/2}/(32 \sqrt{\pi})\sim T^{3/2}$ (dotted line in Fig.~\ref{fig::viscosity}(b)).

At lower temperatures, where the distribution function is no longer Gaussian, we can employ the cGMM expansion:
\begin{align}
    n_p \approx\sum_s a_s \exp\Big( - \frac{\beta p^2}{2m \gamma_s^2} \Big).
    \label{eqn:cGMM_visc}
\end{align}
which is valid for any $z = \exp(\beta \mu) \leq z_{\rm max}$. We choose $z_{\rm max} = 10^5$, cf. Fig.~\ref{fig::GMM fitting}(c). This choice limits the lowest temperature in our analysis to be  $T_{\rm min}\approx 0.1 E_F$ (for this estimate, we used $\mu = E_F$). If one is interested in even lower temperatures, then one should increase $z_{\rm max}$ and redo the cGMM fitting following Sec.~\ref{sec:GMM}.
As a first step, we solve for the temperature dependence of the chemical potential $\mu(T)$, which can be obtained by fixing the total fermionic density:
\begin{align}
    n_{\rm tot} = 2\int \frac{d^3 \bm p}{(2\pi)^3} n_{p} = - 2\left(\frac{m T}{2\pi \beta}\right)^{\frac{3}{2}}\xi_{\frac{3}{2}}(-z). \label{eqn:mu}
\end{align}
Here, the factor of $2$ encodes that we have a two-component Fermi fluid. The momentum integration in Eq.~\eqref{eqn:mu} can also be carried out using Eq.~\eqref{eqn:cGMM_visc}, allowing us to determine the chemical potential within the cGMM approximation. We find that within the applicability of the cGMM expansion ($T \gtrsim 0.1 E_F$), the two calculations for $\mu(T)$ coincide (see Fig.~\ref{fig::viscosity}(a)).

The primary challenge we encounter when analyzing Eq.~\eqref{eqn:eta_def} is the evaluation of the denominator which, in the center-of-mass frame, is proportional to: 
\begin{align}
    I = \displaystyle  \int \frac{d^3 \bm P}{(2 \pi)^3} \int \frac{d^3 \bm q}{(2 \pi)^3} \int \frac{d\Omega}{4\pi}   \frac{q \, q_x^2\, q_y^2\, a^2 }{1 + q^2 a^2}  n_p n_{p'}(1 - n_{p_1})(1 - n_{p_1'}). \label{eqn:I}
\end{align}
We normalize this integral using the corresponding classical expression, where one  additionally neglects the momentum-dependence of the scattering cross section:
\begin{align}
    I_{\rm cl} = \displaystyle e^{2\beta\mu}  \int \frac{d^3 \bm P}{(2 \pi)^3} \int \frac{d^3 \bm q}{(2 \pi)^3} \int \frac{d\Omega}{4\pi}   q \, q_x^2\, q_y^2\, a^2 \, \exp\Big\{ -\frac{\beta(p^2 + p'^2)}{2m} \Big\}.
\end{align}
As in Appendix~\ref{appendix:coll_int}, we write $I$ as a sum of four terms $I = I_1 - I_2 - I_3 + I_4$ and plug in the cGMM expansion~\eqref{eqn:cGMM_visc}. Under the summations over the cGMM indices, we will get the following type of expressions, cf. Eq.~\eqref{eqn:exp_gen}:
\begin{align}
   \frac{\displaystyle
   \int \frac{d^3 \bm q}{(2 \pi)^3} \int \frac{d\Omega}{4\pi}   \frac{q \, q_x^2\, q_y^2}{1 + q^2T/T_a}  \int \frac{d^3 \bm P}{(2 \pi)^3}
   \exp \Big\{ - \frac{P^2}{2\gamma_{P}^2} - \frac{q^2}{\gamma_q^2} + \frac{1}{\gamma_P} (A \bm q + B \bm q') \cdot \bm P \Big\}
   }{\displaystyle
   e^{2\beta\mu}  \int \frac{d^3 \bm q}{(2 \pi)^3} \int \frac{d\Omega}{4\pi}   q \, q_x^2\, q_y^2\, \int \frac{d^3 \bm P}{(2 \pi)^3}  \exp\Big\{ - \frac{P^2}{4} - q^2 \Big\}
   }. \label{eqn:ref_visc}
\end{align}
Note that we rescale all the momenta as $\beta p^2/m \to p^2$. Integration over $\bm P$ yields, cf. Eq.~\eqref{eqn:coll_int_P}:
\begin{align}
    \frac{\displaystyle\int \frac{d^3 \bm P}{(2 \pi)^3}
   \exp \Big\{ - \frac{P^2}{2\gamma_{P}^2} - \frac{q^2}{\gamma_q^2} + \frac{1}{\gamma_P} (A \bm q + B \bm q') \cdot \bm P \Big\}}{\displaystyle\int \frac{d^3 \bm P}{(2 \pi)^3} \exp\Big\{ - \frac{P^2}{4}\Big\}} = \Big(\frac{\gamma_P}{\sqrt{2}}\Big)^3 \exp \Big\{  - C_1 q^2 + C_2 \bm q \cdot \bm q' \Big\}.
\end{align}
Integration over the solid angle $\Omega$ yields:
\begin{align}
   \Big(\frac{\gamma_P}{\sqrt{2}}\Big)^3 \int \frac{d\Omega}{4\pi}  \exp \Big\{  - C_1 q^2 + C_2 \bm q \cdot \bm q' \Big\} = \Big(\frac{\gamma_P}{\sqrt{2}}\Big)^3 e^{-C_1 q^2} \frac{\sinh(C_2 q^2)}{C_2 q^2}.
\end{align}
Plugging these results in, we rewrite Eq.~\eqref{eqn:ref_visc} as:
\begin{align}
     \frac{e^{-2\beta\mu}}{3}\Big(\frac{\gamma_P}{\sqrt{2}}\Big)^3
   \int_0^\infty dq \frac{q^7}{1 + q^2T/T_a} e^{-C_1 q^2} \frac{\sinh(C_2 q^2)}{C_2 q^2}.
\end{align}
This one-dimensional integral, together with the summation over the cGMM harmonics, is easy for numerical evaluations on a single laptop. Figure~\ref{fig::viscosity} shows the computed temperature dependence of the shear viscosity $\eta(T)$: Our result reproduces that of Ref.~\cite{massignan2005viscous}, demonstrating that the cGMM expansion allows for efficient and accurate computation of transport coefficients of strongly-interacting quantum many-body systems.

\end{document}